\def\figI{I}

\def\figflag{I}
\input epsf

\documentstyle[12pt]{report}

\begin{document}

\pagenumbering{roman}

\begin{center}

\huge A Study 

\huge of 

\huge Two Dimensional String Theory

\end{center}

\vspace{10mm}

\begin{center}
by
\end{center}

\begin{center}
Ulf H. Danielsson
\end{center}

\vspace{10mm}

\begin{center}

A Dissertation \\
Presented to the Faculty \\
of Princeton University \\
in Candidacy for the Degree \\
of Doctor of Philosophy 
\end{center}

\vspace{15mm}

\begin{center}
Recommended for Acceptance \\
by the Departement of \\
Physics \\
June 1992

\end{center}

\newpage

\begin{center}

\end{center}
\vspace{4cm}
\begin{center}

{\em Till Karolina och Oskar}

\end{center}

\newpage

\tableofcontents

\newpage

\addcontentsline{toc}{section}{Abstract}

\section*{Abstract}

This thesis is a study of two dimensional noncritical string theory.
The main tool which is used, is the matrix model.

There are several chapters. After a general introduction there follows
an introduction to the Liouville model where the fundamental issues of
its formulation are discussed. In particular, the special states are
introduced. Then, in chapter three, some calculations of partition
functions on genus one are given. These use field theory techniques. The
results are compared with the matrix model. In chapter four the matrix
model itself is introduced. Some of the concepts and relations which are
used in later chapters are explained. Chapters five and six include
comments on two important subjects: nonperturbative issues and string
theory at finite radius. Chapter seven is devoted to zero momentum
correlation functions as calculated in the matrix model. One important
result is a set of recursion relations. Chapter eight extends the
treatment to nonzero momentum. The main result is a clear
identification of the special states. The chapter also includes some
comments on the Wheeler de Witt equation. Chapter nine introduces the
matrix model $W_{\infty}$ algebra. This organizes the results of
previous chapters. In particular, a simple derivation of the genus zero
tachyon correlation functions is given. Chapter ten extends the results
of chapter nine to higher genus. It is seen how a deformation of the
algebra is responsible for much of the higher genus structure.
Some very explicit formulae are
derived. Then, in chapter eleven, the Liouville and matrix model
calculations are compared. Finally, chapter twelve is devoted to some
general conclusions.

The advisor for this work has been Prof. David Gross.

\newpage

\addcontentsline{toc}{section}{Preface}

\section*{Preface}

The thesis consists of several papers and also unpublished material. The
papers are listed in the references as \cite{lic,art1,art2,art3}.
\cite{lic} is, strictly speaking, not a part of this thesis but some
limited use of it is made in chapter two.
Chapter three consists of \cite{art1} and some additional material.
Chapters five and six include some unpublished notes.
Chapters seven and eight consist mainly of \cite{art2} with some minor
contributions from \cite{art3} to chapter eight.
Chapter nine is most of \cite{art3}.
Chapter ten consists of material not published elsewhere.
Chapter eleven is the remainder of \cite{art3}.
The work has in part been supported by a Harold W. Dodds Fellowship.

There are several persons I would like to thank.
First of all I want to thank my advisor David Gross for all his
invaluable help and all the things I have learnt from him. 
I am also grateful to Igor Klebanov for several discussions and
explanations.
Among others at Princeton University who have been of great help to me
are Curtis Callan, Mark Doyle, Andrew Felce, Jose Gonzalez, 
David Lowe, Miguel Martin-Delgado and Michael Newman.

At the University of Stockholm I would like to thank Patrik Johansson,
Anders Karlhede and Bo Sundborg.

Finally I would like to thank Hector Rubinstein at the University of Uppsala
for
all his support and guidance during the past several years. In
particular I would like to thank him for the best advice of all, to go
to Princeton.

\newpage

\chapter{Introduction}

\pagenumbering{arabic}

The outstanding problem of todays physics is the unification of quantum
mechanics and general relativity, quantum gravity. It is widely
believed that such a theory would also provide a unification of all
forces and therefore give a unified picture of all physics. 
The putative new theory would hopefully
also predict a multitude of new and fascinating phenomena. Needless to
say, such a theory does not exist today. 

During the past decade
or two however, a very promising candidate, in fact the only existing candidate,
has been developed. This is string theory.
In string theory several of the problems inherent in quantum gravity
seem to be resolved. In particular the nonrenormalizability of the
perturbative expansion is taken care of by a Planck scale cutoff given
by the size of the string.

Unfortunately a phenomenologically acceptable string theory does not so far
exist. The heterotic string, the most promising candidate, gives a
reasonable gauge group which by symmetry breaking can be reduced to the low
energy $SU(3) \times SU(2) \times U(1)$ which we see. However, there
does not seem to be any unique way of compactifying the 10 space time
dimensions typical of a critical super string, down to the four dimensions
of the real world. For a general introduction, see
\cite{gsw1,gsw2} and references therein.

If we believe in the promise of string theory, but are discouraged by the
lack of immediate success of the critical theories, we should consider
alternatives. Such an alternative is provided by the noncritical string
theories. These string theories rely on solving two dimensional quantum
gravity which in some cases turns out to be a feasible exercise. This
invalidates the reason for restricting oneself to the critical
dimensions. Unfortunately, technical obstacles stop us from directly
constructing a four dimensional string theory. The best we can do so
far, is to obtain a string theory describing a string moving in a two
dimensional space time. This theory lacks a lot of the complications of
higher dimensions but is complicated enough to contain interesting
physics.

Remarkably there also exists a new tool, the matrix models, which allow
us to solve these theories exactly to all orders of string perturbation
theory.
This two dimensional string theory, in particular its matrix model
version, will be the subject of this thesis.

Given the simplicity of its formulation, it is very surprising that 
the two dimensional noncritical string theory was not studied long ago.
While the precise connection between this theory and the Liouville mode
quantum gravity approach requires some technical machinery, there are no
difficulties establishing it as a perfectly consistent two dimensional
string theory.

The important point to realize is that a nontrivial background, in this
case a linear dilaton field, can change the effective central charge and
hence the critical dimension. This is of course just the string theory
way of describing the conformal field theory construction with a
background charge.

More generally from the string theoretic point of view, one needs to
solve the $\beta$ equations. These are precisely the requirement that
the world sheet theory is scale invariant, no conformal anomaly. Without
a dilaton field they reproduce the Einstein equations for the metric.
This is true at tree level for the world sheet $\sigma$ model. In
general, there are corrections at higher orders in $\alpha '$, where
$\sqrt{\alpha '} \sim$ size of the string. In two
dimensions we really need the nontrivial dilaton background to get a
consistent theory. But as was shown in \cite{svart,wi2} this is not the only
possibility. There is a one parameter family of solutions describing a
black hole background, the parameter being the mass of the black hole.

Clearly the two dimensional string theory can serve as a laboratory for
testing many of the ideas of string theory. It might give some insight
into the deeper issues of how string theory gets around the problems of
quantum gravity. In the end this could turn out to be the single most
important contribution of string theory. Even if string theory turns out
{\it not} to be the correct theory of quantum gravity realized in
nature, we might learn a lot by studying how it resolves the quantum
gravity enigma. If nothing else, this solution, even if it is wrong, can
serve as an important source of inspiration in the work towards the
correct theory. 

There are many mysterious aspects of quantum gravity and it is so far
unclear in what sense they are resolved by string theory. One of the
most interesting ones is the problem of black holes and the loss of
quantum coherence. It appears as if important quantum phase information
can get lost without a trace into a black hole. After all, it is claimed
that a black hole has no ``hair" \cite{col}. As a result a pure state can evolve
into a mixed state. This breaks the unitary time evolution of quantum
mechanics. It is important to note that one cannot simply say that the
information is there somewhere inside of the black hole. Eventually the
black hole may evaporate due to the Hawking process revealing nothing. The
Hawking radiation is also supposed to be purely random and thermal
containing no information. How is this mystery solved?

It has been proposed that black holes in fact {\it do} have a lot of
quantum hair \cite{col}. 
This is not forbidden by the classical no hair theorems.
Clearly this would have the potential of solving the quantum coherence
problem. It has also been proposed, \cite{ellis}, that there is extra 
hair in the two dimensional string theory which we are about to
study. The hair would be associated with the $W_{\infty}$ symmetry which
we will encounter at several places throughout this thesis. If this is
true also for higher dimensional string theories as claimed in \cite{ellis},
string theory would have solved one more of the fundamental issues of
quantum gravity. In some sense this would be an even more remarkable and
far reaching achievement than just taking care of the nonrenormalizability.

However, so far none of these speculations have any secure basis. In
fact, it may very well be that the solutions to problems like quantum
coherence have nothing or very little to do with strings. In particular
it is conceivable that the loss of quantum coherence is just what it is.
A sign that quantum mechanics as we know it breaks down, as suggested
by Hawking \cite{haw}. 
Depending on taste, this could be an even more exciting
perspective. Before getting to deep into string theory, let us briefly
remind ourselves what this could mean.

We have come to accept the strange duality which
exists in quantum mechanics as a rather fundamental aspect of nature.
The unitary time evolution on the one hand, and the somewhat magical,
certainly nonunitary, collapse of the wave function at the moment of
observation on the other hand. The second process defies a well defined
physical description. When does it really take place? The philosophical
discussions about this seem endless. It would clearly be desirable to
have a {\em physical} understanding of this. Is it then too farfetched to
suggest that a
nonunitary quantum mechanics, the
nonunitarity supplied by gravity as suggested by the loss of quantum
coherence above, could give a physical process for the collapse of the
wave function? Rather than just trying to device ways of getting rid of
the apparent nonunitarity, we should carefully ask ourselves whether we
might not need it after all. The point to be made is that it is very
likely that
string theory may provide a solution to
the problem of reconciling quantum mechanics
and gravity, but the question is whether that is what we want to
do! Is this really the solution which
nature has chosen?

Clearly the quest for quantum gravity is a very open one. Even if string
theory is our best candidate for the moment, we should be open minded.
It is not obviously true that strings
really have the capacity to solve {\em all} deep conceptual problems
which confronts us in e.g. quantum gravity. 
However, string theory clearly deserves the attention it
receives for whatever clues it might give. With this in mind let us embark
on the careful study of two dimensional string theory which is the
subject of this thesis.

\chapter{The Liouville Model}

\section{Introduction}

In this chapter we will review the two dimensional noncritical string from
the field theoretic point of view. In a later chapter an introduction to
the matrix model approach will follow. 

We will also consider the 
theory as a model of two
dimensional world sheet quantum gravity. For this we will briefly
review some different approaches. Historically this was also the way in
which one first solved the model. Only later the focus has been more on
the string theoretic interpretation. Ironically it is really only the
quantum gravity interpretation which need a nontrivial justification, as
we will see.

Towards the end of the chapter we will discuss the ``special states''.
These are probably one of the most interesting aspects of the theory. Since
they are remnants of the usual excited string modes in higher dimensional
theories, they could potentially tell us a lot about stringy phenomena.
We will encounter them again and again throughout this thesis.

\section{The Way to Solve it}

Our focus will be on
the gravity coupled $c=1$ model 
and its
target space interpretation as a two dimensional string theory.
We will however begin by considering 
the more general case of a minimal model with central charge 
$c\leq 1$ coupled to
gravity. More precisely we will be discussing induced Liouville
gravity.

The action for a minimal model coupled to Liouville gravity, in complex
coordinates $z=\sigma _{1}+i\sigma _{2}$ and conformal gauge
$h=\rho \left( \begin{array}{cc}
            0 & 1  \\
            1 & 0
        \end{array} \right)$,
is given by
\begin{equation}
\frac{1}{2\pi} \int (\partial X \bar{\partial} X +
\frac{i\alpha}{4}\sqrt{h} RX +\partial \phi
\bar{\partial} \phi -\frac{Q}{4}\sqrt{h}R\phi ). \label{a1}
\end{equation} 
$X$ is some matter field, while $\phi$, as we will see in a moment,
is the gravitational field. Recall that in two dimensions the metric has
only one independent component. $R$ is the world sheet curvature while
$\alpha$ and $Q$ are background charges for matter and gravity
respectively. Due to the ``$i$'' in the action, the model is unitary only
for
some special values of $\alpha$. Good introductions to these and other related
issues in conformal field theory may be found in \cite{card} and \cite{gi1}.
$\alpha$ is adjusted to give the appropriate value of the
central charge of the minimal model we are considering, while $Q$ is
tuned in such a way that the total central charge of both matter and
gravity adds up to $26$. 
Hence it is also possible to interpret this
as a consistent string theory. $X$ and $\phi$ are then some target space
coordinates. The curvature term describes the coupling to a linear
background dilaton field.

Given the action (\ref{a1}) the stress energy tensor, defined by
$T_{\alpha \beta} = \frac{4\pi }{\sqrt{h}} \frac{\delta S}{\delta
h^{\alpha \beta}}$, is easily seen to be
\begin{equation}
T_{zz} = -\frac{1}{2}(:\partial X \partial X: -i\alpha \partial ^{2} X)
-\frac{1}{2}(:\partial \phi \partial \phi :+ Q \partial
^{2} \phi)
\end{equation}
by varying the action with respect to the metric. The terms linear in
the fields come from varying the world sheet curvature and some following
partial integrations.
From this the Fourier modes of $T_{zz}$, i.e. the Virasoro generators 
$L_{n}$'s, are extracted as
\begin{equation}
L_{n} = \frac{1}{2} \sum _{-\infty}^{\infty} :(\alpha _{n-m} \alpha _{m}
+\beta _{n-m} \beta _{m}): -\frac{1}{2}\alpha (n+1)\alpha _{n}
-\frac{i}{2}Q(n+1)\beta _{n}.
\end{equation}
$\alpha _{n}$ and $\beta _{n}$ are the matter and gravity oscillators
respectively.
In particular we have
\begin{equation}
L_{0}= \frac{1}{2}(p^{2}+p_{\phi}^{2}-\alpha p -iQp_{\phi})+
\sum _{m > 0} (\alpha _{-m} \alpha _{m}
+\beta _{-m} \beta _{m} ). \label{a4}
\end{equation}

The energy momentum tensor can also give us the relation between
$\alpha$, $Q$ and $c$, which we need. For simplicity of notation let us
just consider the matter part. The propagator is given by
\begin{equation}
<X(z)X(w)>=-\log (z-w) ,
\end{equation}
and the corresponding expression for the antiholomorphic part.
It is a simple exercise using Wick contractions to derive the operator
product
\begin{equation}
T(z)T(w) = \frac{\frac{1}{2} (1-3\alpha ^{2})}{(z-w)^{4}} +...
\label{a6}
\end{equation}
from which one can read 
\begin{equation}
c=1-3\alpha ^{2}. \label{a7}
\end{equation}
Similarly one finds for the Liouville part
\begin{equation}
c_{L}=1+3Q^{2}.
\end{equation}
Since we need $c+c_{L}=26$, this fixes $Q$ to
\begin{equation}
Q=\sqrt{\frac{25-c}{3}}. \label{a9}
\end{equation}

In the original works, \cite{dav1,di1,kni1,poly2}, 
this action was mostly interpreted as
describing two dimensional world sheet quantum gravity coupled to some
matter. Only later it became fashionable to think of it as a 
noncritical string theory. The latter interpretation is clearly justified
without further calculations. We have achieved quantum scale invariance,
the hallmark of a string theory, and later we will also see how to
construct scattering amplitudes. To really make the quantum gravity
connection, some further work is however needed. Let us briefly review
the different steps.

The starting point is to note that the classical scale, or
Weyl, invariance is broken quantum mechanically. The need to regularize
introduces a potential dependence on some scale which only goes away in
the critical dimension. The breaking of scale invariance is measured by
the trace of the energy momentum tensor, $T_{z\bar{z}}$. Let us for
simplicity focus on
$c=1$ where there is no background charge. Classically one
would certainly expect the trace of the energy momentum tensor to be zero, 
but the quantum mechanical answer is in
fact
\begin{equation}
T_{z\bar{z}} = -\frac{c}{24} \rho R \label{a10}
\end{equation}
where $R$ is the world sheet curvature. Let us give a simple, heuristic
but perhaps
illuminating derivation of this result.

The key point is the properties under coordinate transformations of
$T=T_{zz}$ (or $\bar{T}=T_{\bar{z}\bar{z}}$). For $c=1$
the matter part of $T_{zz}$ is just
$-\frac{1}{2} :\partial  X(z) \partial  X(z):$
Due to the normal ordering a coordinate
transformation $z \rightarrow \tilde{z}(z)$ has the following effect:
\begin{equation}
-\frac{1}{2}\lim _{\epsilon \rightarrow 0} 
(\partial _{z} \tilde{z}(z+\epsilon /2)
\partial
_{z} \tilde{z}(z-\epsilon /2) \partial  X(\tilde{z}(z+\epsilon /2))
\partial  X(\tilde{z}(z-\epsilon /2)) -\frac{1}{\epsilon ^{2}}).
\end{equation}
The $\epsilon$ limit and subtraction is the conventional point splitting
version of normal ordering. Some simple algebra then leads to
\begin{equation}
T(z) = (\partial _{z} \tilde{z})^{2}
\tilde{T}(\tilde{z}(z)) - \frac{1}{12} \{ \tilde{z},z \}.
\label{a12}
\end{equation}
The last term is the Schwarzian derivative defined by
\begin{equation}
\{ \tilde{z} ,z \} = \frac{\partial ^{3} \tilde{z}}{\partial z} -
\frac{3}{2} \left( \frac{\partial ^{2}\tilde{z}}{\partial z}
\right)^{2}. \label{a13}
\end{equation}
We now try to reproduce this transformation property {\em classically}
by adding an extra term to $T_{zz} =-\frac{1}{2} \partial X \partial X$.
This term must be
\begin{equation}
-\frac{1}{12} \{ f,z \}
\end{equation}
for some function $f =f(z, \bar{z})$. It is easy to show that indeed
\begin{equation}
\{ f,z \} = (\frac{\partial \tilde{z}}{\partial z})^{2} \{ f, \tilde{z}
\} +\{ \tilde{z} ,z \} .
\end{equation}

We next need the conventional law of conservation of the energy
momentum tensor
\begin{equation}
\nabla _{\bar{z}} T +\nabla _{z} T_{z \bar{z}} =0 \label{a14}
\end{equation}
The extra term in  $T$ 
must be accompanied by a similar extra term in $T_{z \bar{z}}$. Otherwise 
(\ref{a14})
would in general hold not hold. It is easy to see that an extra term in
$T_{z \bar{z}}$ of the form
\begin{equation}
\frac{1}{12}  
\partial \bar{\partial} \log \partial f
\label{a15}
\end{equation}
would do the job. But this is just $-\frac{1}{24} \rho R$ 
in conformal gauge where $R=-2\rho ^{-1} \partial \bar{\partial} \log
\rho$ with the
conformal factor $\rho = \partial  f$! 

In the case of general $c$
the expressions are multiplied by the central charge $c$. This can be
seen from (\ref{a6}) which upon contour integration against the
parameter $f$ produces the
infinitesimal version $f'''$ of the Schwarzian derivative. The net
effect is that we have generated gravitational contributions to the
energy momentum tensors given by
\begin{equation}
T^{G}_{zz} = - \frac{c}{12} \{ f,z \}
\end{equation}
and
\begin{equation}
T^{G}_{z\bar{z}} =-\frac{c}{24} \rho R .
\end{equation}

That $R$ is the only thing that $T_{z \bar{z}}$ can be proportional to is
really clear already from dimensional analysis and coordinate invariance.
The precise coefficient, however, is provided by the above argument.
We will come back briefly to the different expressions for the energy
momentum tensor above when discussing some different approaches to two
dimensional quantum gravity, but let us first see what to do with (\ref{a10}).

The central charge in (\ref{a10}) receive contributions both from the matter
fields and the ghost fields. This gives
\begin{equation}
T_{z \bar{z}} = \frac{26-c}{24} \rho R
\end{equation}
To get scale invariance without $c=26$ we in some sense need $R=0$. The
consistent way to achieve this is by thinking of $R=0$ as an equation of
motion for the two dimensional metric. As an aside one may note that
this ``induced gravity'' is different from ordinary Einstein gravity.
There the action is simply $\int \sqrt{h} R$ which in two dimensions is a
topological invariant, the Euler characteristic. The action of the
induced gravity is instead, in a general gauge,
\begin{equation}
\frac{26-c}{48\pi} \int d^{2}xd^{2}y \sqrt{h(x)} \sqrt{h(y)} R(x)
\frac{1}{\Delta} R(y).
\end{equation}
Hence nonlocal! After some initial despair one finds that the action is
local in conformal gauge which therefore, clearly, is a very sensible
choice. We then get
\begin{equation}
\frac{26-c}{48\pi} \int d^{2}\sigma \sqrt{\hat{h}} 
(\frac{1}{2} \partial ^{a} \phi \partial _{a} \phi
+\hat{R} \phi ) \label{a22}
\end{equation}
where $\rho = e^{\phi }$ with the metric on the form $h=\rho \hat{h}$.
$\hat{h}$ is some fixed background metric, e.g. with constant
curvature. For convenience, one usually rescales the field
$\phi$ to obtain a more standard kinetic term
\begin{equation}
\frac{1}{8\pi} \int d^{2} \sigma \sqrt{\hat{h}}  
(\partial ^{a} \phi \partial _{a} \phi
-Q_{c}\hat{R}\phi ) \label{a19}
\end{equation}
where
\begin{equation}
Q_{c} =\sqrt{\frac{26-c}{3}}. \label{a24}
\end{equation}
Henceforth we will drop the hats.
In string theory language we have fixed the $\alpha '$ for the Liouville
coordinate as $\alpha ' = 2$.
So far so good. Now the tricky part comes, which is the reason for why
the model was not solved until quite recently. The action above is the
{\it classical} action. To quantize we need to do the path integral. A path
integral needs a measure, and the measure needs the world sheet metric
for its definition. But the world sheet metric is a dynamical variable!
Clearly a very confusing situation. The trick is to make a change of
variables in the measure and define it with respect to a new fixed
background metric. The change of variables presumably leads to a
Jacobian. The assumption in \cite{di1} was that the only thing that happens is a
renormalization (finite) of the different couplings in (\ref{a19}). 
In particular,
one needs, for consistency, that $Q_{c}$ as given 
by (\ref{a24}) renormalizes into
$Q$ as given by (\ref{a9}). This change of $26 \rightarrow 25$  can also be
understood as coming from the new quantum contribution to the 
central charge from
the Liouville mode. Independence of the background metric translates
into the requirement of scale invariance in the theory (\ref{a1}).

From many points of view the argument above is purely hand waving,
although it is hard to imagine what else could happen. More rigorous
approaches however do exist and are described in the literature \cite{hok}.

Although this is the common way to do things nowadays, the first solution
was not formulated in the conformal gauge but in a peculiar left-right
asymmetric gauge in \cite{kni1,poly1}. For completeness we will give a short
description of this. This will also throw some further light on the
expressions for the energy momentum tensor. 
The treatment will be
very brief and interested readers are referred to \cite{kni1,poly1}. 
The reasons
for including this section are mainly historical.

\section{Another Solution}

As we already have seen the energy momentum tensor of the Liouville
theory is given by
$$ 
T_{--}^{G} 
= \frac{26-c}{12} \left( \frac{\partial ^{2} \rho}{\rho}-\frac{3}{2}\left( 
\frac{\partial 
\rho}{\rho} \right) ^{2}\right)
$$
\begin{equation}
T_{-+}^{G} 
=-\frac{26-c}{12} \partial _{-} \partial _{+} 
\log \rho .
\end{equation}
Both these expressions can be obtained by varying (\ref{a22}).
We have temporarily changed to a Minkowski metric to agree with the
conventions in \cite{kni1,poly1}.
If we write $\rho (x^{-},x^{+}) = \partial _{-} f(x^{-},x^{+})$, the $T_{--}$
component is the Schwarzian derivative of a new field $f$. Some simple
studies of the coordinate transformations involved show that $f$ can be
thought of as a coordinate transformation which connects conformal gauge
with a gauge where the metric looks like
\begin{equation}
\left(
\begin{array}{cc}
0 & 1/2 \\
1/2 & h_{++}
\end{array}
\right)
\end{equation}
with
\begin{equation}
h_{++} (f(x^{-},x^{+}),x^{+}) = -\partial _{+} f
\end{equation}
This is the left-right asymmetric gauge used in \cite{kni1,poly1} 
where the Liouville
theory was solved for the first time.

A crucial ingredient in this solution is an apparent $SL(2,{\bf R})$ symmetry.
One way to see this symmetry is to note that $T_{--}$ is invariant under
precisely such coordinate transformations in $f$, with coefficients
being arbitrary functions of $x^{+}$, since it is a Schwarzian
derivative. In other words, invariance under
\begin{equation}
f \rightarrow \frac{a(x^{+})f+b(x^{+})}{c(x^{+})f+d(x^{+})}.
\end{equation}
 
The $SL(2,{\bf R})$ structure gives rise to Kac-Moody currents which
are helpful when one attempts to solve the theory. They are the
following components of the metric field $h_{++}$
\begin{equation}
h_{++}
(x^{-},x^{+})=J^{+}(x^{+})-2J^{0}(x^{+})x^{-}+J^{-}(x^{+})(x^{-}) ^{2}
\end{equation}
subject to the equations of motion. Note that the equations of motion,
$T_{--}=0$, imply, but are not equivalent to, $\partial _{f}^{3} 
h_{++} (f,x^{+}) =0$. One can also show that the components $J$ 
transform as they should under the symmetry.

The weights of the $SL(2,{\bf R})$ algebra is then related, through Ward
identities, to the gravitational dimensions. The results obtained agree
with the conformal gauge approach.

One way to understand the difference between the two different gauges is
to note the difference between constraints and equations of motion.
In conformal gauge, the equation of motion is $T_{-+} =0$. In the
left-right asymmetric gauge it is instead the less restrictive
$T_{--}=0$. $T_{-+}=0$ is in this case a constraint which we finally
have to apply to get the correct physical states. They are hence of
highest weight with respect to the $SL(2, {\bf R})$.

The basic trick of \cite{kni1,poly1}
is not so much the different gauge choice. The
$SL(2,{\bf R})$ symmetry can be faked directly in the conformal gauge
\cite{lic}.
One only needs to make a field redefinition of the form $\rho = \partial
_{-} f$. Substitutions like this are something one can contemplate in more
general theories. The resulting theory has
higher derivatives and looks very different from the original one.
Equivalence is guaranteed in the end only by applying appropriate
constraints. Two dimensional gravity is an example where the higher
derivative theory in some sense is simpler to study, thanks to the new
symmetry appearing. 

In the following we will leave this method of solution aside and stick
with the conformal gauge treatment which has become the standard way to
do things.

\section{Correlation Functions}
 
Let us now limit ourselves to the case of main interest in this thesis,
$c=1$. From formulae (\ref{a7},\ref{a9}) 
we see that $\alpha$ is just zero (obviously)
and $Q=2\sqrt{2}$ in the usual convention with $\alpha' _{\phi} =2$. For
reference we write down the action again
\begin{equation}
\frac{1}{4\pi} \int d^{2} \sigma \sqrt{h}
(\frac{1}{\alpha '}\partial ^{a} X \partial _{a} X + 
\frac{1}{2} \partial ^{a} \phi
\partial _{a} \phi -\sqrt{2} R \phi ).
\end{equation}

Let us now see how to construct the operators of the theory. As usual in
string theory, states are created by vertex operators sitting on the
surface. These need to be integrated to yield string scattering
amplitudes. Consistency therefore requires that they have conformal
dimension $(1,1)$. We therefore need the $L_{0}$ generator in (\ref{a4}) for
$c=1$.

If we apply this to a state without oscillators (the general case will
be studied later) i.e. a tachyon, we obtain the condition
\begin{equation}
\frac{\alpha ' p^{2}}{4} -\frac{k}{2} (2\sqrt{2} +k) = 1 \label{a27}
\end{equation}
with solutions
\begin{equation}
k=-\sqrt{2} \pm \frac{\sqrt{\alpha '}\mid p \mid}{\sqrt{2}} .
\end{equation}
Compared to (\ref{a4}) we have redefined $p_{\phi} = -ik$.
We see that there are two solutions. Quite surprisingly these two
different solutions are of very different character \cite{sei1}. We will
come back to this later. For now we simply choose the positive sign and
denote this choice as the ``right dressing''. The ``wrongly dressed'' states
will be briefly discussed elsewhere.

The explicit calculations of tachyon correlation functions are really
very simple \cite{gr5}. Essentially, one may borrow the well established results
from the critical string. Things are in fact even simpler here, and a
lot more explicit calculations can be done.

The main new ingredient, which also is responsible for many of the
peculiar features of two dimensional string theory, is the very
restrictive kinematics. Let us assume that we want to compute an N-point
tachyon correlation function. To do so, we need to impose momentum
conservation. The matter part is just the standard
\begin{equation}
\sum _{i=1}^{N} p_{i} =0.
\end{equation}
The Liouville part receives a correction due to the coupling to the
world sheet curvature and becomes
\begin{equation}
\sum _{i=1}^{N} (-\sqrt{2}+\frac{\sqrt{\alpha '}\mid p_{i} \mid }{\sqrt{2}} ) 
=2 \sqrt{2} (g-1) .
\label{a30}
\end{equation}
If we furthermore assume that
\begin{equation}
p_{1}<0 \mbox{, and } p_{i}>0 \mbox{ for } i \neq 1
\end{equation}
we find 
\begin{equation}
\frac{\sqrt{\alpha '}\mid p_{1} \mid}{\sqrt{2}} = \sqrt{2} (g-1 +N/2) .
\end{equation}
The momentum of the single negative chirality tachyon is hence fixed!

An important complication arises if there is a world sheet cosmological
constant. In string theory we would say that there is a nontrivial
tachyon background which the string couples to. The form of this term is
determined by conformal invariance to be
\begin{equation}
\Delta e^{\alpha \phi}
\end{equation}
where $\alpha =-\sqrt{2}$ for $c=1$ and $\alpha =  
-\frac{Q}{2}+\sqrt{\frac{Q^{2}}{4}-2}$ for a general
minimal model. This follows from (\ref{a27}), the condition that the operator is
(1,1).

This means that Liouville theory with a cosmological constant is not a
free theory on the world sheet, but in fact interacting with 
vertices of infinite order.
Even though a straightforward solution of the theory
which produces all amplitudes unambiguously has ,as yet,
not been obtained, there exist
simple arguments and analytical continuations which give all
amplitudes. The best verification that these arguments make sense is
of course the agreement with the matrix model.

The basic effect of the cosmological term is that it can modify the
momentum conservation law (\ref{a30}) above. 
One insertion of the cosmological term
injects an amount $\alpha$ of Liouville momenta. 
Hence, if we break (\ref{a30})
by $q$ the amplitude can still be nonzero and will be proportional to
\begin{equation}
\Delta ^{q/\alpha} .
\end{equation}
This makes perfect sense in the case of $n=q/\alpha$ an integer. Provided
we by hand insert $n$ extra punctures we can make all computations in
the free $\phi$ (and $X$) theory. We must of course remember also to do
the zero mode integration
\begin{equation}
\int _{-\infty}^{\infty}  d \phi e^{Q(1-g)\phi -\Delta e^{\alpha \phi}}
e^{k\phi}=\frac{1}{\alpha} \Gamma (\frac{Q(1-g)+k}{\alpha}) \Delta
^{\frac{Q(1-g)+k}{\alpha}}
\end{equation}
which give the dependence on $\Delta$. The left hand side is obtained
using $\int \sqrt{h}R =8\pi (1-g)$. We note that for genus one,
without any insertions, Liouville momentum is conserved and the
integration gives rise to a volume
\begin{equation}
\mid \frac{1}{\alpha} \log \Delta \mid . \label{a36}
\end{equation}
This will be used in the next chapter.

A serious problem, however, arises for noninteger $n$. How do we insert
a fractional number of punctures? Luckily it is possible to argue, as
has been done in \cite{gou1}, that the full answer can be obtained by
``analytically'' continuing the integer answer.

Although we should be grateful, and perhaps surprised, that it is
possible to get away with a trick like this, it would clearly be more
satisfying if the calculation could be done directly. This, however,
would require an understanding of path integrals beyond the present state
of the art. Unless, of course, one uses the matrix model.

For future reference, let us give the explicit expression for the
tachyon correlation function as obtained in \cite{frku,gr5}. In principle the
calculation is very simple. Just compute the Veneziano like integral
with the N tachyons and the appropriate number of screening charges.
The answer is
\begin{equation}
\prod _{i=1}^{N} \frac{\Gamma (1-\sqrt{\alpha '}\mid p \mid  )}
{\Gamma (\sqrt{\alpha '}\mid p \mid )}
\frac{d ^{N-3}}{d \mu ^{N-3}} \mu ^{\sum 
\frac{\sqrt{\alpha '}\mid p_{i} \mid}{2} -1} . \label{a41}
\end{equation}
The general integral is in fact a bit tricky and one needs some
analytical argument to easily derive it.

In the above equation we have changed $\Delta$ to its Legendre conjugate
$\mu$. This means that
we are considering 1PI rather than connected amplitudes with respect to
the puncture. In chapter 4 we will consider this in more detail and
find, among other things, that $\mu \sim \frac{\Delta}{\log \Delta}$.
For generic momentum the transition is hence very simple. One
simply absorbs the external leg $\frac{1}{\log \Delta }$ for each screener into
$\Delta$ and replaces it by $\mu$. The $\frac{1}{\log \Delta}$ is
provided by the $\frac{1}{\Gamma (0)}$ associated with the zero momentum
punctures in analogy with (\ref{a41}). At zero momentum we should also
remember to amputate external zero momentum tachyon legs. For, e.g., the
two point function we then get $\log \mu$ rather than $\frac{1}{\log
\Delta}$.
We will come back to this on several
occasions where we also will see that one must be much more careful when
at zero momentum.

\section{The Special States}

If we regard the $c=1$ quantum gravity as 
a critical string theory in a two dimensional
space time with the Liouville field as the extra dimension 
\cite{gr2,po1}, the na\"{\i}ve
expectation would be that the massless tachyon completely exhausts the
spectrum. A simple light cone gauge argument would indicate that,
since there are no transverse dimensions, 
there are no physical excitations except for the center of mass of the string--the 
``tachyon''.
Indeed, in the computation of the one loop partition
function (i.e. genus 1) using continuum methods as in the next chapter, 
only the tachyon seems to
contribute \cite{be1,art1}.
However it turns out that for certain discrete values of the momenta there
exists new states and new nontrivial operators.
In the context of the matrix model these states were first seen
in the calculation of the puncture operator two point function \cite{gr4}. They also
appear in the external legs of multi point puncture operators \cite{gr5,poly1}.
In conformal field theory it is well known that such special states appear for
$c=1$.
In this section we review this story and show how the special states 
appear in $c=1$ conformal 
field theory.

From the point of view of string theory it is not surprising that
there are other degrees  of freedom. After all, if this theory is a
two dimensional theory of space time the two dimensional
metric should be a dynamical degree of freedom. In two dimensions
there are of course no propagating gravitons, yet there are global, 
or topological, degrees
of freedom associated with the metric. Also, in string theory the 
string coupling is a dynamical degree of freedom. This corresponds to the 
zero momentum component of the dilaton.
The full set of special states, with nonvanishing momentum as well, presumably 
correspond to topological degrees of freedom of the 
two dimensional string theory. 
They are the physical
remnants of all the massive modes of the string in higher dimensions. 
If we wish to  be able to construct the most general solution of
two dimensional
string theory we must be able to excite these modes.

A $c=1$ conformal field theory has extra primary states 
which are closely related to the null states of this theory.
Null states are zero norm states that are created by descendents 
of the primary field $e^{ipX}$ for special values of the momenta $p$. 
These special values are quantized in units of ${1\over \sqrt{\alpha'}}$.  
The relevant primary fields are labeled by two integers, $r$~and $s$,
as in the minimal models. 
The momenta take the values $p= {r-s\over \sqrt{\alpha'}}$, 
and the field $e^{ipX}$ has conformal dimension
\begin{equation}
h _{pr} = \frac{(r-s)^{2}}{4} = \frac{\alpha 'p^{2}}{4} .
\label{dim}
\end{equation}
The null states are descendents at level $rs$, hence have conformal dimension
\begin{equation}
h _{null} = \frac{(r+s)^{2}}{4} \label{dim0}.
\end{equation}
The states have both a left and a right moving component, with $\bar{h} =h$,
but we will usually suppress the left moving  component.
In the minimal models all primaries have descending null states. 
This is not
the case for $c=1$ where  the momenta, $p$, in the uncompactified case, 
need not be equal to the values given by
(\ref{dim}), but can take any value.

The reason for the existence of new primary states is that the null states are
not only null (zero norm), but in fact vanish identically. 
The map between states and conformal transformations, i.e. the Virasoro
generators $L_{-n}$, degenerates. Hence there exist new states which cannot
be obtained by conformally transforming the primaries above and 
 are therefore new primary states.
Recall that the descendants of a primary are precisely those states which can
be generated by acting with conformal transformations on the primary.
One way of constructing the new states explicitly is to use the following trick.
For a given level we solve the null state equation for a state with momentum
$p$. Doing so we also obtain a relation between $p$ and the central charge $c$.
Rather than choosing a $p$ such that $c=1$   and obtaining a state 
which vanishes identically, we keep $c\neq 1$. Thereby we can isolate the zero
and extract the new primary. Let us illustrate the procedure by two examples.

The simplest
example is obtained by choosing $r=s=1$ and $p=0$. The relevant null state is
\begin{equation}
L_{-1} |p> = p \alpha _{-1} |p>,
\end{equation}
which vanishes at $p=0$. If we divide by $p$~and then set $p=0$~we obtain
 the new level 1 primary,
\begin{equation}
\lim _{p \rightarrow 0} \frac{1}{p} L_{-1} \mid p> = \alpha _{-1} \mid 0>
\label{ex1}.
\end{equation}
A slightly less trivial example is $r=2$, $s=1$ with $p={1\over\sqrt{2}}$
($\alpha ' =2$). The null state is
\begin{equation}
(L_{-2} - \frac{3}{2(p^{2}+1)} L_{-1}^{2})|p> =  
(p^{2} -\frac{1}{2})\frac{p}{p^{2}+1} (\alpha _{-2} -\frac{1}{p} 
\alpha _{-1} \alpha _{-1})|p>
\end{equation}
and the new level 2 primaries are given by
\begin{equation}
\lim _{p \rightarrow \pm{1\over\sqrt{2}}} \frac{1}{p \mp {1\over\sqrt{2}}}
(L_{-2}
-\frac{3}{2(p^{2}+1)} L_{-1}^{2} ) |p>  
= \pm \frac{2+\sqrt{2}}{6} 
( \alpha _{-2} \mp  \sqrt{2} \alpha _{-1} \alpha _{-1} )
| \pm {1\over\sqrt{2}} >
\label{a43}
\end{equation}

The extra primary states can also be understood in terms of SU(2) multiplets.
If the $c=1$  CFT is compactified on a circle with radius $R$ then the momentum will be
quantized in units of $1/R$. When  $R$ is equal to self dual radius, where the theory
is invariant under $R \rightarrow \alpha ' /R$,  there exists an  extra SU(2)
symmetry. The allowed values of the momentum are
precisely the ones discussed above, namely integer
multiples of ${1/ \sqrt{\alpha '}}$. The primary states, with a given conformal dimension, will
arrange themselves into
SU(2) multiplets. Thus in addition to the discrete momentum
tachyon states there will exist, for these values of the momentum,
additional states which fill these out to full SU(2) multiplets.
These states are primary then for any circle for which the momentum is allowed, and certainly for
the real line for which all momenta are allowed.
A state $(r,s)$, with conformal dimension 
according to (\ref{dim0}),
belongs to a multiplet of dimension $r+s+1$. This enables one to derive a
systematic construction of the new primary states. The states of the
$n+1$ dimensional $SU(2)$ multiplet may be constructed by acting on the
highest weight state with the $SU(2)$ lowering operator. The highest weight
state
is the tachyon $ e^{\frac{ni}{\sqrt{\alpha '}} X(z)}$ and the lowering operator
is
$e^{-\frac{2i}{\sqrt{\alpha '}} X(z)}$.
As an example, the next to highest state is given by
$$
\oint \frac{dw}{2\pi i} :e^{-\frac{2i}{\sqrt{\alpha '}} X(w)}:
:e^{\frac{ni}{\sqrt{\alpha '}} X(z)}: =  
\oint \frac{dw}{2\pi i} \frac{:e^{\frac{ni}{\sqrt{\alpha '}} X(z) -
\frac{2i}{\sqrt{\alpha '}} X(w)}:}{(z-w)^{n}}
$$
\begin{equation}
=\frac{1}{(n-1)!} \frac{\partial ^{n-1}}{\partial w^{n-1}}
:e^{\frac{ni}{\sqrt{\alpha '}}X(z) - \frac{2i}{\sqrt{\alpha '}} X(w)}:
|_{w=z}. \label{a44}
\end{equation}
Equations (\ref{ex1}) and (\ref{a43}), with $n=2$ and $n=3$ respectively,
are easily verified. $n=2$ gives
$:\partial X(z):$, corresponding to the state
$\alpha _{-1}|0>$.
$n=3$ gives
\begin{equation}
:\left[ \partial ^{2} X(z) - \frac{i}{\sqrt{\alpha '}}(\partial X(z))^{2}
\right] e^{\frac{i}{\sqrt{\alpha '}} X(z)} :|0>,
\end{equation}
corresponding to the state
$(\alpha _{-2} - \sqrt{2} \alpha _{-1} \alpha _{-1} ) | 1/\sqrt{\alpha '}>$.
   
When we couple the CFT to quantum gravity the states will be 
gravitationally dressed, \cite{dav1,di1,kni1,poly2}. 
This means that a state with conformal dimension $h$ is multiplied by a factor
$e^{\beta \phi _{L}}$ where $\phi _{L}$ is the Liouville field and 
$\beta$ is such that the overall conformal dimension is one. The new
gravitational
scaling dimension and $\beta$ are given by the well known formulae,
\begin{equation}
\beta = -\frac{Q}{2} +\sqrt{\frac{Q^{2}}{4} -2+2h},\,\,\,\,\,
d = 1 - \frac{\frac{Q}{2} -\sqrt{ \frac{Q^{2}}{4}-2+2h}}
{\frac{Q}{2} - \sqrt{\frac{Q^{2}}{4} -2}},
\end{equation}
where $Q$ is the background charge, $Q=2\sqrt{2}$ for $c=1$.
The expression for $d$ just comes from comparing the scaling of the
specific operator with the scaling of the metric according to
$d=1-\frac{\beta}{\alpha}$.
From this we get the dressed dimensions
$d = \frac{\sqrt{\alpha '}p}{2}$ 
for the tachyon primaries and $d = \frac{r+s}{2}$
for the null descendants, of course still with the same momenta
$\sqrt{\alpha '}p= {r-s}$.
In the case of zero momentum which will be our main concern chapter 7,
$r=s$, and we find $d = r$
for our new zero momentum primary states. These are also the dimensions found
in \cite{gr6} by considering the correlation functions of zero momentum
operators on the sphere using the matrix model.

As already mentioned, the special states can be classified using their
$SU(2)$ quantum numbers $J$ and $m$. For example, the dilaton is
$J=1,m=0$ and the state in (\ref{a43}), $J=3/2,m=1/2$. In figure 2.1 a schematic
picture of all special states is provided. The states at the edges with
$m=\pm J$ are the special tachyons. The $SU(2)$ algebra clearly only
connects states with the same $J$ quantum numbers. An example was given
above in (\ref{a44}). However, there exists a much larger algebra , the
$W_{\infty}$, which connects {\it all} states and of which $SU(2)$ is only 
a small subgroup.
This was shown in \cite{kle1} and \cite{wi1}. In \cite{kle1}
 with similar
methods as above for $SU(2)$.

As we will see in later chapters there is also a $W_{\infty}$ algebra in
the matrix model. This we will use extensively in some explicit
calculations in later chapters.

\begin{figure}
\ifx\figflag\figI
\epsfbox{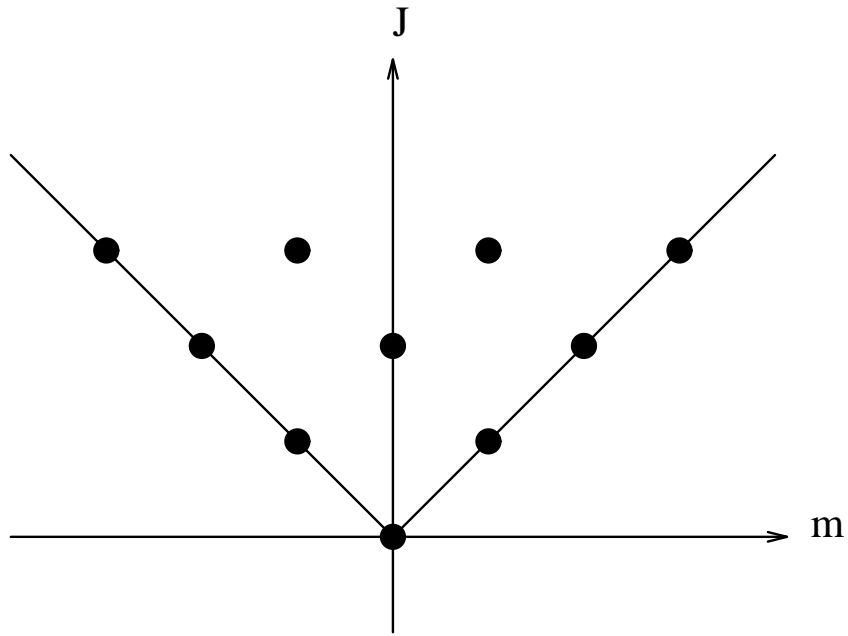}
\else
\vspace{2in}
\fi
\caption{The special states.}
\end{figure}

\section{Summary}

In this chapter we have looked at some basic properties of the two
dimensional noncritical string. We have investigated its spectrum
discovering the special states. We have also considered some simple
correlation functions on the sphere and in that context noted the
peculiar conservation laws for Liouville momentum. In the next chapter
we will take a natural next step by making some calculations not on the
sphere but on the torus.

\chapter{The Genus 1 Partition Function}

\section{Introduction}

In this chapter we will use continuum methods to calculate the genus one
partition function for some low dimensional string theories. Genus one
is particularly simple since there are no need for any screening
charges.
We will limit ourselves to $c=0$ and $c=1$. From the conformal field
theory point of view the first one corresponds to pure gravity while the
second one describes gravity coupled to a single bosonic field. In
string theory we would say that we had a one dimensional space time in
the first case and a two dimensional space time in the second case.

In \cite{be1} similar calculations were done for any minimal model. We will 
however use a more
physical approach and also clarify a few points regarding $c=0$, ``The Theory
of Nothing''.

While doing these calculations we will encounter some quite nontrivial
integrals involving $\eta$ functions. In the past such objects have
occurred in the theory for the critical bosonic string. Due to the
presence of the tachyon however, the particular integrals have all been
divergent and therefore no explicit evaluations have been possible.
For $c \leq 1$ the corresponding integrals are convergent and hence it
is possible to really do all calculations. We will therefore provide
some details and examples of such calculations in two appendices.

One motivation for doing a calculation like this is to compare it with the
matrix model, we will do so and find perfect agreement. Details of the
matrix model calculations can be found in a later chapter.

\section{String Regulation}

Let us consider a scalar particle theory in $d$ dimensions. Starting
with the point particle theory we will construct a theory of strings.
We will calculate the one loop contribution to the vacuum energy.
Later we will be interested in $d=1$ and $d=2$ where we may compare with
calculations using matrix models. This is possible since what we are 
calculating is, in fact, just the genus 1 partition function for two
dimensional gravity.  

The one loop contribution to the vacuum energy is given by 
\begin{equation}
E_{1} = - \log \int D \phi e^{-S} = \frac{V}{2} \log \det \Box = 
\frac{V}{2} Tr \log \Box = \frac{V}{2} \int \frac{d^{d}p}{(2\pi)^{d}} 
 \log \Box
\end{equation}
where $\Box = p^{2} + m^{2}$ is the inverse propagator and $V$ the volume 
of space. 

We will now regard this as the low energy limit of a string theory. The 
embedding of the string world sheet in space time we take to be described
by matter fields of central charge $c$ and also, for noncritical $c$, the
Liouville mode. The Liouville mode will appear as an extra dimension,
\cite{gr2,po1}. The volume will be the length of space in the 
Liouville direction.
As we have seen,
the volume is finite due to interactions governed by the cosmological
constant and given by (\ref{a36}). There are several problems with this
expression.
One is that the momentum integral is obviously UV-divergent. Another problem is that it is
positive while the matrix models, as we will see in a later chapter,
give a negative answer. All of this is resolved
by recalling that we actually are doing string theory. By giving the particle a
finite size the integral is effectively cut off through the new symmetry which
appears: modular invariance.

If we rewrite:
\begin{equation}
E_{1} = \frac{V}{2} Tr \log \Box = 
\frac{V}{2} \lim_{s\to 0} \frac{\partial}{\partial s} \{ \frac{1}{\Gamma (-s)}
\int_{0}^{\infty} dt t^{-s-1} Tr e^{-t \Box} \}
\end{equation}
$t$ can be thought of as the length around the loop. Since we are dealing with
strings, the trace (and $\Box $) will include higher excited states.
For $c \leq 1$ there are no extra field degrees of freedom corresponding
to such excitations. As we will see below in the case of $c=1$, they
cancel in the path integral. 
This is consistent with a na\"{\i}ve light cone argument where, for $d=2$, there
indeed are no transverse degrees of freedom. However, as has already been
explained, there are some special states at discrete momenta.

As usual we must assure $L_{0} - \bar{L}_{0}=0$ for the excited states
and therefore:
$$
E_{1} = \frac{V}{2} \lim_{s\to 0} \frac{\partial}{\partial s} 
\{ \frac{1}{\Gamma (-s)}
\int \frac{dt}{t^{s+1}} \frac{d\phi}{2\pi} \int \frac{d^{d}p}{(2\pi )^{d}}
e^{-t(p^{2} + m^{2})} Tr' e^{-t\Box +i\phi ( L_{0} - \bar{L}_{0})} \}
$$
$$
= \frac{V}{2} \lim_{s\to 0} \frac{\partial}{\partial s} \{ \frac{1}{\Gamma (-s)}
\int \frac{d^{2} \tau}{(\pi \alpha ' \tau _{2})^{s}\tau _{2}} 
\int \frac{d^{d}p}{(2\pi )^{d}}
e^{- \pi \alpha ' p^{2} \tau _{2}}
$$ 
\begin{equation}
\times e^{-\frac{\pi}{6} (2-d) \tau _{2}}
|\prod_{n=1}^{\infty} (1-q^{n})(1-\bar{q}^{n})|^{2(2-d)} \} 
\end{equation}
where $q=e^{2\pi i \tau}$. We have renamed $\phi = 2\pi \tau _{1}$ and
$t=\pi \alpha ' \tau _{2}$ where $\tau =\tau _{1} +i\tau _{2}$.
We have used that $L_{0} + \bar{L} _{0} -2 $ is the closed
string propagator. The only difference from a particle is that we have
several excitations to sum over. The normalization is fixed by
$L_{0}+\bar{L}_{0} =\frac{\alpha ' p^{2}}{2}+...$. 
The necessary background charges, both for matter
(if $c<1$) and for gravity, only contribute to the zero mode parts of 
$L_{0}$ and $\bar{L}_{0}$,
as is clear from (\ref{a4}).
They are responsible for
giving an $m^{2}$ such that modular invariance is obtained. 
Also, the infinite products are the usual contributions from excited
states, ghosts giving the exponent 2 and matter + gravity the $-d$.

The region of integration is originally the infinite strip
$\{ -1/2 \leq Re \tau \leq 1/2; 0 \leq Im \tau < \infty \}$. The integral
is, however, modular invariant. Fixing this invariance may be done by
choosing just one fundamental region to integrate over. The usual choice
is $F = \{ -1/2 \leq Re \tau \leq 1/2; | \tau | \geq 1 \}$.  This 
will make the integrals finite. We may therefore take the s-limit first:
\begin{equation}
E_{1} = - \frac{V}{2} \int _{F} \frac{d^{2} \tau}{\tau _{2} } \int 
\frac{d^{d}p}{(2\pi )^{d}} e^{-\pi \alpha ' p^{2} \tau _{2} }
|\eta (q)|^{2(2-d)}.
\end{equation}
This term is the one coming from taking the derivative of $\Gamma (-s)$, 
hence the minus
sign. The other term, from $(\pi \alpha '\tau _{2})^{s}$, 
is proportional to $s$ and 
hence zero for
any finite cutoff. We finally obtain
\begin{equation}
E_{1} = - \frac{V}{2} \frac{1}{(4\pi ^{2} \alpha ' )^{d/2}} \int _{F}
\frac{d^{2} \tau}{\tau _{2}^{1+d/2}}|\eta (q)|^{2(2-d)}. \label{b5}
\end{equation}

\section{A Theory of Something}

For $c=1$ we have $d=2$ and therefore according to (\ref{b5})
\begin{equation}
E_{1} = -\frac{V}{8 \pi ^{2} \alpha '} \int _{F} \frac{d^{2}\tau}{\tau _{2}
^{2}}
= - \frac{V}{24\pi \alpha '}.
\end{equation}
Now $\alpha '=(\alpha _{m}' \alpha _{\phi } ')^{1/2}$, i.e. $\alpha _{m} '$ 
for matter and $\alpha _{\phi } '$ for gravity. We also have, implicitly,
$V \propto (\alpha _{\phi} ')^{1/2}$ to get correct dimensions. To be able to compare with the matrix model
results we put $\alpha _{m} ' = 1/4$, $\alpha _{\phi } '=2$ and use the 
result (\ref{a36})
\begin{equation}
V = - \frac{1}{\sqrt{2}} \log \Delta
\end{equation}
where $\Delta$ is the cosmological constant to get
\begin{equation}
E_{1} = \frac{1}{24\pi } \log \Delta
\end{equation}
in agreement with \cite{gr2} apart from a factor $1/2$ due to
doubling of the free energy not accounted for there, \cite{be1}. This
will be explained at the end of the next chapter.

This is probably the simplest example of how strings regulate a theory.
As already stated, the only physical state is a massless tachyon.
Furthermore, it's only the tachyon which propagates around our loop.
Contrary to the critical string no excited states are needed to provide
modular invariance.
We have a string regulation of a massless particle in two dimensions.

This gives a possibility to give a hand waving argument for introducing
the string.
The invariance which turns into modular invariance is inversion
of the length around the loop $t \to \alpha ' /t$. We can consistently
omit $t< (\alpha ') ^{1/2}$. To get this invariance we must, however,
introduce an extra parameter
$\phi$ describing the particle. It must transform as 
$\phi \to \frac{\alpha '}{t^{2}} \phi$ to give invariance. 
The consistent way of putting this together is modular invariance.
The omission of small $t$ is identical to restricting the integration to the 
fundamental region introduced previously.

\section{$\eta$ Function Integrations}

This section is a mathematical interlude where we will consider the 
integral (\ref{b5}) in more detail. The
trivial case $d=2$, where there are no $\eta$ function integrations we
have already done above. What about other cases? Below we will show how
to do the integral for $d=1$ and $d=-1$. 

Let us first consider $d=1$.
Our integral is then
\begin{equation}
\int _{F} 
\frac{d^{2} \tau}{\tau _{2}^{3/2}} |\eta (q)|^{2} . \label{enoll}
\end{equation}
The trick is to rewrite the $\eta$ functions in the following way
\begin{equation}
|\eta (q)|^{2} = \frac{1}{2\sqrt{\tau _{2}}}( \sqrt{6} \sum_{n,m} 
e^{-\frac{6\pi}{\tau _{2}} |n+m\tau |^{2}} - 
\sqrt{\frac{3}{2}} \sum_{n,m} e^{-\frac{3\pi}{2 \tau _{2}}
|n+m\tau|^{2}} ).  \label{etaeq}
\end{equation}
This is proven in appendix 3A. Hence we need to consider
\begin{equation}
\sum _{n,m} \int \frac{d^{2}\tau}{\tau _{2}^{2}}
e^{-\frac{\pi x}{2 \tau _{2}}\mid n+m\tau \mid ^{2}} . \label{tjosan}
\end{equation}
The next step is to use the method of \cite{mc1,tan}
used in \cite{be1}. The $n$'s and $m$'s in (\ref{tjosan}) may be thought of as
describing different windings around the torus. Since 
(\ref{tjosan}) is modular invariant we can use modular transformations to put
all the winding in each case around one specific cycle. We effectively trade
one of the sums for a sum over modular transformations. These transformations,
when operating on the fundamental region, cover the full strip
$\{-1/2 \leq Re \tau \leq 1/2 ;0 \leq Im \tau \leq \infty \}$.
We get
$$
\sum _{n,m} \int \frac{d^{2}\tau}{\tau _{2}^{2}}
e^{-\frac{\pi x}{2 \tau _{2}}\mid n+m\tau \mid ^{2}}
$$
\begin{equation}
=\int_{F} \frac{d^{2} \tau}{\tau _{2}^{2}} + 2 \int_{0}^{\infty}
\frac{d\tau _{2}}{\tau _{2}^{2}} \sum_{n>0} e^{\frac{-\pi x}{2 \tau _{2}} 
n^{2}}
=\frac{\pi}{3}+\frac{2\pi}{3x} . \label{b12}
\end{equation}
Putting everything together then gives
\begin{equation}
\int _{F} 
\frac{d^{2} \tau}{\tau _{2}^{3/2}} |\eta (q)|^{2}
=\frac{\pi}{3\sqrt{6}} .
\end{equation}

The next example $d=-1$ is slightly more involved but the techniques are
the same. As shown in appendix 3B it is again possible to rewrite the
$\eta$ functions in such a way that the result (\ref{b12}) can be used.
The result is
\begin{equation}
\int _{F} 
\frac{d^{2} \tau}{\tau _{2}^{1/2}} |\eta (q)|^{6} 
= \frac{1}{6\sqrt{2}} .
\end{equation}
The details of the calculation may be found in appendix 3B.

In this section we have seen some examples of explicitly calculated
$\eta$ function integrations. Presumably there are other cases where a
similar method would work.

\section{A Theory of Nothing}

Let us now use the result of
the $d=1$ calculation to obtain the
vacuum energy for the $c=0$ model, pure gravity.
The volume in
this case equals $- \frac{\sqrt{3}}{2} \log \Delta$ 
as obtained from
(\ref{a36}).
Note that we have 
no factor of $\alpha _{m} '$ here. The result is
\begin{equation}
E_{1} = -\frac{1}{48} \mid \log \Delta \mid ,
\end{equation}
which again agrees with the matrix models apart from a factor 1/2.

Let us compare with the derivation in 
\cite{be1} for $c=0$ i.e. the $(p,q)=(2,3)$ model. There a matter field 
was included
and $E_{1}$ for finite radius $R$ calculated. 
$E_{1} (R/ \sqrt{\alpha '_{m}} =\sqrt{pq} ) - 
E_{1} (R/ \sqrt{\alpha ' _{m}} =\sqrt{p/q})$
then give the $(p,q)$ models following \cite{fr1}, in particular 
the $(2,3)$ one.
In these cases the Dedekind $\eta$-functions cancel since we always
calculate in $d=2$. The complicated
integrals are instead due to having a finite $R$.

For $c=0$ we have an alternative, as shown by our calculation 
above. We may do without the
matter field completely which is really the natural thing for $c=0$. 
In that case the Dedekind functions do not 
cancel (\ref{enoll}), on the other hand
there are no finite radius  to worry about. 
The equivalence is illustrated by the identity (\ref{etaeq}).

So, for $c=0$ we may either have no matter at all, really pure gravity (our
version) or we may have a bosonic field with a background charge giving a net
$c=0$ (as in \cite{be1}). The two approaches are completely equivalent,
and the physical content the same.

In other words, two variants of a theory of nothing.

\section{Summary}

In this chapter we have calculated the genus 1 partition function for
$c=0$ and $c=1$. We have compared the answers with the matrix model
and found agreement. These calculations are basically the only ones
which have been done at genus 1. This illustrates the difficulties
associated with calculations using field theory. The more remarkable are
the achievements of the matrix model to which we turn in the next
chapter.

\newpage

\section*{Appendix 3A}

We shall first prove
\begin{equation}
|\eta (q) |^{2} = \frac{1}{2} \sum_{s,t}
\left[ q^{\frac{3}{2} (\frac{s}{6} +t)^{2}}
\bar{q}^{\frac{3}{2} (\frac{s}{6} -t)^{2}} -
q^{\frac{3}{2} (\frac{s}{2} + \frac{t}{3})^{2}}\bar{q}^{\frac{3}{2}
(\frac{s}{2} - \frac{t}{3})^{2}} \right] . \label{bevis}
\end{equation}
First we note that
\begin{equation}
\eta (q) = q^{1/24} \prod_{n=1}^{\infty} (1-q^{n}) =
\sum_{-\infty}^{\infty} (-1)^{n} q^{\frac{3}{2} (n -\frac{1}{6})^{2}} .
\label{eta}
\end{equation}
This may be obtained using the Jacobi triple product:
\begin{equation}
\prod_{n=1}^{\infty} (1-x^{n})(1+ x^{n-1/2}y)(1+x^{n-1/2}y^{-1})=
\sum_{-\infty}^{\infty} x^{\frac{1}{2} n^{2}} y^{n} ,
\end{equation}
setting $x=q^{3}$ and $y=-q^{-1/2}$ \cite{gi1}. We then get
\begin{equation}
|\eta (q) |^{2} = \sum_{n,m} (-1)^{n+m} q^{\frac{3}{2} (n-1/6)^{2}}
\bar{q} ^{\frac{3}{2} (m-1/6)^{2}} .
\end{equation}
We will prove (\ref{bevis}) directly from this by identifying terms. 
We first look for terms in (\ref{eta}) such that
\begin{equation}
\left\{
\begin{array}{ll}
(\frac{s}{6} +t)^{2} = (n-1/6)^{2} \\
(\frac{s}{6}-t)^{2} = (m-1/6)^{2}
\end{array}        
\right.  .
\end{equation}
We find
\begin{equation}
\left\{
\begin{array}{ll}
t=\frac{n-m}{2} \\
s=3(n+m)-1
\end{array}
\right. \label{s2}
\end{equation}
or
\begin{equation}
\left\{
\begin{array}{ll}
t=\frac{m-n}{2} \\ 
s=-3(n+m)+1
\end{array}
\right.  \label{s3}
\end{equation}
with $n\pm m$ even.

So we have found all terms with $n\pm m$ even in (\ref{eta}) represented. 
The factor $1/2$ in (\ref{bevis}) corrects for the existence of two solutions above.

Analogously we find
\begin{equation}
\left\{
\begin{array}{ll}
t=\frac{3(n+m)-1}{2} \\ 
s=n-m
\end{array}
\right.
\end{equation}
or
\begin{equation}
\left\{
\begin{array}{ll}
t=\frac{-3(n+m) +1}{2} \\ 
s=m-n
\end{array}
\right.
\end{equation}
for $n\pm m$ odd.

All terms in (\ref{eta}) are now accounted for. However, in the first 
term of (\ref{bevis}) we 
have only taken care of terms where $s$ is not even or divisible by 3 
(see (\ref{s2}, \ref{s3})) and
in the second term, terms where $s$ is odd and $t$ not divisible by 3. What 
about the rest? They can be seen to cancel by considering the solutions to
\begin{equation}
\left\{
\begin{array}{ll}
(\frac{s_{1}}{6} +t_{1})^{2} = (\frac{s_{2}}{2} +
\frac{t_{2}}{3})^{2} \\
(\frac{s_{1}}{6} -t_{1})^{2} = (\frac{s_{2}}{2} - \frac{t_{2}}{3})^{2}
\end{array}
\right.
\end{equation}
i.e.
\begin{equation}
\left\{
\begin{array}{ll}
s_{1} = s_{2} \\
t_{2} = 3t_{1} 
\end{array}
\right.
\end{equation}
or
\begin{equation}
\left\{
\begin{array}{ll}
s_{1} = 2 t_{2} \\ 
s_{2} = 2 t_{1}
\end{array}
\right.  .
\end{equation}
(\ref{bevis}) is now proven. Finally the formula for Poisson resummation is
needed
\begin{equation}
\sum _{n=-\infty}^{\infty} e^{-\pi n^{2} a+2\pi nab}=\frac{1}{\sqrt{a}}
e^{\pi ab^{2}}\sum _{m=-\infty}^{\infty} e^{-\frac{\pi m^{2}}{a} -2\pi i
mb}
\end{equation}
see e.g. \cite{gsw2}. 
A resummation on $s$ yields 
the desired result (\ref{etaeq}).

\newpage

\section*{Appendix 3B}

To calculate
\begin{equation}
\int _{F} 
\frac{d^{2} \tau}{\tau _{2}^{1/2}} |\eta (q)|^{6} 
\end{equation}
we again appeal to the Jacobi triple identity as in appendix 3A. In this
case it yields
\begin{equation}
\prod _{n=1}^{\infty} (1-q^{n})^{3} = \lim _{y \rightarrow -q^{1/2}}
\frac{\sum _{-\infty}^{\infty} q^{\frac{1}{2} n^{2}} y^{n}}{1+q^{1/2}/y}
.
\end{equation}
With $y=-\frac{q^{1/2}}{1-\epsilon}$ this becomes 
\begin{equation}
\prod _{n=1}^{\infty} (1-q^{n})^{3} = \lim _{\epsilon \rightarrow 0}
\sum _{-\infty}^{\infty} \frac{1}{\epsilon} \left( \frac{-1}{1-\epsilon}
\right) ^{n}
q^{\frac{1}{2} n^{2}i\frac{1}{2}n}
\end{equation}
and finally
\begin{equation}
\eta (q)^{3} =
\sum _{n=1}^{\infty} (2n+1)(-1)^{n}
q^{\frac{1}{2} (n+\frac{1}{2})^{2}} .
\end{equation}
We then need to prove
$$
\mid \eta (q) \mid ^{6} =
4\sum _{n,m=1}^{\infty} (n+\frac{1}{2})(m+\frac{1}{2})(-1)^{n+m}
q^{\frac{1}{2} (n+\frac{1}{2})^{2}}
\bar{q}^{\frac{1}{2} (m+\frac{1}{2})^{2}}
$$
\begin{equation}
= 2\sum _{s=odd,t}^{\infty} (\frac{s^{2}}{4}-t^{2})
q^{\frac{1}{2} (\frac{s}{2}+t)^{2}}
\bar{q}^{\frac{1}{2} (\frac{s}{2}-t)^{2}} .
\end{equation} 
This can be proven in a similar way as in appendix 3A by identifying terms.

Then we put $q=e^{i\pi \tau}$ and obtain
$$
\mid \eta (q) \mid ^{6} =
2\sum _{t, s=odd}^{\infty} (\frac{s^{2}}{4}-t^{2})
e^{-\pi \tau _{2} (\frac{s^{2}}{2}+t^{2}i\pi \tau _{1} 2st}
$$
$$
= 2\sum _{s=odd}^{\infty} e^{-\pi \tau _{2} \frac{s^{2}}{2}}
(\frac{s^{2}}{4}+\frac{1}{2\pi} \frac{d}{d\tau _{2}}) \sum _{t}^{\infty}
e^{-2\pi \tau _{2} t^{2}+2i\pi \tau _{1}st}
$$
\begin{equation}
= 2\sum _{s=odd}^{\infty} e^{-\pi \tau _{2} \frac{s^{2}}{2}}
(\frac{s^{2}}{4}+\frac{1}{2\pi} \frac{d}{d\tau _{2}}) \left( 
\frac{1}{\sqrt{2\tau _{2}}}
e^{-\frac{\pi \tau _{1}^{2}s^{2}}{2\tau _{2}}} \sum _{n}
e^{-\frac{\pi n^{2}}{2\tau _{2}} -\frac{n\pi \tau _{1} s}{\tau _{2}}} 
\right)
\end{equation}
where we in the last step performed a Poisson resummation. 

From this we find
$$
\int _{F} 
\frac{d^{2} \tau}{\tau _{2}^{1/2}} |\eta (q)|^{6}
$$
\begin{equation}
= \frac{1}{\pi \sqrt{2}} (-\frac{d}{dx} -\frac{1}{2}) 
\sum _{s=odd,n} \int \frac{d^{2}\tau}{\tau _{2}^{2}}
e^{-\frac{\pi x}{2 \tau _{2}}\mid n+s\tau \mid ^{2}} \mid _{x=1} .
\label{b34}
\end{equation}

Let us calculate
\begin{equation}
\sum _{s=odd,n} \int \frac{d^{2}\tau}{\tau _{2}^{2}}
e^{-\frac{\pi x}{2 \tau _{2}}\mid n+s\tau \mid ^{2}}
\end{equation}
The only slight complication is that the sum is only over odd s. Under
modular transformations the measure is invariant and in the exponent we
see that
$$
\tau \rightarrow \tau +1 \mbox{ takes } s \rightarrow s , n  \rightarrow n+s
$$
\begin{equation}
\tau \rightarrow -1/\tau \mbox{ takes } s \rightarrow n , n  \rightarrow -s
.
\end{equation}
We therefore realize
\begin{equation}
\sum _{s=odd,n=even} =\sum _{s,n=odd} = \sum _{s=even,n=odd} .
\end{equation}
We can not, however, get $\sum _{s,n=even}$ in this way. But this
sum is immediately given by (\ref{b12}) to be
\begin{equation}
\sum _{s,n=even} = \frac{\pi}{3} + \frac{\pi}{6x} .
\end{equation}
We therefore find 
$$
\sum _{s=odd,n} \int \frac{d^{2}\tau}{\tau _{2}^{2}}
e^{-\frac{\pi x}{2 \tau _{2}}\mid n+s\tau \mid ^{2}}
$$
\begin{equation}
=\sum_{s=odd, n=even} +\sum _{s,n=odd} = \frac{2}{3} (\sum _{s,n}-
\sum _{s,n =even})=\frac{\pi}{3x} .
\end{equation}
By insertion in (\ref{b34}) we then get
\begin{equation}
\int _{F} 
\frac{d^{2} \tau}{\tau _{2}^{1/2}} |\eta (q)|^{6} 
= \frac{1}{6\sqrt{2}}
\end{equation}

\chapter{The Matrix Model}

\section{Introduction}

Matrix models are new and powerful tools which can be used
to solve a variety of different
models for two dimensional quantum gravity. Hence they provide solutions
of noncritical string theory. To be honest, matrix models have been
around for quite some time as models for random surfaces and two
dimensional quantum gravity.
In the beginning however, one usually limited oneself to a large $N$
limit ($N$ being the dimension of the matrix) such that only surfaces
of spherical topology survived. It was a very important discovery when
it was first realized that an appropriate continuum limit, ``the double
scaling limit'', existed which made it possible to retain {\it all} genus
in a sensible fashion. The first such matrix models \cite{bre1,do1,gr1}
described various
theories of quantum gravity coupled to, not necessarily unitary, matter
with central charge $c <1$. The model which will be of interest to us,
however, is the case $c=1$, the topic of this thesis. 
While $c<1$ was basically a problem of doing
very complicated matrix integrals, something which may be done using
orthogonal polynomials, the $c=1$ case requires a different method 
\cite{ka1,gi2,gr3}.
In this chapter we will review some of these issues.

Unfortunately the applicability of the matrix model seems to stop at
$c=1$. This is similar to the techniques of field theory in previous
chapters. It is presumably related to the fact that for higher values
of $c$ one gets new field degrees of freedom which cannot be encoded in
just a matrix. We should also recall that the presence of a tachyon
makes the $d>2$ theories quite ill behaved. A natural way to proceed
would therefore be to consider super symmetric theories. When this is
being written, however, the progress in such directions has been very
limited.  

\section{The Matrix Model}

Let us start with the $c=1$ string partition function
\begin{equation}
Z = \sum _{g=0}^{\infty} g_{st}^{2(g-1)} \int {\cal D} h {\cal D}
X e ^{-\frac{1}{4\pi
\alpha '}\int _{\Sigma
_{g}} \sqrt{h} \partial ^{a} X \partial _{a} X}
\end{equation}
where $g_{st}$ is the string coupling. We have summed over all genus
$g$. In the path integral we have included an integration over $h$, the
two dimensional world sheet metric. Classically, the action is of course
independent of $h$ due to Weyl and reparametrization invariance. Quantum
mechanically, as we have seen, the measure ${\cal D}X$ has a dependence due to
the need of a regulator.

The basic idea of the matrix model approach is to represent the surface
by a triangulation. The triangulation, or rather its dual, is regarded
as a Feynman diagram of some scalar field theory. The reason to consider
the dual graph is of course that the order of all the vertices is then
the same. It is three in the case of a triangulation although any polygon
would do. The vertices correspond to interactions and the edges to
propagators in the scalar field theory. The matter, i.e. $X$ field, part
of the action becomes in this discretized version
\begin{equation}
\int {\cal D} X e^{-\frac{1}{4\pi \alpha '}
\int _{\Sigma _{g}} \sqrt{h} \partial ^{a} X \partial _{a} X}
\sim \int \prod _{i} dt_{i} e^{-\eta ^{2} \sum _{<ij>} \mid
t_{i}-t_{j} \mid ^{2}} \label{c2}
\end{equation}
$\eta ^{2}$ is $\sim 1/\alpha '$, $<ij>$ means a sum over nearest
neighbors.

In this picture the integration over all metrics, $\int {\cal D} h$, turns into
a sum over all triangulations. From (\ref{c2}) it is clear that we need a space
time theory whose propagator is
\begin{equation}
e^{-\eta ^{2} \mid t_{i}-t_{j} \mid ^{2}} .
\end{equation}
Since the Fourier transform is given by $e^{-\frac{1}{4\eta ^{2}} p^{2}}$, this
requires the very complicated kinetical term $\phi e^{-\frac{1}{4\eta ^{2}}
\frac{\partial ^{2}}{\partial t ^{2}}} \phi$. Such a theory would be
very difficult to handle. Fortunately we do not need this. In the
continuum limit the lattice spacing is going to zero, and the space time
momentum is going to zero in units of one over the lattice spacing. 
In the continuum limit
we are indeed not supposed to see the discreteness of the surface. This
is precisely what we need. In this limit the propagators $e^{-
\frac{1}{4\eta ^{2}}
p^{2}}$ and $ \frac{1}{p^{2}+4\eta ^{2}}$ coincide. It may be noted
that $\frac{1}{p^{2}+4\eta ^{2}}$ is the Fourier transform of
$e^{-\eta \mid t_{i} - t_{j} \mid }$. Hence the model to
investigate in the continuum limit is simply
\begin{equation}
\dot{\phi} ^{2} + 4\eta ^{2} \phi ^{2} + V(\phi )
\end{equation}
where we have added some interactions $V(\phi )$. A $\phi ^{3}$ would
give triangulations. The argument above should not be thought of as a
proof. It is rather a heuristic argument which serves as a motivation
for further study. The models ultimate success and agreement with known
results from Liouville theory is the best justification.

Given a triangulation it is a classical result that the genus $g$ of the
corresponding surface is given by
\begin{equation}
2-2g=V-E+F
\end{equation}
where $V$ is the number of vertices, $E$ the number of edges and $F$ the
number of faces of the triangulation. It is important that our scalar
field theory can measure the genus, otherwise we could not introduce the
string coupling $g_{st}$. The correct way to achieve this is by letting
$\phi$ be not just a single scalar field, but an $N \times N$ hermitean
matrix. The partition function to study is then
\begin{equation}
Z=\int {\cal D} \phi e ^{-\beta \int Tr[\frac{1}{2}\dot{\phi}^{2} +U(\phi )]}
,
\end{equation}
where $\beta$ is $1/\hbar$. From here we see that each face of the
triangulation
is weighted by $\beta$, from the Feynman diagram vertex, each edge by
$\beta$ from the propagator, and finally each vertex or Feynman diagram
loop by $N$. 
This last point is the reason for having matrices. Combining this we
find
\begin{equation}
(\frac{1}{\beta})^{E} \beta ^{F} N^{V} = \beta ^{2-2g}
(\frac{N}{\beta})^{V}
\end{equation}
for the weight factor of some given triangulation. Clearly the string
coupling will be $g_{st} \sim 1/\beta$. If this is to describe a
continuum string theory there must be a continuum limit such that
surfaces with infinite number of vertices, or faces, dominate. From
above we understand that the partition function 
can be seen as a power series in
$\frac{N}{\beta}$. The trick is to adjust $\frac{N}{\beta}$ to some
critical value where this sum diverges. This means that the $V
\rightarrow \infty$ surfaces dominate and we have a continuum limit.
However, we clearly want to extract some sensible, finite answers. That
this is possible is the single most remarkable feature of the matrix
model. By taking $N$ and $\beta$ to infinity in such a way that the
critical point is approached, the resulting divergence may be cancelled
against the $1/\beta$ behavior of the string coupling, renormalizing
everything to finite values. This is the double scaling limit. It is a
nontrivial fact that this is possible to do simultaneously for all
genus.

Now that we have established this equivalence of models, how do we solve
the matrix model? A hermitean matrix $\phi$ may be decomposed into a
diagonal piece with eigenvalues and a unitary part $U$, i.e. $\phi =
U\Lambda U^{\dagger}$. The matrix model measure is similarly decomposed
into these parts
\begin{equation}
{\cal D} \phi = \prod d\lambda _{i} {\cal D} U \Delta ^{2} (\lambda _{i})
\end{equation}
$\lambda _{i}$ are the eigenvalues, and
\begin{equation}
\Delta (\lambda _{i})=\prod _{i<j} (\lambda _{i} -\lambda _{j})
\end{equation}
the Jacobian of the change of variables. It is sometimes called the van
der Mond determinant.

If we integrate over the unitary part $U$ the remaining partition
function is
\begin{equation}
Z=\int {\cal D} \lambda \Delta (\lambda _{i}(t_{2}))
e ^{-\beta \int _{t_{1}}^{t_{2}} dt \sum _{i=1}^{N} [\frac{1}{2}
\dot{\lambda}_{i}^{2}
+U(\lambda _{i})]} \Delta (\lambda _{i}(t_{1})),
\end{equation}
where the eigenvalues $\lambda _{i}$ are fermionic. This procedure is
strictly speaking only valid for the infinite real line (for the $t$
variable). At finite radius the angular variables do not decouple. We
will come back to the finite radius case in a later chapter.

The double scaling procedure is controlled by introducing the critical
parameter
\begin{equation}
\Delta = 1-\frac{N}{\beta} ,
\end{equation}
assuming that we have renormalized in such a way that
$\frac{N}{\beta}=1$ is the critical point. Since $V$ measures the area
of the world sheet we understand from
\begin{equation}
(\frac{N}{\beta})^{V} = (1-\Delta )^{V} \sim e^{-\Delta V}
\end{equation}
that $\Delta$ must be the world sheet cosmological constant.

It is now time to attempt some more explicit calculations. A suitable
first object is the ground state energy $E=E(\Delta )$ of the system.
First we note that a small change in $\Delta$ is the same as a change in
the number of fermions $N$. Clearly, in the ground state, the fermions
are filling up the potential well up to some Fermi energy $\mu$.
Therefore 
\begin{equation}
\frac{\partial E}{\partial \Delta } = \beta \mu .
\end{equation}
We have chosen the zero of the energy to be at the top of the potential,
see figure 4.1. 

\begin{figure}
\ifx\figflag\figI
\epsfbox{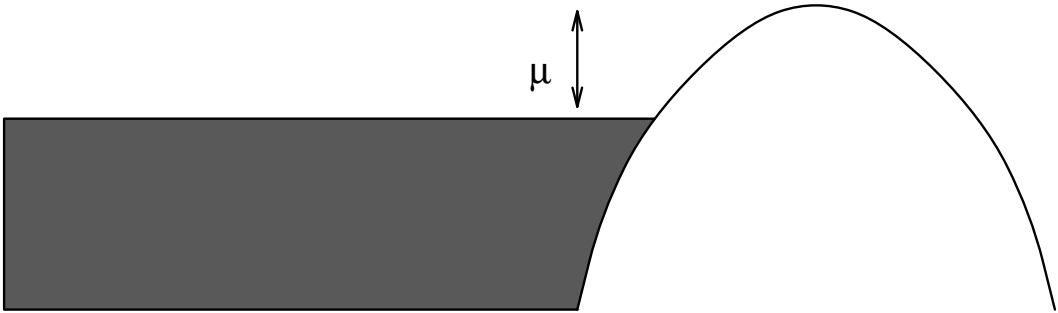}
\else
\vspace{1.5in}
\fi
\caption{The top of the matrix model potential and the Fermi sea.}
\end{figure}

Since $\mu$ has a very clear matrix model interpretation, it
is more natural to work with $\mu$ than with $\Delta$. In this picture
the double scaling limit is obtained by taking $\beta \rightarrow
\infty$ while keeping $\mu \beta$ fixed. In this limit
the only thing that matters is the top of the potential.
The top is magnified more and more as $\beta \rightarrow \infty$ and
$\mu \rightarrow 0$. Even so, it is clear that 
some kind of a cutoff is
needed for the proper definition of the theory. The Fermi sea must have
a bottom and another shore somewhere. The typical case is a potential of
the form \cite{gr4}
\begin{equation}
\frac{1}{4\alpha '} (\lambda ^{2} -\lambda ^{4}) .
\end{equation}
When we expand around the top we find the leading contribution
\begin{equation}
-\frac{1}{2\alpha '} \lambda ^{2} . \label{c15}
\end{equation}
We have redefined $\lambda$ by a shift to the maximum of the potential.
Another type of cutoff which also is relevant for a precise 
nonperturbative definition of the theory, is to put infinite walls on both
sides of the potential. In either case, as the double scaling limit is
approached all higher corrections to (\ref{c15}) become irrelevant and the walls
move off to infinity. This is important for a well defined theory.

Let us now
introduce the Legendre transform $\Gamma (\mu )$ given by
\begin{equation}
E(\Delta ) = \beta ^{2} \mu \Delta - \Gamma (\mu ) .
\end{equation}
By standard arguments $E(\Delta )$ is the generating functional for
connected diagrams, while $\Gamma (\mu )$ generates 1PI (one particle
irreducible) diagrams with respect to the puncture. We will, unless
stated otherwise, consider the 1PI amplitudes. The notation will be
\begin{equation}
<P...P> = \frac{\partial ^{n} \Gamma }{\partial \mu ^{n}} 
\end{equation}
It is very convenient
at this point to introduce the density of states, $\rho (\mu )$, clearly
given by
\begin{equation}
\rho (\mu ) = \frac{\partial \Delta }{\partial \mu }
\end{equation}
Using this, the ground state energy can be written as
\begin{equation}
E(\Delta ) = \int ^{\mu} e \rho (e)de .
\end{equation}
Note that this is consistent with
\begin{equation}
\frac{\partial E}{\partial \mu }
=\frac{\partial E}{\partial \Delta }
\frac{\partial \Delta}{\partial \mu } = \mu \rho (\mu )
\end{equation}
{\em if} we think of $\Delta$ as $\Delta (\mu )$. Also
\begin{equation}
\Delta = \int ^{\mu} \rho (e) de = <P> .
\end{equation}
This is a good point to introduce the Fermi liquid picture which uses
phase space to illustrate the system. In phase space each fermion
follows a trajectory $p^{2}-\lambda ^{2} = e$, where $e$ is its energy.
Hence the system can be described, classically, by figure 4.2. A liquid
rotating in time. The excitations, i.e the tachyons, correspond to
ripples on the surface which rotate, and evolve, with time.

\begin{figure}
\ifx\figflag\figI
\epsfbox{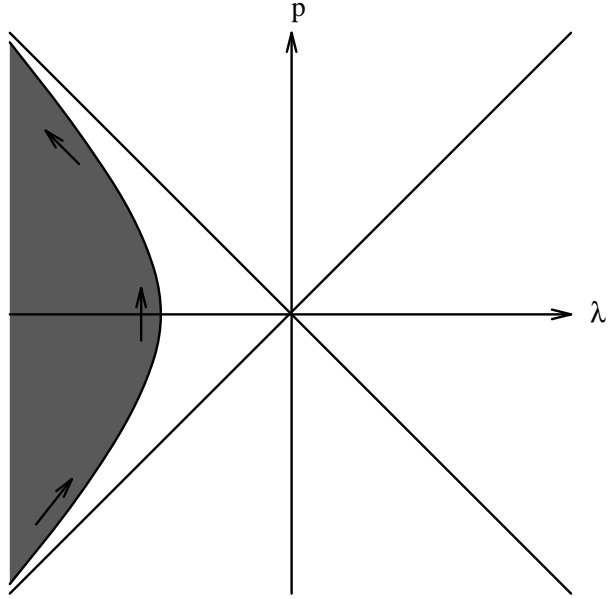}
\else
\vspace{2.5in}
\fi
\caption{Phase space picture of the Fermi sea.}
\end{figure}

With phase space coordinates the puncture one point function can be
written as
\begin{equation}
<P>=\int _{H \leq -\mu} dp d\lambda  .
\end{equation}
The two point function follows from this as
$$
<PP> = \frac{\partial }{\partial \mu} \int dp d\lambda \Theta (-\mu -H)
$$
\begin{equation}
= \int dp d\lambda \delta (-\mu -H) = \oint _{Fermisurface} . \label{c23}
\end{equation}
More generally for some operator $O$ consisting of polynomials in
$\lambda$ and $p$:
\begin{equation}
<OP>=\oint _{Fermisurface} O . \label{c24}
\end{equation}
The two point, or $\rho (\mu )$, can also be written as

\pagebreak

$$
<PP>= \frac{1}{\pi}{\rm Im} \sum _{n} \frac{1}{e_{n}+t_{0}+i\epsilon}
$$
\begin{equation}
=\frac{1}{\pi}{\rm Im} \int _{0}^{\infty} dT 
\sum _{n} <n\mid e^{-T(\mu +H)} \mid n>=
\frac{1}{\pi}{\rm Im} \int _{0}^{\infty} dT 
\int d\lambda G(\lambda , \lambda ;T) .
\label{c25}
\end{equation}
We see that there are several ways of approaching computations of matrix
model correlation functions. One way is to start with the first
expression in (\ref{c25}). 
One may then analytically continue to a right side up
oscillator. This was the way the original calculation, \cite{gr3}, was done. 
The
sum over the energy eigenvalues becomes a sum over the imaginary energy
eigenvalues in a continued right side up. This is also the way in which
we will do computations in this thesis. Another approach is to use the
path integral formulation, i.e. to use the last representation in
(\ref{c25}). This was, for nonzero momentum,
first done in \cite{mo1} and extended in \cite{das3}.

For reference we give the genus expansion of the puncture two point
function as obtained in \cite{gr3}
$$
<PP>= -\frac{1}{\pi}{\rm Im} \sum _{n=0}^{\infty} 
\frac{1}{\frac{i}{2\sqrt{\alpha '}} (2n+1)
-t_{0}}= -\frac{\sqrt{\alpha '}}{\pi} {\rm Re} \psi
(\frac{1+i2\sqrt{\alpha '} t_{0}}{2}) 
$$
\begin{equation}
= \frac{\sqrt{\alpha '}}{\pi} \left(
\mid \log \mu \mid +\sum
_{n=0}^{\infty} B_{2n}(\frac{1}{2}) (-1)^{n} \frac{1}{2n}
\frac{1}{(\sqrt{\alpha '} t_{0})^{2n}} \right) , \label{c26}
\end{equation}
where $B_{2n}(\frac{1}{2}) =(2^{1-2n} -1 )B_{2n}$. 

This is a good point
to reconsider the results of the previous chapter. From the 
relations derived earlier in this chapter, it follows that for genus
one
\begin{equation}
E_{1}= -\frac{1}{24\pi \sqrt{\alpha '}} \mid \log \Delta \mid .
\end{equation}
As we have pointed out this is twice as large as the field theory answer. The
explanation is however simple. Recall the matrix model potential and the
Fermi sea. The expression (\ref{c26}) really presupposes 
that there are {\it two}
Fermi seas. One on each side of the top of the potential. Hence there
are really two copies of the world, perturbatively disconnected. Clearly
this leads to a doubling of the free energy. In the next chapter we
will briefly consider a case where there is only {\it one} world.

There are several additional comments to be made about the genus 1
partition functions. This will also, in fact throw some light on how
special the choice of an inverted harmonic oscillator is.

As discussed in chapter 3 the genus 1 partition function is basically
obtained just from the diagram of a single particle closed loop. This we
could write as $\sim \int \partial ^{2}G(x,x)$, where 
$G(x,y) = \log \mid x-y \mid $ is the
propagator at equal times. $x$ is a space coordinate. 
This certainly needs some regularization. One possibility is
a lattice cutoff $\tilde{L}$ in some finite volume $L$. The result would
schematically be
\begin{equation}
E_{1} \sim L(\frac{1}{\tilde{L}^{2}}+\frac{1}{L^{2}}) \label{c28}
\end{equation}
The second term is the $\zeta$ function regularized sum of oscillators
in the volume $L$. In the $L \rightarrow \infty$ limit this term is
dropped. It is only this second term which is universal, the
first one needs some new physics, e.g. the scale of the cutoff, to be
determined. The modular invariance of the last chapter is precisely such
a prescription. The expression (\ref{c28}) looks rather symmetric. In
fact, the first term can {\it also} be thought of as a $\zeta$ function
regularized sum, but now in some ``internal'' space of volume $\tilde{L}$.
We will see in a moment from where this comes in the matrix model.

As noted in \cite{dem}, the $E_{1}$ can be thought of as arising from insisting
that the coincident propagator of the one loop diagram should be defined
by normal ordering in {\it $\lambda$} space. When we coordinate
transform to $\tau$ space, where $\tau$ is the time of flight for
$\lambda$,
we pick up a Schwarzian derivative which
precisely gives $E_{1}$. The map between $\lambda$ and $\tau$ is
$\lambda = \sqrt{\mu } \cosh \tau $ in general. {\it At} the top it is
$\lambda \sim e^{\tau}$. But this reminds us about a similar situation
in conformal field theory where one maps from the plane to the cylinder.
There it is well known that the Schwarzian derivative of the map and an
explicit sum over modes on the cylinder give the same result. In our
case the circumference of the ``cylinder'' is the imaginary period of
$e^{\tau}$, i.e. the time period for oscillations in the continued
harmonic oscillator. This is $\tilde{L}$. As we will see later, this
period is also related to the quantized momentum of the special states.

This discussion began with a promise that it would say something
about the more general models with anharmonic potentials. It is clear
that $\tilde{L}$, the characteristic size of the string, is independent
of the cosmological constant, either $\mu$ or $\Delta$, {\it only} for
the harmonic potential. Hence we have $L \sim \log \mu$ and $\tilde{L} \sim
1$. In general the period depends on the amplitude and we have both $L$
and $\tilde{L}$ $\sim \mu ^{-k}$ for some positive $k$.

Not only does the effective size of the string depend on the
cosmological constant, there is no natural way of keeping the scale of
the string and the scale of space very different. This is only one of
many strange features of the multi critical models.

\section{The Special States}

In the discussion of the Liouville theory in chapter 2, the special
states were introduced.

In the matrix model representation of the theory
it is easy to see some of these degrees of freedom.
In particular, there exist special operators of zero
momentum,
which are represented as time independent
perturbations of the matrix model potential. 
These are the analogs of the scaling operators
in the one dimensional matrix model. They have the physical effect, in the matrix model,
of moving one from one critical point to another.
Their
correlation functions are easily computed on the sphere and for a general genus surface.
In chapter 7 we will investigate these zero momentum 
correlation functions and show that
they obey certain recursion relations which relate all of them to
the puncture two point function. We will consider both one particle irreducible
and connected amplitudes. The recursion relations allow for computation of
any zero momentum correlation function for any genus. 
They are very similar to the recursion
relations for $c<1$ theories and suggest a topological interpretation.
We will also rewrite the recursion relations in the form of constraints on the
puncture one point function. These constraints obey a Virasoro algebra. 

We will then proceed, in chapter 8, to consider the operators of 
nonzero momentum.
The obvious candidate in the matrix model for these operators are the 
time dependent perturbations of the matrix model potential. We will
calculate correlation functions both on the sphere and on a surface of
arbitrary genus of such time dependent perturbations. 
We find that correlation functions of these operators have real poles in the 
Euclidean momentum,
conjugate to the target space variable, at precisely the expected values
of the momenta of the special states. 

What is the meaning of these special operators in the matrix model?
A hint to their meaning is provided by the analysis, \cite{gr6},
of the discrete string. This is the model in which the string is mapped onto
a discrete set of points with equal spacing $\epsilon$.
It was shown that this model was equivalent to the model where the string
is mapped onto a {\em continuous} line, as long as $\epsilon<\epsilon_{\rm cr}$,
at which point a transition appears to take place. Now this model of
triangulated surfaces is the dual of the case where the string is
mapped onto a circle of radius $R={1 \over \epsilon}$ and the transition is
dual to the Kosterlitz-Thouless transition which arises due to 
the condensation of vortices. We will discuss this a little bit more in
chapter 6.  One explanation of the equivalence of the
continuous and discrete strings, given in \cite{gr7}, is to note that  
the discretized line can be replaced by a continuous variable $t$ with a 
periodic potential, $V(t)=\sum_na_n \cos ({2 \pi n\over \epsilon}t)$.
The lowest dimension
operator in this potential, $n=1$, has undressed conformal dimension
$\frac{\alpha '}{4} (\frac{2\pi}{\epsilon})^{2}$ and becomes
relevant (i.e. of undressed conformal dimension one),
when 
$\epsilon =\epsilon_{\rm cr}=\pi \sqrt{\alpha'}$.
This is where the phase transition of the discrete string took place.
The variable $t$ is forced to take values at the minima of $V(t)$ and
the real line is effectively discretized. Now, these
terms in the potential correspond precisely to periodic time dependent
perturbations of the $c=1$ matrix model and the place where the 
transition takes place
corresponds precisely to one of the special values of the
momenta at which the special
states occur. Thus we might conjecture
that the physical meaning of these operators
is that they transform the continuous line into a set of 
discrete points, with a period
given by the inverse of the special values of the momenta --- {\em
spontaneous punctuation}.

As we will see later on, it is not enough to just consider the matrix
eigenvalues when identifying the special states. To really be able to
see the individual special states one needs to invoke the conjugate
momentum. The precise way in which this leads to the $W_{\infty}$
symmetry will be discussed in chapter 9.

Finally it is useful to point out that
one important advantage of the analytical continuation approach advocated
in \cite{art2,art3}, is that the special states are given a 
very clear interpretation. As we
will see in more detail in  chapter 8, they correspond to resonances in
the harmonic oscillator. This is also a reflection of the fact that they
are present for Euclidean momenta. Hence to really see them in a Fermi
liquid picture, we {\em must} use the analytical continuation. This is
illustrated in figure 4.3. Now any special state is associated to some
periodic ripple on the Fermi surface.

\begin{figure}
\ifx\figflag\figI
\epsfbox{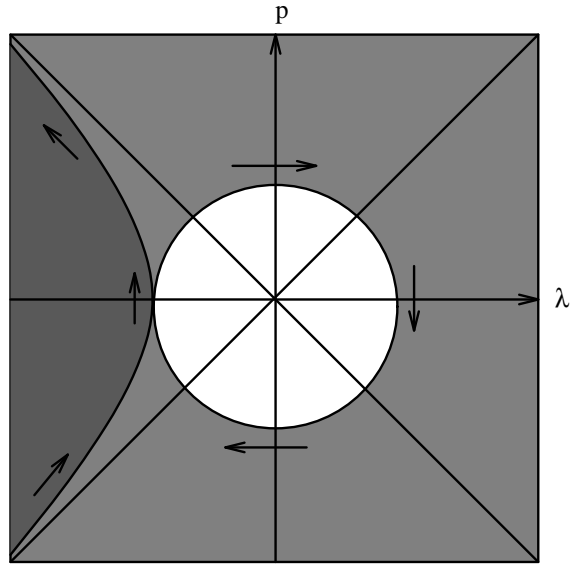}
\else
\vspace{2in}
\fi
\caption{Analytical continuation of the phase space picture.}
\end{figure}

\section{Summary}

In this chapter we have described the $c=1$ matrix model and how it
solves many of the problems of formulating a noncritical string theory.
Several formulas for calculating correlation functions have been
obtained which will be useful in later chapters. We have also
encountered the special states which will be of major importance in
coming chapters.

\chapter{Nonperturbative Effects}

\section{Introduction}

In this chapter we will discuss some of the nonperturbative issues of
two dimensional string theory. This is really the most remarkable
property of the matrix model solution, the possibility of obtaining
nonperturbative results. A general discussion of nonperturbative effects
in string theory \cite{she1} suggests that the $(2g)!$ 
behavior in (\ref{c26}) and the
$e^{-1/g_{st}}$ behavior we will find below are generic for closed
string theory. This was first observed for the $c=1$ model in \cite{gr3}.

\section{One World or Two}

We will consider an explicit and very simple
example as an illustration of what happens nonperturbatively. An
extensive treatment can be found in \cite{mo3}. For our limited purposes we will
use other and simpler methods.

Our starting point, as for most cases in this thesis, will be the
expression (\ref{c26}) for the puncture two point function. Perturbatively each
of the terms in the genus expansion can be obtained using the WKB
approximation. Since the expansion turns out not to be Borel summable,
the summation is not perturbatively determined. Additional information
is needed. Roughly speaking, we have the freedom to add terms of the form
$e^{-1/g_{st}}$, where $g_{st}$ is the string coupling. Such terms do
not show up in the small $g_{st}$ expansion. The two point function as
given by (\ref{c26}) therefore includes additional nonperturbative information
which seems to have been provided by the matrix model. However, as we
will see, the choice is not unique.

The following is a simple modification of the matrix model which leads to a
nonperturbatively different result. Instead of having
just the upside down harmonic oscillator potential, we imagine having an
infinite wall erected at the top of the potential. One might say that
one has one rather than two worlds. Clearly the WKB
expansion, which only feels the potential up to the classical
turning point, will be insensitive to this. The wall can only be felt
nonperturbatively. The expression for the puncture two point function
can be easily adjusted to describe this situation. The effect of the
wall is to impose a new boundary condition by forcing the wave functions
to be zero at the position of the wall. Hence, rather than summing over
all energy eigenstates, we sum only over the odd ones. Therefore, the two
point function in the case of the wall is given by
\begin{equation}
<PP> = \frac{1}{\pi} {\rm Re} \sum _{n=odd} 
\frac{1}{\frac{1}{2\sqrt{\alpha '}}(2n+1) +i\beta \mu} = 
-\frac{\sqrt{\alpha '}}{2\pi }
{\rm Re} \psi (\frac{3+2\sqrt{\alpha '}\beta \mu i}{4}) .
\end{equation}
By the above reasoning we know that this must perturbatively be
precisely half of the result without a wall. 
This can be checked explicitly by using the same kind of expansion as in
(\ref{c26}) and the
Bernoulli polynomial identity $B_{2n}(1/2) = 4^{n} B_{2n}(3/4)$. However,
nonperturbatively there may be a difference. Let us calculate it. By
using the $\psi$ function identities:
$$
\psi (2z) = \frac{1}{2} (\psi (z) +\psi (z+\frac{1}{2})) + \log 2 \\
$$
\begin{equation}
\psi (1-z) = \psi (z) +\pi \cot \pi z ,
\end{equation}
one easily shows
\begin{equation}
\psi (\frac{1+z}{2}) = \psi (\frac{3+z}{4}) + \log 2 - \frac{\pi}{2} \cot
(\pi \frac{1+z}{4}) + \frac{1}{2}(\psi (\frac{3-z}{4}) -\psi
(\frac{3+z}{4})) .
\end{equation}
With $z=2\sqrt{\alpha '} \beta \mu i$ imaginary it follows
\begin{equation}
{\rm Re} \psi (\frac{1+2\sqrt{\alpha '}
\beta \mu i}{2})= {\rm Re} \psi (\frac{3+2\sqrt{\alpha '}\beta
\mu i}{4}) + \log 2 - \frac{\pi}{2 \cosh \pi \sqrt{\alpha '} \beta \mu}
.
\end{equation}
The nonperturbative nature of the difference is obvious. Expanding it
for small $e^{-1/g_{st}}$ we find
\begin{equation}
-\pi \sum _{n=0}^{\infty} (-1)^{n} e^{-(2n+1)\pi \sqrt{\alpha '}
\beta \mu} . \label{d5}
\end{equation}
This nonperturbative ambiguity can also be understood from an integral
representation point of view. We note that
\begin{equation}
\psi '(\frac{1+z}{2}) = \psi '(\frac{1+z}{4}) +\frac{1}{2} (\psi
'(\frac{3+z}{4})-\psi '(\frac{1+z}{4})) .
\end{equation}
Using
\begin{equation}
\psi '(z) = \int _{0}^{\infty} \frac{te^{-zt}}{1-e^{-t}}
\end{equation}
one can, by a change of variables $t \rightarrow -t$, prove that 
for $z =2\sqrt{\alpha '} \beta \mu i$
$$
\frac{2}{\sqrt{\alpha '}} \frac{d}{d\beta \mu}
{\rm Re} \psi (\frac{1+z}{4}) =  {\rm Re}  \int _{0}^{\infty}
\frac{t e^{-\frac{1+z}{4}t}}{1-e^{-t}}
$$
\begin{equation}
\frac{2}{\sqrt{\alpha '}} \frac{d}{d\beta \mu}
{\rm Re} \psi (\frac{3+z}{4}) = {\rm Re}  \int _{0}^{-\infty}
\frac{t e^{-\frac{1+z}{4}t}}{1-e^{-t}} .
\end{equation}
The integrands are the same, but the contours differ. One is the
real line from $0$ to $+\infty$, the other one the real line from $0$ to
$-\infty$. If we rotate one
contour into the other, we pick up residues from the poles along the
imaginary axis, this precisely gives (\ref{d5}).

Hence we have established the nonperturbative nature of the difference
between one and two worlds.

\section{Instantons}

It would be nice to have a more physical picture of the nonperturbative
string theory. We will try to achieve this below.

Nonperturbative effects are in general associated with instantons. 
In this case we
have eigenvalue fermions at the Fermi surface 
tunneling through the upside down potential.
Without the wall, the instanton action for one fermion tunneling through
is obtained from the Euclidean action $\beta \int dt (\frac{1}{2}
\dot{\lambda}^{2} -\frac{1}{2\alpha '} \lambda ^{2})$ evaluated for the
classical tunneling solution of energy $\mu$.
We also need to subtract the action for
a fermion {\em not} doing any tunneling, i.e. just sitting at $\mu =
\frac{1}{2\alpha '} \lambda ^{2}$. The result is
\begin{equation}
\beta \int _{-\sqrt{2\alpha '\mu}}^{\sqrt{2 \alpha '\mu}} 
d\lambda \sqrt{2\mu - \frac{1}{\alpha '} \lambda ^{2}} = \sqrt{\alpha '}\pi
\mu \beta .
\end{equation}

Now, for two worlds the instantons correspond to tunnelings
through the potential and back again. The leading instanton
contribution to the propagator, $G(\lambda ,\lambda )$, has then the
weight
\begin{equation}
e^{-2\sqrt{\alpha '} \pi \mu \beta} .
\end{equation}
Considering the integral representation of $Re \psi (\frac{1+z}{2})$ we
indeed note poles leading to such contributions.

Let us turn to the case of {\it one} world. We write the two world
propagator as a sum over even and odd (i.e. one world) parts 
$G(\lambda , \lambda ) 
=G_{e}(\lambda , \lambda ) 
+G_{o}(\lambda , \lambda )$.
Then 
$G(\lambda , -\lambda ) 
=G_{e}(\lambda , \lambda ) 
-G_{o}(\lambda , \lambda )$ and from that
$G(\lambda , \lambda ) 
-2G_{o}(\lambda , \lambda ) =
G(\lambda , -\lambda )$.
The leading contribution to
$G(\lambda , -\lambda )$ is a tunneling solution through the
potential barrier, but {\it not} back again and hence of order
\begin{equation}
e^{-\sqrt{\alpha '} \pi \mu \beta} .
\end{equation}
This verifies and illustrates the calculations in the previous section.

\section{Summary}

Clearly much remains to be done concerning the nonperturbative aspects
of the two dimensional string. The important contribution of the matrix models
is a laboratory where we can check our understanding of nonperturbative
string physics. As we have seen above, the nonperturbative effects have
a rather simple interpretation in terms of single eigenvalue tunneling.
What remains to be obtained is a clear picture of what happens in a {\it
string} language. 
One possibility, which has been raised in the past, is
that by clarifying this issue, one could eventually find the 
``real meaning'' of string
theory. Perhaps this will involve concepts which transcend the
traditional genus expansion.

\chapter{String Theory at Finite Radius}

\section{Introduction}

In this chapter we will consider the two dimensional string theory with
compactified time on some radius $R$. In other words, string theory at
finite temperature. This has been extensively treated
in \cite{low1}. 
We will however use another and very simple approach to reproduce
some of these results. It is our feeling that the method below may
provide some useful insights into the basics of the matrix model
description of string theory.

As mentioned in chapter 4, it is not really correct to ignore
the angular parts of the matrix integral for finite $R$, \cite{gr6,gr7}. 
The angular
parts are related to vortices. These vortices are found to cause a phase
transition at a radius of twice the self dual radius, this is the
Kosterlitz-Thouless transition. As described in \cite{gr6,gr7} the
transition is caused by operators 
$\cos (\frac{R}{\alpha '} (X_{R}-X_{L}))$,
which create vortices and become relevant at
$R = 2\sqrt{\alpha '}$.
Below this radius the vortices destroy
the string world sheet.
Therefore, the calculation which we will consider below, 
assumes automatically that vortices are excluded for
some reason. It is not clear if this is a physically reasonable
assumption. 

In the presence of vortices the target space duality of string theory
under $R \rightarrow \alpha '/R$ is explicitly broken. One might
comment, \cite{ovr}, that there is a dual object to vortices, ``spikes'', which
tend to condense for {\it large} $R$, more precisely 
$R>\frac{1}{2}\sqrt{\alpha '}$. The spikes are created by 
$\cos (\frac{\alpha '}{R} (X_{R}+X_{L}))$. This is the
punctuation phenomenon we discussed at the end of chapter 4.
If we include such spikes, duality
is restored but to the price of having an instable theory
for any radius! Luckily, spikes may be forbidden simply by imposing 
translational
invariance since the operators which excite them, break this
invariance. This is not true for the vortices.

The fact remains, however, that duality {\it is} broken. Clearly, this
could have far reaching consequences also for higher dimensional string
theories. The argument leading to the Kosterlitz-Thouless transition is,
after all, a very general one. It is just, as usual, a matter of
competition between energy and entropy. To see this, one has to introduce
a cutoff, like the triangulation of the matrix model. As the cutoff
is removed, the energy of a vortex goes to infinity but at the same time
the number of possible vortices also rises. The temperature decides
which one will win.

Hence we might learn from the matrix model two dimensional string theory
to be careful when we apply concepts like duality to draw general string
theory conclusions. We might be using a principle not realized in a
realistic theory.

With this in mind, let us now give a very simple finite radius
calculation.

\section{A Simple Example}

Let us derive a finite radius expression for the puncture two point
function. To do that, let us consider the fermionic field theory action
on finite radius
\begin{equation}
S = \int _{0}^{2\pi R} dt d\lambda [\frac{1}{\beta} \dot{\psi} ^{\dagger} \psi
+ \frac{1}{\beta ^{2}}
\frac{\partial \psi ^{\dagger}}{\partial \lambda}
\frac{\partial \psi}{\partial \lambda}+ U(\lambda )\psi ^{\dagger} \psi ]
.
\end{equation}
The path integral,
\begin{equation}
Z= \int {\cal D} \psi {\cal D} \psi ^{\dagger} e ^{-\beta S} ,
\end{equation}
is easily done by expanding $\psi$ in eigenfunctions of 
$\frac{\partial}{\partial t}$ in time and eigenfunctions of the
harmonic oscillator in the $\lambda$ coordinate. The eigenvalues are
$\frac{2m+1}{2R}$ for the time part since we need anti periodic boundary
conditions for the fermions. For the harmonic oscillator part we have
the standard $\frac{1}{2\sqrt{\alpha '}}(2n+1)$. 
There is also the energy shift $t_{0}=\beta \mu$.
A standard evaluation of the logarithm of the path integral, giving a
logarithm of a determinant and hence a trace of a logarithm of the 
eigenvalues, then yields the following expression for the two point function
\begin{equation}
<PP>= \frac{1}{\pi R}{\rm Im} \sum _{n,m} 
\frac{1}{(\frac{1}{2\sqrt{\alpha '}}
(2n+1)i+\frac{(2m+1)i}{2R}
-t_{0})^{2}} . \label{e3}
\end{equation}
If the sum over $m$ is performed the result of \cite{gr6} is recovered.
Using this expression a general correlation function can be obtained
through perturbation theory in the same way as for the $R=\infty$ case
extensively treated elsewhere in this thesis.

Once this is established, it is also easy to write down a recipe which
converts the ordinary $R=\infty$ two point function into the finite
$R$ case. The point of doing so is that this procedure, as is easily seen,
commutes with the variations giving other correlation
functions. As we will see, this procedure is precisely the one derived
independently in \cite{low1}.

In words, what one has to do is to take one $t_{0}$ derivative and then
sum over all shifts $\frac{(2m+1)i}{2R}$ in the $t_{0}$ variable. The
shift can be written down using the identity
\begin{equation}
e^{z\frac{\partial}{\partial x}} f(x) = f(x+z)
\end{equation}
for any function $f(x)$. We get
$$
(1-e^{-\frac{i}{R}\frac{\partial}{\partial t_{0}}})
\sum _{n,m} \frac{1}{(\frac{1}{2\sqrt{\alpha '}}(2n+1)i +
\frac{(2m+1)i}{2R} -t_{0})^{2}}=
\sum _{n} \frac{1}{(\frac{1}{2\sqrt{\alpha '}}
(2n+1)i +\frac{i}{2R} -t_{0})^{2}}
$$ 
\begin{equation}
=e^{-\frac{i}{2R}\frac{\partial}{\partial t_{0}}}
\sum _{n} \frac{1}{(\frac{1}{2\sqrt{\alpha '}}(2n+1)i -t_{0})^{2}}=
e^{-\frac{i}{2R}\frac{\partial}{\partial t_{0}}}
\frac{\partial}{\partial t_{0}} \sum _{n} 
\frac{1}{\frac{1}{2 \sqrt{\alpha '}}(2n+1)i -t_{0}} .
\end{equation}
Hence
\begin{equation}
\frac{1}{R} \sum _{n,m} \frac{1}{(\frac{1}{2\sqrt{\alpha '}}(2n+1)i +
\frac{(2m+1)i}{2R} -t_{0})^{2}}=
\frac{\frac{1}{R}\frac{\partial}{\partial t_{0}}}
{e^{\frac{i}{2R}\frac{\partial}{\partial t_{0}}}-
e^{-\frac{i}{2R}\frac{\partial}{\partial t_{0}}}}
\sum _{n} \frac{1}{\frac{1}{2\sqrt{\alpha '}}(2n+1)i -t_{0}} .
\end{equation}
This verifies the result of \cite{low1} using a different method.

There are some interesting features of this derivation. Duality is very
explicit in the expression (\ref{e3}). That is, if we do not care about the
vortices as discussed in the introduction.
We can in fact learn a bit about how it
is realized in the matrix model. Apparently duality relies on the fact
that the eigenvalue spectrum of a harmonic oscillator is the same as
for a fermion with anti periodic boundary conditions in a box! Clearly
one cannot expect duality to hold in any matrix model. In general a 
nonharmonic potential would have nonequally spaced eigenvalues and hence
break duality.

Curiously, by the same reasoning, many correlation functions will break
duality. This is not surprising for nonzero momentum correlation
functions. However, the above expressions suggest this to be the case
even at {\it zero} momentum for the special states.

\section{Summary}

In this chapter a simplified derivation of some results for finite
radius has been given. We have indicated how duality is realized in the
matrix model. An important observation is that na\"{\i}ve duality is not
expected to hold in a general matrix model. 

\chapter{Zero Momentum Correlation Functions}

\section{Introduction}

A first exercise in the two dimensional string theory is the
calculation of zero momentum correlation functions. Clearly this is the
simplest thing to do. Such calculations will be described in this
chapter. We will, apart from some comments
towards the end of the chapter, limit ourselves to powers of the matrix
model eigenvalue. The conjugate momentum will be introduced in later
chapters. One of the main results of this chapter will be a set of
recursion relations for these correlation functions.

\section{The Gelfand-Dikii Equation}

Let us briefly recall the basics of the matrix model approach to $c=1$ quantum
gravity or two dimensional string theory as described in chapter 4. We will limit ourselves to the uncompactified case.
The partition function is given by
\begin{equation}
\int {\cal D} \phi e^{-\beta \int Tr (\frac{1}{2}\dot{\phi}^{2} + U(\phi )) dt},
\label{1}
\end{equation}
where $\phi$ is a hermitean $N \times N$ matrix. 
As shown in \cite{bre2}, the angular variables may be integrated out to 
leave only the $N$
eigenvalues. They correspond to $N$ fermions in the potential $U$. Below
we will consider correlation functions of precisely these eigenvalues.

As we have seen, the density of states, $\rho$, in the potential, is the
same as the 
1PI puncture two point 
function and hence a
very interesting and basic quantity in the theory.
In the coordinate basis it can be written as
\begin{equation}
\rho (\mu _{F}) = \frac{1}{\pi \beta} 
{\rm Im} \int d\lambda <\lambda \mid \frac{1}{\hat{H} - \mu _{F} -i\epsilon} 
\mid \lambda>
\end{equation}
By investigating some properties of this expression, we will be able to
extract information about zero momentum correlation functions of the
matrix eigenvalue variable.

We begin by defining the function
\begin{equation}
R(\lambda ) = <\lambda \mid \frac{1}{- \frac{1}{2} \nabla _{\lambda }^{2} 
+U(\lambda )}
\mid \lambda >,
\label{5}
\end{equation}
i.e. the diagonal component of the resolvent. This function obeys the
nonlinear Gelfand-Dikii equation \cite{ge1}:
\begin{equation}
-2RR ^{\prime \prime} + (R^{ \prime} )^{2} + 8 U(\lambda ) R^{2} =4.
\label{6}
\end{equation}
Here $\lambda$ is shifted so that the maximum of the potential is at the origin.
We have also shifted $U$ by
$\mu _{F} = \mu _{c} - \mu $.
In the double scaling limit we expand $U(\lambda )$ around
its maximum  and rescale the coordinate.  
If $U(\lambda ) 
\sim \lambda ^{n}$,
then  $\lambda \sim \beta ^{\frac{2}{n+2}}$.  
In the following it is assumed that 
all quantities are rescaled in the appropriate fashion.

The correlation functions
we are going to calculate will be correlation functions of time independent
perturbations, $\lambda ^{k}$, around the top of the potential. 
The corresponding
operators we call $O_{k}$. They are essentially the operators
${1\over N} {\rm Tr}[\phi -\phi_c]^k$. To  perform the calculations we include
in $U$ general perturbations $\lambda ^k$ coupled to sources $t_{k}$, which 
we later will vary.
Taking one derivative of (\ref{6}) gives the linearized version of
the Gelfand-Dikii equation,
\begin{equation}
-\frac{1}{2} R ^{\prime \prime \prime} + 4UR^{ \prime} +2U^{\prime} R =0.
\label{7}
\end{equation}
This formula is also easily verified by using
$R(\lambda ) = \sum _{n=0}^{\infty} 
\frac{ \psi ^{\dagger} _{n} (\lambda ) \psi _{n} (\lambda ) }{E_{n}}$,
and the Schr\"{o}dinger equation for the energy eigenfunctions 
$\psi _{n} (\lambda )$.
Note that we are dealing with an inverted potential and the $E_{n}$'s are
therefore, in general, complex.
It is this linearized version of the Gelfand-Dikii equation which we will
exploit in the following in order to get information about 
the correlation functions of the theory.

The resolvent $R(\lambda )$ determines the density of states,
\begin{equation}
\rho (\mu ) = \frac{1}{\pi} {\rm Im} \int  R(\lambda ) d\lambda ,
\label{9}
\end{equation}
which is also  the puncture two point function $<PP>$. Other
two point functions can be obtained by taking the moments of $R$,
\begin{equation}
< O _{k} P> = \frac{1}{\pi} {\rm Im} \int  \lambda ^{k} R(\lambda ) d\lambda .
\label{10}
\end{equation}
These amplitudes are one particle irreducible (1PI) with respect 
to the puncture.

Some comments on the integration region are in order. There are
essentially two different cases depending on whether we have one or two
worlds as explained in chapter 4. From the point of
view of the WKB approximation we should integrate only up to the top of of the potential, {\sl
i.e.} 
from $-\infty$ to zero in the case of one world. With two worlds, we
should
integrate all the way from
$-\infty$ to $+\infty$. This is effectively
a doubling of the system.
At a critical point, where we just have a leading even power contribution
to $U(\lambda )$, $R(\lambda )$ is clearly even. 
This immediately implies that, with two worlds,
(\ref{10}) is zero for $k$ odd. For the quadratic
critical point this is
true also with one world. 
The above integral for an odd operator turns out to be real
and therefore gives no contribution to the imaginary part.

\section{1PI Recursion Relations}

We will now use the linearized Gelfand-Dikii equation to derive recursion
relations for the 1PI amplitudes.
We start with (\ref{7}) for a general potential
\begin{equation}
U(\lambda ) = \sum _{p=0} ^{\infty} t_{p} \lambda ^{p}.
\end{equation}
The $t_{p}$'s will act as sources for the $O_{p}$ operators.
We will later restrict ourselves to the quadratic critical point
\begin{equation}
t_{i} =0 \hspace{1em} \forall i \neq 0,2  \hspace{2em}
 t_{0}= \beta \mu \hspace{2em} t_{2} = -\frac{1}{2\alpha '} < 0.
\label{12}
\end{equation}
We then compute the $\lambda ^{k+1}$ moment of (\ref{7}), obtaining
\begin{equation}
-\frac{1}{2} (k+1)k(k-1) <O_{k-2} P> +
2\sum_{p=0} ^{\infty} t_{p} (2k+p+2)<O_{k+p}P>=0.
\label{f11}
\end{equation}
To obtain this, we have made repeated partial integrations and 
discarded boundary terms at infinity which do not contribute to 
the singular part of the correlation 
functions. In the case of one rather than two worlds, one might
worry about boundary terms at $\lambda =0$. However these are real, as
follows immediately from the WKB expansion of $R(\lambda )$ in terms of
$\sqrt{U(\lambda )}$. Hence, they do not contribute 
to the imaginary parts of the
integrals, which give the correlation functions.
Using $<O_{p}...> = \frac{\partial}{\partial t_{p}} <...>$ we
may take $t_{p}$ ($p>0$) derivatives of (\ref{f11}) to derive
\begin{equation}
\begin{array}{c}
-\frac{1}{2} (k-1)k(k+1)<O_{k-2} O_{k_{1}} \dots O_{k_{p-1}}P> \\ \\
+4t_{0} (k+1)<O_{k} O_{k_{1}}\dots  O_{k_{p-1}} P> 
+4t_{2} (k+2)<O_{k+2} O_{k_{1}} \dots O_{k_{p-1}} P> \\ \\
+2\sum_{i=1}^{p-1} (2k +k_{i} +2) <O_{k+k_{i}} O_{k_{1}}...\hat{O}_{k_{i}} 
...O_{k_{p-1}}P>= 0. \label{14}
\end{array}
\end{equation}
For simplicity we have chosen the quadratic critical point of (\ref{12}). 
When we take the 
derivatives, $t_{0} = \beta \mu$ is kept constant. This is because 
$\beta \mu$ is the nonamputated puncture one point function and 
we are considering
1PI amplitudes. 

With this recursion relation any zero momentum correlation function for
arbitrary genus may be expressed in terms of the puncture
two point function.
Using the fact that the genus g contribution to the correlation function depends
 on
$t_0$ according to
$<O_{k_{0}}...O_{k_{p-1}} P>_{g} \sim t_{0}^{\frac{1}{2} 
\sum_{i=0}^{i=p-1} k_{i}+1-p-2g}$,
we can write a formula where we
exhibit the explicit genus dependence:
\begin{equation}
\begin{array}{c}
-\frac{1}{2} (k-1)k(k+1)<O_{k-2} O_{k_{1}} ... O_{k_{p-1}}P>_{g-1} \\ \\
+4t_{0} (k+1)<O_{k} O_{k_{1}} ... O_{k_{p-1}}P>_{g}
+4t_{2} (k+2)<O_{k+2} O_{k_{1}} ... O_{k_{p-1}}P>_{g} \\ \\ 
+2 \sum _{i=1}^{p-1} (2k +k_{i} +2) <O_{k+k_{i}}O_{k_{1}}...\hat{O}_{k_{i}}
...O_{k_{p-1}} P>_{g}=0, \label{15}
\end{array}
\end{equation}
where the first term is understood to be zero for $g=0$.

These recursion relations are similar to the ones that exist for $c<1$. The genus mixing
first term looks like a contribution from a pinched handle, while the last term
corresponds to contact terms between operators. What is lacking
are the quadratic terms which arise when the surface is pinched into two separate surfaces.
The reason these are absent is that we have considered 1PI amplitudes with respect
to the puncture which precisely excludes such terms. The corresponding
recursion relations for connected amplitudes will be derived in the 
next section.

It is straightforward to check that the results for the sphere, 
obtained in \cite{gr2} using different methods:
\begin{equation}
<O_{k_{0}} O_{k_{1}} ... O_{k_{p-1}} P>_{0} =
 - \mid \ln \mu \mid \left( \begin{array}{c}
                      2\Sigma \\ \Sigma
                     \end{array}  \right)
\frac{1}{\pi \sqrt{2|t_{2}|}} \frac{1}{(-4t_{2})^{\Sigma}}
\frac{\partial ^{p-1}}{\partial t_{0} ^{p-1}} t_{0} ^{\Sigma},
\label{rek0}
\end{equation}
for even $k_{i}$'s, satisfy (\ref{15}), where $\Sigma = \frac{1}{2}\sum _{i=0}
^{p-1} k_{i}$.
Note that the first term of (\ref{15}) does not
contribute. 
The factors of $t_{2}$ assure that $t_{2}$ derivatives generate $O_{2}$
insertions. Note that, in using the recursion relations, we have to fix the
normalization $<PP>_{0} = - \frac{1}{2\pi} |\ln \mu |$ for $t_{2} =-2$
($\alpha ' = 1/4$).
The formula is also true for odd $k_{i}$'s, unless $2\Sigma$ is odd, in
which case the correlation functions are zero. This follows from (\ref{14}) for
$k=-1$, $p=1$ which states that $<O_{1} P> =0$, as we discussed in the
previous section.

Let us explicitly evaluate a simple example of a correlation function at
arbitrary genus. We consider the operator $O_{2}$. 
If we put $k=0$ in (\ref{15})
the first term vanishes, i.e. no genus mixing, and we are left with
\begin{equation}
\begin{array}{c}
8t_{2} <O_{2} O_{k_{1}}...O_{k_{p-1}}P>_{g} \\ \\
+4t_{0} <O_{k_{1}}...O_{k_{p-1}}PP>_{g} 
+ (4(p-1) +4 \Sigma )<O_{k_{1}}...O_{k_{p-1}}P>_{g}=0,
\end{array}
\end{equation}
i.e.
\begin{equation}
2t_{2} <O_{2} O_{k_{1}}...O_{k_{p-1}}P>_{g}
=-(p-1 + \Sigma + t_{0} \frac{\partial}{\partial t_{0}})
<O_{k_{1}}...O_{k_{p-1}}P>_{g}.
\end{equation}
Since $<O_{k_{1}} ... O_{k_{p-1}} P>_{g} \sim t_{0} ^{\Sigma +2-p-2g}$
we find
\begin{equation}
2t_{2} <O_{2} O_{k_{1}}...O_{k_{p-1}}P>_{g}
=-(2 \Sigma +1-2g)<O_{k_{1}}...O_{k_{p-1}}P>_{g}.
\end{equation}
This expression is similar to the dilaton equation obtained in \cite{dij1} for
topological gravity.
In particular
\begin{equation}
<\overbrace{O_{2}...O_{2}}^{n}>_{g} = 
\bigl(\frac{-1}{2t_{2}}\bigr)^{n} \prod _{p=1}^{n} (2p-1-2g)<>_{g}.
\end{equation}
Note that $<O_{2}>_{g} = \frac{1}{2t_{2}}(2g-1)<>_{g}$, which suggests
that $O_{2}$ is almost, but not quite, the dilaton.

\section{Connected Recursion Relations}

Let us now derive recursion relations for connected amplitudes. This is not
more difficult than in the 1PI case. 
In the case of the 1PI (with respect to the puncture) amplitudes 
the generating functional, $\Gamma$, depends
on the nonpuncture sources $t_{k}$ ($k \geq 1$) and the nonamputated puncture
one point function $t_{0} = \beta \mu$. $t_{0}$ is kept constant while the
other $t_{k}$'s vary. The generating functional for connected
amplitudes,
the vacuum energy $E$, depends instead on the 
$t_{k}$ ($k \geq 1$) and $\Delta$, the
puncture source, i.e. the cosmological constant. In this case $\Delta$ is
kept constant when the $t_{k}$'s vary and $t_{0}$ is a function
of the $t_{k}$'s. $\Delta$ is also the 
amputated one point function. $\Gamma$ and $E$ are related by the Legendre
transform
\begin{equation}
E(\Delta )=\beta ^{2} \Delta \mu - \Gamma (\mu ).
\end{equation}
We have
\begin{equation}
\beta \Delta = \frac{\partial \Gamma}{\partial (\beta \mu )}, \hspace{2em}
\beta \mu = \frac{\partial E}{\partial (\beta  \Delta )},\hspace{2em}
\frac{\partial ^{2} E}{\partial t_{k} \partial (\beta \Delta )} =
- \frac{\partial \mu}{\partial \Delta} \frac{\partial ^{2} \Gamma}{\partial
t_{k} \partial (\beta \mu )},
\end{equation}
as a consequence of
\begin{equation}
0= \frac{d\mu}{dt_{k}} = \frac{\partial \mu}{\partial t_{k}} +
\frac{\partial \mu}{\partial \Delta} \frac{\partial \Delta}{\partial t_{k}}.
\end{equation}
The 1PI and connected two point functions are simply proportional to each other.
The same recursion relation is valid. The difference comes when we take 
$t_{k}$ derivatives to obtain higher point functions. For connected diagrams
$t_{0} =\beta \mu$ is no longer to be kept constant. The resulting recursion
relations are no longer linear but rather quadratic
\begin{equation}
\begin{array}{c}
{-\frac{1}{2} (k+1)k(k-1) <O_{k-2} O_{k_{1}} ... O_{k_{p-1}} P>_{c}
+2t_{2} (2k+4) <O_{k+2} O_{k_{1}} ... O_{k_{p-1}} P>_{c} } \\ \\ 
+ 4(k+1)\sum _{i\in k}
<O_{i_{1}}...O_{i_{n}}P>_{c} <O_{k} O_{i_{n+1}} ... O_{i_{p-1}} P>_{c}
\\ \\  
+2 \sum _{i=1}^{p-1} (2k +k_{i} +2) <O_{k+k_{i}} O_{k_{1}} ... \hat{O}_{k_{i}}
...O_{k_{p-1}} P>_{c}  =  0 . \label{25}
\end{array}
\end{equation}
Note that $t_{0} = <P>_{c}$.
We recognize an extra term suggestive of a surface pinching into two. If we
extract the explicit genus dependence we find
\begin{equation}
\begin{array}{c}
-\frac{1}{2} (k+1)k(k-1) <O_{k-2} O_{k_{1}} ... O_{k_{p-1}} P>_{c,g-1} \\ \\
+4(k+1)\sum _{g_{1}+g_{2}=g}\sum _{i\in k}
<O_{i_{1}}...O_{i_{n}}P>_{c,g_{1}} 
<O_{k} O_{i_{n+1}} ... O_{i_{p-1}} P>_{c,g_{2}}\\ \\ 
 +4t_{2} (k+2) <O_{k+2} O_{k_{1}} ... O_{k_{p-1}} P>_{c,g} \\ \\
+ 2 \sum _{i=1}^{p-1} (2k +k_{i} +2) <O_{k+k_{i}} O_{k_{1}} ... \hat{O}_{k_{i}}
...O_{k_{p-1}} P>_{c,g}  =  0. 
\end{array}
\end{equation}
This structure is very similar to the one encountered in $c<1$ theories
\cite{dij1}.

All connected amplitudes may also be built by using the 1PI building blocks.
Let us consider a simple example. Consider
\begin{equation}
<O_{k} P>_{c} = -\frac{\partial \mu}{\partial \Delta} <O_{k} P> =
-\frac{1}{\rho} <O_{k} P>.
\end{equation}
The $1/\rho$ prefactor simply adds an external propagator as it should. Now
take a $\frac{\partial}{\partial t_{l}} \mid _{\Delta = const}$ derivative.
We get
\begin{equation}
\begin{array}{c}
<O_{k}O_{l}P>_{c} = ( \frac{1}{\rho ^{2}} 
\frac{\partial \rho}{\partial t_{l}} + \frac{1}{\rho ^{2}} 
\frac{\partial \rho}{\partial \mu} 
\frac{\partial \mu}{\partial t_{l}} )<O_{k}P>  \\ \\
- \frac{1}{\rho} (<O_{k}O_{l}P> + \frac{\partial \mu}{\partial t_{l}}
<O_{k} PP>). 
\end{array}
\end{equation}
Now use $\frac{\partial \Delta}{\partial t_{l}} + 
\frac{\partial \Delta}{\partial \mu} \frac{\partial \mu}{\partial t_{l}}=0$
to obtain
\begin{equation}
\begin{array}{c}
<O_{k}O_{l}P>_{c}  = 
\frac{1}{\rho ^{2}} <O_{l}PP><PO_{k}> 
-\frac{1}{\rho ^{3}} <O_{k}P><PPP><PO_{l}> \\ \\
-\frac{1}{\rho} <O_{l}O_{k}P> 
+\frac{1}{\rho ^{2}} <O_{l}P><PPO_{k}>. 
\end{array}
\end{equation}
It is easily checked that this expression indeed obeys (\ref{25}) using the
recursion relations for the 1PI amplitudes. The $1/\rho$ factors provide
external and internal propagators where needed.

\section{Virasoro Constraints}

There is an interesting way of rewriting the recursion relations (\ref{f11}),
which explains where they really come from.
They are simply equivalent to
\begin{equation}
L_{k} <P> =0 \hspace{2em} \forall k \geq -1,
\end{equation}
where
\begin{equation}
L_{k} = - \frac{1}{4} (k+1)k(k-1) \frac{\partial}{\partial t_{k-2}}
+\sum _{p=0}^{\infty} t_{p} (2k+p+2) \frac{\partial}{\partial t_{p+k}}.
\end{equation}
It is straightforward to verify that the $L_{k}$'s obey a Virasoro algebra
\begin{equation}
[L_{n},L_{m}] = (n-m) L_{n+m}.
\end{equation}
The Virasoro generators can be understood as representing the algebra of
coordinate transformations in the Liouville coordinate, i.e.
$L_{k-1} \sim \lambda ^{k} \frac{\partial}{\partial \lambda }$.
Let us see how this works. In the full quantum treatment, it is
convenient to use the matrix eigenvalue $\lambda$ and its conjugate
momentum $p$. The generator of the coordinate transformation is then
$p\lambda ^{k}$. One can convince oneself that one need not worry
about different orderings in this case. Acting on the Hamiltonian we
find two parts,
\begin{equation}
[p\lambda ^{k} ,U(\lambda )]= \lambda ^{k} \frac{\partial U}{\partial
\lambda}
\end{equation}
and
\begin{equation}
[p\lambda ^{k}, p^{2}] =k(\lambda ^{k-1} p^{2} +p^{2} \lambda
^{k-1})-\frac{k}{2} (k-1)(k-2)\lambda ^{k-3} .
\end{equation}
When this is evaluated at the Fermi surface in accordance with the
variational prescription of (\ref{c23}) and (\ref{c24}), the
$\frac{p^{2}}{2}$ gives a $U(\lambda )$. Hence
\begin{equation}
<(-\frac{1}{4}k(k-1)(k-2)\lambda ^{k-3}
+ 2k \lambda ^{k-1} U(\lambda ) +\lambda ^{k}
\frac{\partial U}{\partial \lambda} )P> =0
\end {equation}
which is precisely equal to  (\ref{f11}).

Unfortunately, the $L_{-1}$ constraint for the quadratic critical point is
quite trivial
\begin{equation}
L_{-1}<P> =2t_{2} <O_{1}P>=0 .
\end{equation}
It does not seem to yield a string equation determining $<PP>$, unlike
the case for the $c<1$ models, \cite{dij1}.

A natural generalization is obviously to consider general
transformations $p^{n}\lambda ^{m}$. As far as $\lambda ^{k}$
correlation functions goes, this will not give anything new. The constraints
one obtains are simply linear combinations of the old ones. Although,
of course, they represent the full algebra not only the Virasoro part.

\section{Summary}

In this chapter we have seen how to obtain recursion relations for 
both 1PI and connected
zero momentum correlation functions. With the help of these, any zero momentum
correlation function for any genus may be calculated, once we know the
puncture two point function. The recursion relations were furthermore found
to correspond to coordinate transformations in the Liouville coordinate.
This was the reason why our recursion relations could be expressed as
constraints obeying a Virasoro algebra.
The presence of the Virasoro algebra, and as we already have noted, a
full $W_{\infty}$ algebra, is of utmost importance. We will in chapter
nine and ten take full advantage of this algebra.

By
adding time independent operators to the potential the critical behavior 
may change. 
The recursion relations and their Virasoro representation should be useful
for studying such other critical points, different from 
the ordinary quadratic. They
should also describe the flows between different critical points.
Although it is
not clear how to identify these new theories, it is easy to extract some
information about them.
The Virasoro constraints which we have derived are valid for any 
potential and provide relations between different correlation functions. 

As we have seen, the recursion relations do not determine $<PP>$. 
This has to be specified
independently, before the recursion relations can determine
all other correlation functions.
In general, when we consider some nonquadratic 
potential, more data need to be specified.
For a theory with a $\lambda ^{2n}$ as the leading term
in the potential, we will need to specify $<PP>$, $<PO_{1}>$,...,$<PO_{2n-2}>$.
The correlation $<PO_{2n-1}>$ is always zero. The scaling of $<PP>$ 
can be derived directly from
the WKB approximation and is given by $\mu ^{1/2n-1/2}$.

So much for zero momentum correlation functions. In the next chapter we
will generalize to nonzero momentum. While doing so, the importance of the
special states will become even clearer.

\chapter{Nonzero Momentum Correlation Functions}

\section{Introduction}

Let us now consider the matrix model and its special states,
also for nonzero momentum.
 
As we have seen the natural candidates for special state correlation 
functions are
correlation functions of powers of the matrix model eigenvalues. Such
objects were studied
in \cite{art2} and the expected poles for discrete momenta were
found. In this chapter we will
extend some of the results of \cite{art2} and the previous chapter, for Ward
identities. In particular, we will find that the recursion relation 
(\ref{f11}) is
in fact the Wheeler de Witt equation for the two dimensional world sheet
quantum gravity.

The recent developments revealing the $W_{\infty}$ symmetries
indicate however that this is not the whole story. One should also
consider correlation functions involving powers of the conjugate
momentum. This we will do in the next chapter, where we will
show the existence of this
$W_{\infty}$ symmetry which will greatly simplify the subsequent
calculations.

\section{The Wheeler de Witt Equation}

Ward identities for correlation functions are in general obtained by changes of
variables in the path integral. Examples of such Ward identities were obtained
in the last chapter from simple coordinate changes in the matrix eigenvalue. 
They can be
thought of as generated by commutators, or classically,
i.e. on the sphere,
Poisson brackets with $p\lambda ^{m}$. 
They obviously obey a Virasoro algebra, which is part of a $W_{\infty}$
algebra generated by all monomials $p^{n} \lambda ^{m}$. We may also introduce
time dependence and consider generators with certain momenta $q$, i.e.
$p^{n} \lambda ^{m} e^{iqt}$.
Let us, as an example, make a $p \lambda ^{k} e^{iqt}$ variation of the 
one puncture function. 

As we saw in chapter three, the two puncture function is schematically given by
\begin{equation}
<PP> = \frac{1}{\pi}
{\rm Im} \int_{0}^{\infty} dT \int d\lambda  G(\lambda , \lambda ;T) ,
\end{equation}
where the calculation is done at the Fermi surface. We
will shift its energy to zero, hence putting the Fermi energy as a
constant term in the
potential. $G$ is the path integral given by
\begin{equation}
G(\lambda _{1},\lambda _{2}) = \int _{\lambda _{1}}^{\lambda _{2}}
[dp d\lambda] e^{-\beta \int dt(p\dot{\lambda }-\frac{1}{2}p^{2}+U(\lambda
))}, \label{ban}
\end{equation}
in Euclidean time,
where $U(\lambda)=\sum _{p} t_{p} \lambda ^{p}$ is the potential.
The variation of the partition function would have involved a sum over
all states in the Fermi sea up to the Fermi level. By  inserting a
puncture, i.e. taking a derivative with respect to the Fermi energy as
in the last chapter, we
restrict ourselves to the Fermi surface. Next we perform the variation
of $<P>$. The measure, as given by (\ref{ban}), is invariant under the change of
variables. (Clearly, we are not supposed to differentiate with respect
to $t$ when changing variables in the measure.) 
The change in the action gives rise to the following identity
among two point functions
\begin{equation}
< (\int dt (iq \lambda ^{k}p +k\lambda ^{k-1}p^{2}+
\frac{\partial U}{\partial \lambda} \lambda ^{k})e^{iqt}) T>=0 .
\end{equation}
The puncture is now a tachyon, T,  carrying away the momentum.
The piece with a single $p$ is evaluated by integrating over $p$ in the
path integral, obtaining a $\dot{\lambda }$ which then is partially
integrated. We then switch to a Hamiltonian formulation, remembering
that we should use Weyl ordering.
We finally obtain
\begin{equation}
-\frac{1}{4}k(k-1)(k-2)<O_{k-3}P>_{g-1}+    \label{rec}
\sum _{p} (2k+p)t_{p}<O_{p+k-1}P>_{g} =0,
\end{equation}
to all genus at zero momentum, as in the last chapter, and
\begin{equation}
\sum _{p} (2k+p)t_{p}<O_{p+k-1}T>_{q,g}
+ \frac{q^{2}}{1+k}<O_{k+1}T>_{q,g} =0
\end{equation} 
on the sphere for nonzero momentum. The combined case is somewhat more
complicated due to the need of taking an imaginary part and an effective
imaginary shift $iq$ in $t_{0}$, relevant for higher genus. We will
not consider this any further.
In these equations $q$ indicates momentum and $g$ genus. We have introduced the
notation $O_{k}$ for $\lambda ^{k}$. 
The different genus for the
first term in (\ref{rec}) 
is due to a $p$, $\lambda$ commutator which arises when we
want to evaluate the $p^{2}$ against the wave function of the Fermi
surface. This gives an $\hbar$, which is the same as the
genus coupling constant.
Following \cite{mo1}
we may define the loop operator given by
\begin{equation}
w(l)=e^{l\lambda } . \label{loop}
\end{equation}
It corresponds to cutting out a hole in the surface with a boundary of
length $l$. The reason is the following. If we insert a power $n$ of the
original matrix variable $\phi$ on the surface this creates a little hole,
the length of the boundary being proportional to $n$ (the number of legs) and
the lattice spacing $a$. For
fixed $n$ the length clearly shrinks to zero in the double scaling
limit. To get a finite length we must also take $n$ to infinity.
Introducing $\lambda$ as the eigenvalues $m$ of $\phi$
expanded around the top of the potential,
we find in the double scaling limit
\begin{equation}
m^{n} \sim (1+a\lambda )^{l/a} \rightarrow e^{l\lambda} .
\end{equation}
We may then Fourier transform to obtain a differential equation in the
loop length. We get
\begin{equation}
[\sum _{p} t_{p} (l^{2} \frac{\partial ^{p}}{\partial l ^{p}}
+\frac{p}{2}l\frac{\partial
^{p-1}}{\partial l^{p-1}})+t_{0}l^{2} - l^{4}
]<w(l)T>=0 
\end{equation}
at zero momentum and
\begin{equation}
[\sum _{p} t_{p} (l^{2} \frac{\partial ^{p}}{\partial l ^{p}}
+\frac{p}{2}l\frac{\partial
^{p-1}}{\partial l^{p-1}})+t_{0}l^{2}
+q^{2}]<w(l)T>=0 \label{WdW}
\end{equation}
on the sphere.
The $l^{4}$ term is of order one higher in the string coupling and
do not contribute on the sphere. $t_{0}$ is the Fermi energy $\mu$ with
the appropriate number of $\beta$'s absorbed. 
In the case of the usual harmonic oscillator potential, where
$t_{0}=\beta \mu$, the 
resulting equations are in fact the Wheeler de  Witt equations obtained in 
\cite{mo1}. There they were inferred from the explicitly computed two
point function.

On the sphere this is one of
the most striking verifications of the equivalence of the Liouville and
matrix models. At zero momentum, the 
Wheeler de Witt equation is
just the Fourier transform of the Gelfand-Diiki equation for the resolvent
for the Schr\"{o}dinger operator. This was the way in which the zero momentum
version of (\ref{rec}) was derived in the last chapter. If we want to be
careful we need to rescale $\lambda$ by $\sqrt{-t_{2}}$
to get a dimensionless $l$, see section 9.4.
This is needed for the exact correspondence
between the above result and the mini super space canonical quantization
of Liouville theory. Recall that, \cite{gr4}, $t_{2}=-\frac{1}{2\alpha
'}$. 
From (\ref{WdW}) one might try to draw some
conclusions about the Liouville theory correspondence to the more
general potentials above. Clearly the last term, which corresponds to
the matter piece, does not change while we change potential. Instead, it
is the piece which would be expected to arise from a canonically
quantized kinetical term for the Liouville mode, which gets modified.
Hence one is led to the conclusion that these more general models
(however with $p$ independent potentials)
may correspond, not to modifications of the matter theory, but rather to 
different theories for the Liouville part. This is also consistent with
the point of view for which this paper will argue, that the special
states must be represented using both $\lambda$'s and $p$'s, not just
the $\lambda$'s.

\section{Two Point Function on the Sphere}

So far we have seen how to derive some simple recursion relations for
various correlation functions of the matrix eigenvalue. We will now turn
to some other methods for calculating correlation functions. We will
obtain results for all genus, but we will begin by
deriving an expression for the two point function of
special operators with nonzero momentum on the sphere.
 
We will use the same method as in \cite{gr4}.
The two point function is then given by
\begin{equation}
<O_{r}O_{s}>
= \frac{1}{(2\pi )^{2}} 
\int_{{-T\over 2}}^{{T\over 2}} d\tau _{1} d\tau _{2} O_{r}(\tau _{1}) O_{s}(\tau _{2})
\sum _{n=1}^{\infty} \frac{n^{2} \omega ^{3}}{p^{2} + n^{2}\omega ^{2}}
\cos n \omega \tau _{1} \cos n \omega \tau _{2}, \nonumber \\
\label{sfar}
\end{equation}
where $T= 2\sqrt{\alpha '}| \log \mu |$ is the period of 
the classical motion and 
$\omega = 2\pi / T$. $\tau$ is the classical time of flight variable,
which for the present purposes may be though of as
the Liouville mode. 
We have chosen $\tau$ to be zero at the classical turning point. This is the
natural choice since the critical behavior arises from this region.
The classical equation of motion at the Fermi surface is given by
\begin{equation}
\frac{d\lambda}{d\tau} = \sqrt{2(\mu _{F}- U(\lambda )} \label{diffekv}\,\,{\rm for} \,
U(\lambda ) = 
\frac{1}{4 \alpha '} (\lambda ^{2} - \lambda ^{4}).
\end{equation}
If we expand about the top of the potential, $\lambda = x + \frac{1}{\sqrt{2}}$,
keeping only the quadratic term, we get
\begin{equation}
\frac{dx}{d\tau} = \sqrt{\frac{x ^{2}}{\alpha '} -2\mu}
\Rightarrow x (\tau ) = \sqrt{2\alpha '} \mu ^{1/2} \cosh
\frac{\tau}{\sqrt{\alpha '}}.
\label{klass}
\end{equation}
Hence
\begin{equation}
O_{r} = x ^{r} = (2\alpha ')^{r/2} 
\mu ^{r/2} \cosh ^{r} \frac{\tau}{\sqrt{\alpha '}}.
\label{forst}
\end{equation}
If we now tried to use (\ref{forst}) in (\ref{sfar}) and perform the Fourier
transforms we would get in trouble. As $\mu \rightarrow 0$ and 
$T \rightarrow \infty$, (\ref{forst}) diverges at the limits of the integration.
The reason is that our approximation is good only up to 
$|\tau| = T/4$ and breaks down thereafter. One way to take care of this is to
simply put in a cutoff. The critical behavior should not 
depend on the nature of the cutoff.
Let us however be slightly more careful and
study the exact solution of (\ref{diffekv}) in terms of elliptic functions.
With our convention for $\tau$, the solution is
\begin{equation}
\lambda (\tau ) = \lambda _{-} cd (\frac{\lambda _{+}}{\sqrt{2\alpha '}}\tau )
\hspace{1in}
\lambda _{\pm} ^{2} = \frac{1}{2} \pm 2\sqrt{\alpha '} \mu ^{1/2}.
\end{equation}
The function $cd$ is the same as $sn$ translated a quarter of a period, and can
be
written as
$cd(u) = \frac{cn(u)}{dn(u)}$,
where for small $\mu$ the elliptic functions $cn$ and $dn$ are given by
\cite{abr1},
\begin{equation}
cn(u) \sim \frac{1}{\cosh ^{2} u} - \frac{m_{1}}{4} ( \tanh ^{2} u -
\frac{u \sinh u}{\cosh ^{2} u} ),
\end{equation}
and
\begin{equation}
dn(u) \sim \frac{1}{\cosh ^{2} u} + \frac{m_{1}}{4} ( \tanh ^{2} u +
\frac{u \sinh u}{\cosh ^{2} u } ) ,
\end{equation}
where
$m_{1} = 1 - (\frac{\lambda _{-}}{\lambda _{+}})^{2}$.
Consider the expansion
$cd(u) \sim \frac{1 - \frac{m_{1}}{4} ( \sinh ^{2} u - u \tanh u )}
{1 + \frac{m_{1}}{4} ( \sinh ^{2} u + u \tanh u )}$,
valid to only first order in $m_{1}$.
For $\frac{m_{1}}{4} (\sinh ^{2} u + u \tanh u ) < 1$, we may expand the
denominator to get
\begin{equation}
cd(u) \sim 1 + \frac{m_{1}}{4} - \frac{m_{1}}{4} \cosh 2u .
\end{equation}
From this we recover (\ref{klass}), but now we see that it has limited validity.
As $\mu$ and $m_{1}$ go to zero, the expansion can not be trusted above
$|\tau | = T/4$. 
Performing the Fourier transforms up to $|\tau | = T/4$ we obtain
\begin{eqnarray}
(\frac{\alpha '}{2})^{r/2} \mu ^{r/2} \sum _{k=0} ^{r} \left( \begin{array}{c}
                                     r \\ k
                                 \end{array} \right)
\frac{\frac{4}{\sqrt{\alpha '}}(r/2 -k) 
\sinh (r/2 -k)\frac{T}{2\sqrt{\alpha '}}}
{\frac{4}{\alpha '}(r/2-k)^{2} +n^{2}\omega ^{2}} (-1)^{n/2} ,
\end{eqnarray} 
for $n$ even and
\begin{eqnarray}
(\frac{\alpha '}{2})^{r/2} \mu ^{r/2} \sum _{k=0} ^{r} \left( \begin{array}{c}
                                     r \\ k
                                 \end{array} \right)
\frac{2n \omega \cosh (r/2 -k)\frac{T}{2\sqrt{\alpha '}}}
{\frac{4}{\alpha '}(r/2-k)^{2} +n^{2}\omega ^{2}} (-1)^{(n-1)/2} ,
\end{eqnarray}
for $n$ odd. To compare with previous results, we assume the same contribution
from both tops of the potential (around $\tau =0$ and $\tau = T/2$). This 
doubles the Fourier transforms. We then need to rewrite the sum over $n$
as a contour integral, i.e.
$$
{\frac{1}{4\pi ^{2}} (\frac{\alpha '}{2})^{(r+s)/2}
\mu ^{(r+s)/2} \oint \frac{dz}{4i}
\frac{4z ^{2}}{p^{2} +z^{2}} 
\sum _{k=0}^{r} \sum _{l=0}^{s} 
\left( \begin{array}{c} r \\ k \end{array} \right)
\left( \begin{array}{c} s \\ l \end{array} \right) }
$$
$$
 \times \left( \cot \frac{\pi z}{2\omega} \frac{16(r/2 -k)(s/2-l)}{(4(r/2-k)^{2} +
\alpha 'z^{2})(4(s/2-l)^{2} +\alpha ' z^{2})} \right.
$$
$$ 
\times \sinh\left[ ({r \over 2}-k){T\over 2 \sqrt{\alpha '}}\right] 
\sinh\left[ ({s \over 2}-l){T\over 2 \sqrt{\alpha '}}\right]
$$
$$
 - \tan \frac{\pi z}{2\omega}
\frac{4\alpha ' z^{2}}{(4(r/2-k)^{2} +
\alpha 'z^{2})(4(s/2-l)^{2} +\alpha 'z^{2})}
$$
\begin{equation} \left.
\times \cosh\left[ ({r \over 2}-k){T\over 2 \sqrt{\alpha '}}\right] 
\cosh \left[ ({s \over 2}-l){T\over 2 \sqrt{\alpha '}}\right] \right) ,
\label{strulputt}
\end{equation}
where the contour wraps around the real axis. This is obtained by using
\begin{equation}
\sum _{n=even} ^{\infty} H(n)=\frac{1}{4i} \oint dz \cot (\frac{\pi z}{2})
H(z)\hspace{.1in} 
\sum_{n=odd} ^{\infty} H(n)=-\frac{1}{4i} \oint dz \tan (\frac{\pi z}{2}) H(z),
\end{equation}
and a rescaling $z \rightarrow z/\omega$.

We can rotate the contour in (\ref{strulputt}) to
wrap around the imaginary axis and evaluate the expression by picking out
the poles. The $z= \pm ip$ poles correspond to tachyon contributions while
the other poles are associated with the special operators we are concerned with.
There are both simple and double poles of this kind. The simple poles do not
give rise to any singularity in $\mu$.
From the
double poles we get $|\ln \mu |$ terms from derivatives of the
$\cot$ and $\tan$ factors. We finally obtain
\begin{equation}
(\frac{\alpha '}{2})^{(r+s)\over 2} \frac{\sqrt{\alpha '}}{\pi}
\mu ^{(r+s)\over 2} |{\rm ln} \mu | 
\sum _{k=0}^{r}
\left( \begin{array}{c} r \\ k \end{array} \right)
\left( \begin{array}{c} s \\ {(s+r)\over 2} -k \end{array} \right)
\frac{4({r\over 2}-k)^{2}}{\alpha 'p^{2} -4({r\over 2}-k)^{2}} \label{g0} ,
\end{equation}
where we have assumed $r\leq s$. Note that the $k=r/2$ term only 
contributes at $p=0$,
where the above expression indeed agrees with previous results for the zero
momentum special operators. Indeed, when $p=0$ we find
\begin{equation}
- (\frac{\alpha '}{2} )^{(r+s)/2} \frac{\sqrt{\alpha '}}{\pi} 
\mu ^{(r+s)/2} |\ln \mu | \left( \begin{array}{c}
                                   r+s \\ (r+s)/2

                                 \end{array} \right) ,
\end{equation}
in agreement with (\ref{rek0}) for $t_{2} =- \frac{1}{2\alpha '}$.
We will return later to a discussion of (\ref{g0}) and in particular the
significance of the poles.

One might note that the rotation of contours is precisely related to the
analytical continuation to a right side up oscillator which will be the
basis for the calculation in the next section.

\section{Two Point Function for all Genus}

We will now derive a formula for the special operator two point function which is valid
for all genus. The starting point will be the puncture two point function
\begin{equation}
<PP> = -\frac{1}{\pi } {\rm Im} \sum _{n=0}^{\infty} \frac{1}{E_{n} - t_{0}},
\end{equation}
where $E_{n}$ are the energy eigenvalues of the upside down potential and
hence imaginary. We will be especially interested in the quadratic
case where
$E_{n} = \frac{1}{2 \sqrt{\alpha '}}(2n +1)i$ and $t_{0} = \beta \mu$.
However, the following derivation is 
valid for the general case. We will conclude this section with a
few comments on the changes that occur when we have an arbitrary potential.

Correlation functions of arbitrary, in general time dependent, operators
are obtained by  adding these to the Hamiltonian and 
evaluating the energy eigenvalues
$E_{n}$ to the appropriate order in perturbation theory. This is a well defined
prescription even for time dependent perturbations.
To do this, we evaluate the Euclidean matrix element
\begin{equation}
<n| T(e^{-\int_{-T/2}^{T/2} H(t)}) |n> \label{ham} ,
\end{equation}
where $H(t) = H_{0} +V(t)$ with $H_{0}$ is the unperturbed inverted harmonic
oscillator Hamiltonian and $V(t)$ is some time dependent perturbation which we
will take as $\epsilon _{1} e^{ip_{1}t} O_{k} + 
\epsilon _{2} e^{ip_{2}t} O_{l}$. If we expand to second order in
$\epsilon _{1}$ and $\epsilon _{2}$ we get
$$
\epsilon _{1} \epsilon _{2} \int_{-{T\over 2}}^{T\over 2} d t_{2}
\int_{{-T\over 2}}^{t_{2}} dt_{1} 
<n|O_{k}|m><m|O_{l}|n> 
$$
\begin{equation}
\times 
e^{-E_{n}({T\over 2} -t_{2}) -E_{m}(t_{2} -t_{1}) - E_{n}(t_{1} +{T\over 2} ) +i
p_{1}t_{1}
+ip_{2}t_{2}}.
\end{equation}
To do the integrals we extract a common factor $e^{-E_{n}T}$ and recall that
the $E_{n}$'s are imaginary. Dropping oscillating terms, we find
in the limit
$T \rightarrow \infty$ (for momentum conservation $p_{1} +p_{2} =0$)
\begin{equation}
\epsilon _{1} \epsilon _{2} e^{-E_{n}T} \frac{T}{E_{m} -E_{n} +ip_{1}}.
\end{equation}
Since we have two different
time orderings we obtain for (\ref{ham})
\begin{equation}
e ^{-E_{n}T} \left[ 1+ \epsilon _{1} \epsilon _{2}
 T \sum _{m=0}^{\infty} <n| O_{k} |m><m| O_{l}|n> 
\frac{2(E_{m}-E_{n})}{p^{2} +(E_{m}-E_{n})^{2}}\right] .
\end{equation}
Interpreting this as
$e^{-E_{n}T} (1 +\epsilon _{1} \epsilon _{2} \delta E_{n} T ) \sim
e^{-(E_{n} - \epsilon _{1} \epsilon _{2} \delta E_{n})T}$,
we find the perturbation in $E_{n}$.

The above result then allows us to evaluate the second order variation of the
two puncture correlation function to be

\pagebreak

$$
<PPO_{k}O_{l}> = -\frac{1}{\pi }
\frac{\partial ^{2}}{\partial \epsilon _{1} \partial
\epsilon _{2}} {\rm Im} \sum _{n=0}^{\infty} \frac{1}{E_{n}-t_0}
| _{\epsilon _{1}=
\epsilon _{2} =0}
$$
\begin{equation}
 =\frac{1}{\pi } {\rm Im} \sum_{n,m} 
\frac{1}{(E_{n}-t_0)^{2}} <n| O_{k}|m><m| O_{l} |n>
\frac{2(E_{m} -E_{n})}{p^{2} + (E_{m} - E_{n} )^{2}},
\end{equation}
or after removing one puncture,
\begin{equation}
<PO_{k}O_{l}> = \nonumber \\
- \frac{1}{\pi } {\rm Im} \sum _{n,m} \frac{<n|O_{k}|m><m|O_{l}|n>}
{(E_{n} -t_{0} )(E_{m} - t_{0} )} \frac{(E_{n} -E_{m})^{2}}
{p^{2} + (E_{m} -E_{n})^{2}}. \label{gall}
\end{equation}
(Recall that the $E_{n}$'s are imaginary.)
Let us now limit ourselves to the quadratic critical point, i.e. the
inverted harmonic oscillator, and compare with our previous results
for the sphere. 

The matrix elements are polynomials in $n$. On the sphere,
only the highest power of $n$ survives since this term will give the highest
power of $\mu \beta$ after the sum is performed. Using $<n|\lambda |m> =
(i\frac{\sqrt{\alpha '}}{2})^{1/2} 
(\delta _{n,m+1} \sqrt{n} + \delta _{n+1,m} \sqrt{m})$ the leading terms of
all the matrix elements form a Pascal triangle, hence
\begin{equation}
<n| O_{r} |n+k> \sim \left( \begin{array}{c}
                          r \\ (r-k)/2
                         \end{array} \right) n^{r/2} 
(\frac{i\sqrt{\alpha '}}{2})^{r/2},
\end{equation}
for $k= -r, -r+2,..., r$, otherwise zero.
In the spherical limit we can replace the sum over $n$ by an integral.
We get
\begin{equation}
{\rm Im} \sum _{n=0} ^{\infty} 
\frac{(in)^{(r+s)/2}}{\frac{1}{2\sqrt{\alpha '}}(2n+1)i - \beta \mu }  
\sim  {\rm Re} 
\left[ \beta ^{(r+s)/2}(\sqrt{\alpha '})^{(r+s)/2+1} \int _{0}^{\infty} dx
\frac{(ix)^{(r+s)/2}}{x+i\mu} \right] .
\end{equation}
The integral is defined by taking a suitable number of derivatives and keeping
only the part singular in $\mu$. With even $r+s$ we get
\begin{equation}
(\mu \beta )^{(r+s)/2 }(\sqrt{\alpha '})^{(r+s)/2+1} |\ln \mu |.
\end{equation}
Odd $r+s$ give zero, confirming that odd-even correlators are zero.
Putting all of this together we recover (\ref{g0}).

As we have indicated (\ref{gall}) is valid for a general potential. 
The poles have a simple explanation. They appear at the values of momenta that can excite
transitions between the energy levels in the upside down potential. In the case of
a quadratic maximum these are linearly spaced and lead to integer values (in units of
${1 \over \sqrt{\alpha'}}$) of the momentum.
If we, however, consider a critical
theory generated by a potential whose leading term at the maximum is 
higher than 
quadratic, the situation will be more complicated.
The energy levels in the upside down potential will now
no longer be linearly spaced. The resonant values of the momenta will be differences
of these levels, and in general we would expect a strange transcendental set
of discrete momenta. What conformal field theory could produce such special physical
states?

\section{Correlation Functions from Path Integrals}

In \cite{mo1}, a generating functional for all correlation functions was
calculated using path integrals. In this section we will indicate how our
results for the special operators agree with this approach. We will
consider both the sphere and later all genus.

Let us first state the
results of \cite{mo1}. We write them as
$$
<O_{r}O_{s}P> = \frac{\partial^{r+s}}{\partial^{r} z_{1}\partial^{s} z_{2}}
{\rm Im} \int _{0}^{\infty} d \xi 
\frac{ e^{i\beta \mu \xi +i/2 \coth (\xi /2) 
(z_{1}^{2} + z_{2}^{2})}}{\sinh(\xi /2)} 
$$
$$
\times \left[ 2\pi e^{-i\pi |p|/2} \frac{\sinh(|p|\xi /2)}{\sin \pi |p|}
J_{|p|}(\frac{z_{1}z_{2}}{\sinh(\xi /2)}) \right.
$$
\begin{equation} \left.
+\sum _{k=0}^{\infty} \frac{4i^{k}k}{k^{2} - p^{2}}
J_{k}(\frac{z_{1}z_{2}}{\sinh(\xi /2)}) 
\sinh (k\xi /2)\right] _{z_{1} = z_{2} =0}.
\label{pust}
\end{equation}
The derivation of this expression is very straightforward. One simply
needs the propagator of the harmonic oscillator. One then need to Fourier
transform to momentum space and also make an integration against the
loop operators, \cite{mo1}. 

The term which will be of interest to us is the last one. It corresponds
to the contribution from the special states.
Let us first consider the spherical case. We expand (\ref{pust}) for small
$\xi$ and pick out the term of order $q=(r+s)/2$ in $z_{1}$ and $z_{2}$.
A small $\xi$ expansion will clearly give the string perturbation
series.
Doing so we obtain
$$
{\rm Im} \left[ (z_{1}z_{2})^{q} i^{q} \int _{0}^{\infty} d\xi 
\frac{e^{i\beta \mu \xi}}{\xi ^{q}} \right.
$$
\begin{equation} 
\times \sum _{n+2m+l=q} \sum _{k=0}^{n} \frac{1}{k!(n-k)!m!(l+m)!}
z_{1} ^{2k-n} z_{2} ^{n-2k} \left. \frac{4l^{2}}{p^{2} -l^{2}} \right] .
\label{exp}
\end{equation}
The integral is defined by taking a suitable number of derivatives and
neglecting terms analytic in $\mu$. For fixed powers of both $z_{1}$ and
$z_{2}$ we get
$$
4\frac{z_{1}^{r} z_{2}^{s}}{r!s!} \frac{r+s}{2} 
(\beta \mu ) ^{(r+s)/2 -1} |\ln \mu |
$$
\begin{equation}
\times \sum _{k=0}^{r} \left( \begin{array}{c}
                             r \\ k 
                             \end{array} \right)
                             \left( \begin{array}{c}
                             s \\ (s+r)/2 -k
                             \end{array} \right)
\frac{4(r/2-k)^{2}}{p^{2}-4(r/2-k)^{2}} .
\end{equation}
Hence we find agreement with our results for the sphere apart from
an unimportant normalization factor. Let us also check
the formula for general genus in a few cases. For simplicity we will limit
ourselves to $r=s$. In that case we get
$$
{\rm Im} \left[ (\frac{iz_{1} z_{2}}{2})^{r}  
\int _{0}^{\infty} e^{i\beta \mu \xi} 
\sinh (\frac{k\xi}{2}) \frac{4k}{p^{2}-k^{2}} \right.
$$
\begin{equation} 
\times \sum _{l=0}^{\frac{r-k}{2}} \frac{1}{l!(r-l)!}
\left( \begin{array}{c}
            r-l \\ (r-k)/2 -l
       \end{array} \right)
\left( \begin{array}{c}
            r-l \\ (r+k)/2 -l 
       \end{array} \right)
\left. ( \frac{1}{\sinh \xi /2})^{r+1-2l}\right].
\end{equation}
We may identify the harmonic oscillator matrix elements by
expanding the inverse powers of $\sinh \xi /2$ in powers of $e^{-n\xi}$
using
\begin{equation}
(\frac{1}{\sinh \xi /2})^{r} = 2^{r} e^{-r\xi /2}
\sum _{n=0} ^{\infty} \frac{(r+n-1)...(n+1)}{(r-1)!} e^{-n\xi}.
\end{equation}
For $k=r$, which is the term with the highest momentum pole, we find
\begin{equation}
 {\rm Im} \left[
(\frac{iz_{1}z_{2}}{2})^{r} \int _{0}^{\infty} 
e^{i\beta \mu \xi} \sinh (\frac{r\xi}{2})
\frac{4r}{p^{2}-r^{2}} (\frac{1}{\sinh \xi /2})^{r+1}\right] .
\end{equation}
To relate this to our previous formulae we use the recursion relation for Hermite polynomials
$ H_{n+1}(\lambda ) = 2xH_{n}(\lambda ) -2nH_{n-1}(\lambda )$, 
and also the expression for
the normalized wave functions $\psi _{n} (\lambda ) = 
\frac{1}{\pi ^{1/4} (n!)^{1/2} 2^{n/2}}
H_{n}(\lambda ) e^{-\frac{1}{2} \lambda ^{2}}$.
From which
it follows that
\begin{equation} 
<n|\lambda ^{r}|n+r> = (\frac{(n+r)...(n+1)}{2^{r}})^{1/2}(-\alpha
')^{r/4} .
\end{equation}
We then obtain (with $\alpha ' =1$), 
\begin{equation}
2^{r+3} {\rm Im}\left[  \frac{(z_{1}z_{2})^{r}}{r!r!}
\int _{0} ^{\infty} e^{i\beta \mu \xi} (1-e^{-r\xi}) 
\sum _{n=0}^{\infty} <n|\lambda ^{r}|n+r>^{2} 
e^{-(n+1/2)\xi} \frac{r}{p^{2}-r^{2}}\right] . 
\end{equation}
Similarly, for  $k=r-2$ we find
$$
2^{r+3}
{\rm Im}\left[ (iz_{1}z_{2})^{r} \int _{0}^{\infty} e^{i\beta \mu \xi} 
\sinh \left( \frac{(r-2)\xi}{2}\right) \frac{r-2}{p^{2} -(r-2)^{2}}
\right.
$$
\begin{equation} \left.
\times
\left( \frac{r^{2}}{r!}( \frac{1}{\sinh {\xi\over 2}}) 
^{r+1} + \frac{r-1}{(r-1)!}
 ( \frac{1}{\sinh{\xi \over 2} }) ^{r-1} \right) \right] .
\label{trudellutt}
\end{equation}
Now using  
\begin{equation}
<n|\lambda ^{r}|n+r-2> = 
(\frac{(n+r-2)...(n+1)}{2^{r}})^{1/2} r(n+\frac{r-1}{2})
(-\alpha ')^{r/4}
\end{equation}
which again is a result of the Hermite recursion
relation,
we can  rewrite (\ref{trudellutt}) as
$$
2^{r+3}
{\rm Im}\left[ \frac{(z_{1}z_{2})^{r}}{r!r!}  \int _{0} ^{\infty} e^{i\beta \mu \xi}
(1 - e^{-(r-2)\xi}  ) \right. 
$$
\begin{equation} \left.
\times 
\sum _{n=0}^{\infty} <n|\lambda ^{r}|n+r-2>^{2} e^{-(n+1/2)\xi} 
\frac{r-2}{p^{2} - (r-2)^{2}} \right] .
\end{equation}
To compare with the previous expression, we rewrite
our formula (\ref{gall}) as
$$
\frac{2}{\pi} \sum _{k} {\rm Im} \left[ \int _{0}^{\infty} e^{i\beta \mu \xi}
(e^{-\frac{k}{\sqrt{\alpha '}} } -1) \right.
$$
\begin{equation} \left.
\times 
\sum _{n=0}^{\infty} <n|O_{1}|n+k><n+k|O_{2}|n> 
e^{-\frac{(2n+1)}{2\sqrt{\alpha '}}\xi} 
\frac{\sqrt{\alpha '} k}{\alpha ' p^{2} - k^{2}}\right] ,
\end{equation}
which indeed, with $\alpha ' =1$, agrees with the examples above.

\section{Identification of States}

In this section we will compare the matrix model results with the predictions
of the continuum theory. We will try to identify the matrix model operators
which can be used to excite the continuum theory special states.

Consider a correlation function $<O_{n}O_{n}>$ in the matrix model.
According to (\ref{g0}) the scaling dimension of the $O_{n}$
operator is $d= \frac{n}{2}$. Furthermore, we have noted the presence
of  poles at momenta 
$p=\pm \frac{k}{\sqrt{\alpha '}}$, 
where $k=-n,-n+2,...,n$. These combinations of
scaling dimensions and momenta are the same as those  which we find in the continuum 
theory. As was explained in the introduction, there exist in the continuum theory
operators with dimension
$d=\frac{r+s}{2}$ and momenta $p=\pm \frac{r-s}{\sqrt{\alpha '}}$ 
corresponding to
level $rs$ descendants of tachyons  at that momentum.
We are therefore tempted
to identify the $p=\pm \frac{k}{\sqrt{\alpha '}}$ pole of
$<O_{n}O_{n}>$ as coming from  the level $\frac{1}{4} (n^{2}-k^{2})$ descendant
of the $p=\pm \frac{k}{\sqrt{\alpha '}}$ tachyon. 
It is reasonable to assume that 
the matrix model operators $O_{n}$ with these momenta can be used to
excite the continuum special states according to the above identification.
The zero momentum states, however, differ from the above scheme. 
There are no poles at
zero momentum. On the other hand, it is only for zero momentum which the
term in (\ref{gall}) with diagonal harmonic oscillator matrix 
elements contribute.

Let us illustrate the above procedure by a couple of examples. The $O_{2}$ 
operator has scaling dimension $d=1$ and the $<O_{2}O_{2}>$ correlation 
function a pole
at $p=\pm \frac{2}{\sqrt{\alpha '}}$. In the continuum theory this combination
of dimension and momentum corresponds to the special tachyon
operator $e^{\pm i\frac{2}{\sqrt{\alpha '}}X}$.
A more interesting example
is $O_{4}$. It has scaling dimensions $d=2$ and its two point correlation 
function poles at $p=\pm \frac{2}{\sqrt{\alpha '}}$ and 
$p =\pm \frac{4}{\sqrt{\alpha '}}$. The pole of highest momentum is again 
one of the tachyons with descending null states, this time
$e^{\pm i\frac{4}{\sqrt{\alpha '}}X + \sqrt{2} \phi _{L}}$,
but the other one would seem
to be related to the level 3 descendant of momentum 
$p=\pm \frac{2}{\sqrt{\alpha '}}$. We can also consider odd operators.
$O_{3}$ has scaling $d=3/2$ and, in addition to a tachyon pole, 
a pole at $p=\pm \frac{1}{\sqrt{\alpha '}}$,
which can be identified as the level 2 special state given by (8). 

We may also consider correlation functions between different matrix model
operators, i.e. $<O_{n}O_{m}>$ with $n \neq m$. The number of poles is now
determined by the lowest dimension
operator. Since we have nonzero correlation
functions between {\it different} operators there is substantial mixing.
This means that the identification of matrix model operators
and continuum theory special states is not one to one. This problem will
be resolved in the next section where we introduce the matrix model
generators of the $W_{\infty}$ symmetry.

\section{Summary}

In this and the previous chapter
we have studied the special operator correlation functions
of $c=1$ quantum gravity. For nonzero momentum, the matrix model
correlation functions of time dependent perturbations
of the potential about its maximum are found to have poles 
at the special values of momenta predicted
by Liouville theory. The operators also have the correct scaling dimensions.
This allow us to identify the matrix model operators which may be used to
excite any of the special states in Liouville theory.

We have also derived expressions for the
two point correlation functions at nonzero momentum. 
The correlation functions which we have derived are found to have 
poles which, for a general
time independent potential, are given by 
energy level differences in an anharmonic oscillator. In the harmonic case the
levels are equally spaced and precisely coincide with the special momenta
of Liouville theory. In the general case the meaning of our results are
not clear. 

We could also imagine adding time dependent pieces to the potential. Clearly,
this will drastically change the physics if the time dependence is such that
we sit on top of one of the momentum poles.
We have, in chapter 4,  speculated that this leads to models in which the target
space is punctuated--{\it i.e.} 
the continuous line is transformed into a set of discrete
points with spacing related to the special values of momenta. 
It would be very interesting to explore the structure of these theories.

\chapter{The Matrix Model $W_{\infty}$ Symmetry}

\section{Introduction}

In previous chapters we have seen how to compute certain correlation
functions of the matrix model eigenvalue. We have done both direct
calculations and derived recursion relations. There is nothing wrong with
deriving recursion relations in the way we have done. In particular, one
can certainly extend the results to correlation functions involving
powers of the conjugate momentum.
However, it is more convenient
to make use of the large set of symmetries in the theory, the
$W_{\infty}$. In this chapter we will see how this is realized in the
matrix model.

After we have done that we will use the algebra to calculate correlation
functions. We will also compare with the results obtained in previous
chapters. 

The calculations will focus on the sphere. Higher genus are reserved for
the next chapter.

\section{The $W_{\infty}$ Algebra}
 
As shown by Witten in
\cite{wi1} 
it is convenient to change
basis to $(\lambda -p )^{n} (\lambda + p )^{m}$ (at $\alpha ' =1$) 
rather than dealing with
the $\lambda$ and $p$ separately. For certain time dependent 
prefactors, we then 
obtain transformations which leave the action invariant. These
transformations are generated by, in Minkowski time,
\begin{equation}
W^{r,s} = (\lambda +p )^{r} (\lambda -p )^{s} e^{qt}  \label{w}
\end{equation}
with $q=r-s$.
These obey the $W_{\infty}$ algebra
\begin{equation}
\{W^{r_{1},s_{1}},W^{r_{2},s_{2}}\}=2(r_{2}s_{1}-r_{1}s_{2})W^{r_{1}+r_{2}-1,
s_{1}+s_{2}-1}, \label{alg}
\end{equation}
classically generated by the Poisson brackets $\{ \lambda , p\} =1$.
For a general momentum $q$ in (\ref{w}) we find, when acting on the Minkowski 
action 
$S=\int (p \dot{\lambda} -\frac{1}{2} (p^{2}-\lambda ^{2}))$,
\begin{equation}
\{W^{r,s},S\}=(r-s-q)W^{r,s} . \label{svar}
\end{equation}
Hence a symmetry for appropriate discrete values of imaginary momentum. This is
equivalent to saying that
$W=W(p,\lambda,t)$ is
a solution of
\begin{equation}
\frac{dW}{dt} = \frac{\partial W}{\partial t} + \{H,W\} =0 .
\end{equation}
Expressed in terms of the initial conditions, $p_{0}$ and $\lambda
_{0}$, we have
$W= W(p_{0}, \lambda _{0})$, i.e. any time
independent function of the initial conditions. The generators (\ref{w}) are
then simply obtained through evolution in time.
Hence the transformations can be understood as time independent canonical
transformations of the initial conditions.

A more convenient way of labeling the generators is through their
$SU(2)$ quantum numbers $J=(r+s)/2$ and $m=(r-s)/2$.
With the definition 
\begin{equation}
W_{J,m} = (\lambda +p )^{J+m}(\lambda -p )^{J-m} e^{2mt}, \label{h5}
\end{equation}
one gets
\begin{equation}
\{W_{J_{1},m_{1}},W_{J_{2},m_{2}}\}=4(m_{2}J_{1}-m_{1}J_{2})
W_{J_{1}+J_{2}-1,m_{1}+m_{2}} . \label{alg2}
\end{equation}

As in the previous chapter, we will continue to 
Euclidean time, i.e. $t \rightarrow it$, and
consider the system as a right side up oscillator continued to imaginary
frequency. The latter effectively means
$p \rightarrow ip$ and
an extra $i$ in the structure constant of (\ref{alg2}) and the
eigenvalue of (\ref{svar}). As already emphasized several times, this
analytical continuation has several advantages. Contrary to the
case above the $W_{\infty}$ elements now act like step operators
capable of exciting the system. The operators (\ref{h5}) however, 
make a step in
imaginary energy and in fact take you out of the Hilbert space. This has
also been discussed in \cite{wad}.

The continuation also leads to the possibility of drawing pictures of
the special states. This was mentioned at the end of chapter 4.
Recall figure 4.3. With the identification of the special states above
as $W_{\infty}$ generators, let us consider a couple of examples. The
actual special state perturbations would look like
\begin{equation}
(p^{2}+\lambda ^{2})^{J-m} ((\lambda +ip)^{J+m}+(\lambda -ip)^{J+m}) .
\end{equation}
Both signs of the momentum need to be included. In figure 9.1 we
find pictures of the $J=3/2$ and $J=2$ special tachyons.

\begin{figure}
\ifx\figflag\figI
\epsfbox{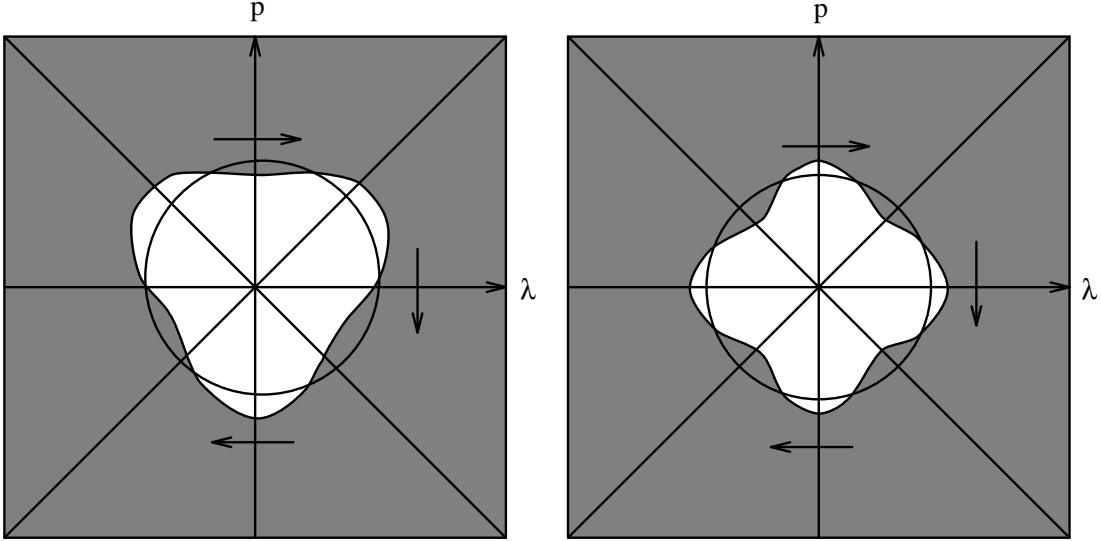}
\else
\vspace{3in}
\fi
\caption{$J=3/2$ and $J=2$ special tachyons.}
\end{figure}

\section{Correlation Functions from $W_{\infty}$}

In this section we will calculate correlation
functions involving the operators $W_{\infty}$ as defined in the previous
section. The $W_{\infty}$
symmetries will help us organize the Ward
identities. As an example, let us start with the two point function.
Using
\begin{equation}
<PP>=-\frac{1}{\pi} {\rm Im} \sum _{n=0}^{\infty} \frac{1}{E_{n}-t_{0}}
\end{equation}
and
simple perturbation theory we get
\begin{equation}
<W_{1}W_{2}P>=-\frac{1}{\pi}{\rm Im} \sum _{timeord} \sum _{n,k} 
\frac{<n \mid W_{1} \mid k>
<k \mid W_{2}
\mid n>}{E_{n}-t_{0}}\frac{1}{ip_{1}+E_{k}-E_{n}} .
\end{equation}
Since the $W$'s are of the form (\ref{h5}), continued to Euclidean time,
they are simply raising or lowering
operators in the continued harmonic oscillator. 
This means that only a few of the matrix elements are
nonzero. Since $W$ raises by $2m=r-s$ we get
\begin{equation}
<W_{1}W_{2}P>=-\frac{1}{\pi} {\rm Im} \sum _{n} \frac{<n\mid [W_{1},W_{2}] \mid
n>}{E_{n}-t_{0}}\frac{1}{i(p_{1}-2m_{1})} . \label{kom2}
\end{equation}
We have reduced the two point function to a one point function using the 
commutation relations. If we restrict ourselves to the sphere, use
the algebra given by (\ref{alg})
and directly calculate the one point
function we get
\begin{equation}
<W_{1}W_{2}>=- \frac{4(m_{2}J_{1}-m_{1}J_{2})}{p_{1}-2m_{1}}\frac{1}{\pi} 
\frac{\mu
^{J_{1}+J_{2}}}{J_{1}+J_{2}} \mid \log \mu \mid 2^{J_{1}+J_{2}-1}\label{2p}
\end{equation}
or equivalently
\begin{equation}
<W_{1}W_{2}>=\frac{-2m_{1}}{2m_{1}-p_{1}}\frac{1}{\pi} (2\mu ) ^{J_{1}+J_{2}}
\mid \log \mu \mid . \label{2pp}
\end{equation}

The simplest way to obtain the one point function on the sphere is to use 
the classical Fermi liquid picture introduced by Polchinski \cite{po3}. 
As we saw in chapter three, we simply need to
do the phase space integral 
\begin{equation}
<W>=\int dp d\lambda W(p,\lambda ) .
\end{equation}
This was also discussed in \cite{wad}.
To make everything well defined, we need to introduce an extra puncture,
i.e. take a derivative with respect to the cosmological constant. Doing that,
the integral over the whole Fermi sea becomes just an integral over the Fermi
surface
\begin{equation}
<WP>=\oint W(p, \lambda ) .
\end{equation}
The integral is to be performed along a hyperbola $\frac{\lambda
^{2} - p^{2}}{2}=\mu$. Although the answer is really infinite, we know that the
piece nonanalytic in $\mu$ is $\frac{1}{\pi} \mid \log \mu \mid$ for
$<PP>$, i.e. when we only integrate $1$. The other zero momentum
operators simply involve integrations of 
$(\frac{\lambda ^{2} -p^{2}}{2})^{n}=\mu
^{n}$, again constants along the Fermi surface hyperbola, so we get
\begin{equation}
<W^{n,n} P> = -\frac{1}{\pi} (2\mu )^{n} \mid \log \mu \mid  
\end{equation}
and from this (\ref{2p}) follows. Our conventions are such that $\alpha
' =1$. Another approach, which is convenient
when calculating correlation functions of nonzero momentum, is to
continue to the right side up oscillator where the Fermi surface is a
circle. In that case, however, we need to remember to put the Liouville
volume $\mid \log \mu \mid$ in by hand. 
Parenthetically we may note how a general 
correlation function may be obtained in this way. For instance, the two point
function is obtained by perturbing the Hamiltonian, and hence the 
Fermi surface by one of the operators. 
If we integrate the other operator against the change in Fermi surface we get
the correlation. 

Let us now consider the more complicated case of a
three point function. In this case there are six different time
orderings.
Again perturbation theory gives us
$$
\frac{1}{\pi} {\rm Im} \sum _{timeord} \sum _{n,m,k} \frac{<n \mid W_{1}
\mid m><m\mid W_{2}\mid k><k\mid W_{3}\mid n>}{E_{n} - t_{0}}
$$
\begin{equation}
\times \frac{1}{ip_{1} + E_{m}-E_{n}}\frac{1}{ip_{3} +E_{n}-E_{k}} .
\end{equation}
Let us make the sum over time orderings more explicit. We find three
terms of the form
\begin{equation}
\frac{1}{\pi} {\rm Im} \sum _{n} \{ 
\frac{<n\mid (W_{1}W_{2}W_{3}+W_{3}W_{2}W_{1})\mid
n>}{E_{n}-t_{0}} \frac{1}{i(p_{1}-2m_{1})i(p_{3}-2m_{3})} + perm \}.
\end{equation}
This may, after some straightforward manipulations, be rewritten as
\begin{equation}
<W_{1}W_{2}W_{3}>= \frac{<[W_{1},[W_{2},W_{3}]]>}{(p_{1} -2m_{1})(p_{3}+
2m_{1}+2m_{2})}
+\frac{<[W_{2},[W_{1},W_{3}]]>}{(p_{2}-2m_{2})(p_{3}+2m_{1}+2m_{2})} . 
\label{3kom} 
\end{equation}
For the sphere, we now use the algebra (\ref{alg2}) and the explicitly 
calculated one point
function to get
$$
<W_{1}W_{2}W_{3}>=
$$
$$
- \frac{(p_{1}-2m_{1})m_{2}m_{3}J_{1}+(p_{2}-2m_{2})m_{1}m_{3}J_{2} +
(p_{3}-2m_{3})m_{1}m_{2}J_{3}}{(p_{1}-2m_{1})(p_{2}-2m_{2})(p_{3}-2m_{3})}
$$
\begin{equation}
\times \frac{8}{\pi} (2 \mu ) ^{\sum _{n}^{N}
J_{n} -1 } \mid \log \mu \mid .  \label{strul}
\end{equation}

The general higher point function can be obtained recursively from the
three point by use of (\ref{alg}) and (\ref{svar}). To get the $N$
point function with an additional operator $W_{N}$, we vary the $N-1$
point function with $W_{N}$ knowing that the total variation is zero.
The variation consists of two terms. One from varying the action as
given by (\ref{svar}) and a sum of terms from varying the other
operators as given by (\ref{alg2}). Each of the terms in this sum is
obtained by shifting $p_{i} \rightarrow p_{i}+p_{N}$, $m_{i} \rightarrow
m_{i} +m_{N}$ and $J_{i} \rightarrow J_{i}+J_{N+1} -1$. We also need to
multiply with the Clebsch-Gordan coefficient $2(J_{i}m_{N}-J_{N}m_{i})$.
It is an easy exercise to check that (25) is obtained by applying
this procedure to (\ref{2p}) or (\ref{2pp}).

There is one subtlety in the variational
procedure which should be noted. Previously we derived recursion
relations for zero momentum correlation functions by varying $<P>$.
Recursion relations for $<PP>$ are obtained from these 
just by taking a $t_{0}$
derivative. Since the $<P>$ recursion relations involve 
an explicit $t_{0}$ we find an
extra term, $2k <O_{k-1}P>$, in the $<PP>$ recursion relations compared
to na\"{\i}ve expectations.
The reason is that we insisted on rewriting any conjugated momentum $p$
in terms of $\lambda$'s. The relation involves the potential $U$ and
hence depends on the coupling constants, in particular $t_{0}$. The
statement is that although 
$<U(\lambda )P>=<\frac{p^{2}}{2}P>$,
we have in fact
$<U(\lambda )PP> \neq <\frac{p^{2}}{2}PP>$.
In the $W_{\infty}$ recursion relations we are
discussing, we never perform any evaluations like this and hence need not
to worry about these issues. 
For instance, if we want to include extra punctures in the
correlation functions we are free to do so {\it without} extra terms in
the $W_{\infty}$ relations. 

Using this method one can write down several different recursion
relations. One simple example is:
\begin{equation}
<T_{J,J}W_{J_{1},m_{1}}\prod _{i=2}^{N} T_{J_{i},J_{i}}>= 
\frac{4J(m_{1}-J_{1})}{p-2J} <W_{J+J_{1}-1,J+m_{1}}\prod_{i=2}^{N}
T_{J_{i},J_{i}}> . \label{rek}
\end{equation}
We will come back to this relation later, when we compare with the Liouville
model results.

Rather than considering these general expressions, let us look at a
couple of important examples where the form of the general $N$ point
function is particularly simple.

The first example is the $N$ point function of special tachyons. It is
given by
\begin{equation}
<\prod _{n=1}^{N} T_{n}> = \frac{- \prod _{n=1}^{N} 2J_{n}}{\prod
_{n=2}^{N}(2J_{n}-p_{n})}\frac{2}{\pi} \frac{d^{N-3}}{d\mu ^{N-3}} 
(2\mu )^{\sum _{n=1}^{N} J_{n} -1} \mid \log \mu \mid  . \label{tach}
\end{equation}
The quantum numbers have been chosen as
$m_{n}=J_{n}$ for $n>1$ and
$m_{1}=-J_{1}$.
This is just the pole part of the general tachyon correlation function as 
computed both in the matrix model \cite{gr5,mo1} and in the Liouville theory 
\cite{gr5,frku}, up to
a factorized normalization factor. The proof is by varying the three
point. We cannot just vary the three point tachyon correlation
function, since some of the $J$'s are really $m$'s in disguise and $J$
and $m$ vary differently. Instead we start with the general three point
and make an arbitrary number of tachyon variations. A simplification is
that we at each step only have to vary the single negative chirality
tachyon. It is only from there we will get a nonzero 
Clebsch-Gordan
coefficient. Following the
prescription above, performing $N-3$ variations we get
$$
<\prod _{n=1}^{N} T_{n} > =-\left[ (p_{1}-2m_{1} +\sum _{i=4}^{N}
(p_{i}-2m_{i}))m_{2}m_{3}(J_{1}+\sum _{i=4}^{N}J_{i} -N+3) \right.
$$
$$ \left.
+(p_{2}-2m_{2})(m_{1}+\sum _{i=4}^{N} m_{i})m_{3}J_{2}
+(p_{3}-2m_{3})(m_{1}+\sum _{i=4}^{N} m_{i})m_{2}J_{3} \right]
$$
$$
\times \frac{\prod _{i=4}^{N} 4(J_{i}(m_{1}+\sum
_{j=i+1}^{N}m_{j})-(J_{1}+\sum_{j=i+1}^{N}J_{j}-N+i)m_{i})}
{(p_{1}-2m_{1}+\sum _{i=4}^{N} (p_{i}-2m_{i}))\prod_{n=2}^{N}(p_{n}-2m_{n})}
$$
\begin{equation}
\times \frac{8}{\pi} (2 \mu )^{\sum _{n=1}^{N} J_{n} -N+2} \mid \log \mu \mid
.
\label{tok}
\end{equation}
The product in the denominator is the product of all the Clebsch-Gordan
coefficients of the variations. Note that each get shifted by the
successive variations. By the use of momentum conservation and evaluating
the $m$'s as $J$'s, the formula (\ref{tach}) is proved. This derivation
shows how the combinatorial factor from the $\mu$ derivatives is a
consequence of the $W_{\infty}$ symmetry.

Since this derivation is quite illuminating, it might be
reasonable to consider a particular example which is more transparent than
the general derivation above. Let us consider a four point tachyon
correlation function. A sketch of the relevant setup with three vertices
and a one point is given in figure 9.2. The vertices give rise to the factor
$$
(J_{2}m_{1}-J_{1}m_{2})(J_{3}(m_{1}+m_{2})-(J_{1}+J_{2}-1)m_{3})
$$
$$
\times (J_{4}(m_{1}+m_{2}+m_{3})-(J_{1}+J_{2}+J_{3}-2)m_{4})
$$
\begin{equation}
= 2J_{1}J_{2}J_{3}J_{4}(J_{1}+J_{2}+J_{3}+J_{4}-1)
(J_{1}+J_{2}+J_{3}+J_{4}-2) .
\end{equation}
The last factor is then cancelled by the one point function. The
remainder, apart from a factorized part, is the well known $\sum
J_{i} -1$ found by many other methods.

\begin{figure}
\ifx\figflag\figI
\epsfbox{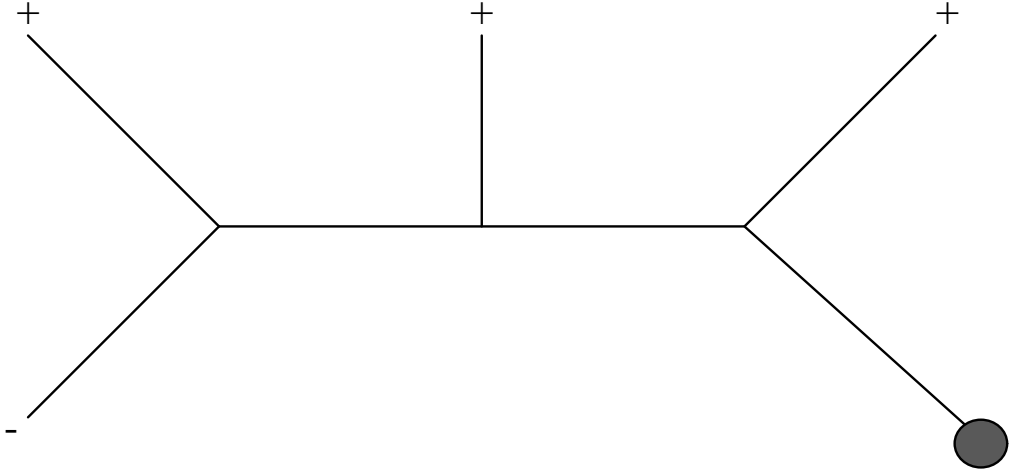}
\else
\vspace{2.5in}
\fi
\caption{Graph for four point tachyon correlation function.}
\end{figure}

If we want to consider the zero momentum operators,
we have to be careful. The Clebsch-Gordan coefficients are zero in this case
but these zeroes precisely cancel the momentum poles and leave a finite result.
We get
\begin{equation}
<\prod _{n=1}^{N} W_{n}>= - \frac{1}{\pi} 
\frac{d^{N-2}}{d \mu
^{N-2}} (2 \mu ) ^{\sum _{n=1}^{N} J_{n}} \mid \log \mu \mid .
\end{equation}
We will use induction for the proof. We find
$$
<\prod _{n=1}^{N+1}W_{n}> = - \sum_{k}4(J_{N+1}m_{k}-J_{k}m_{N+1}) 
\frac{1}
{p_{N+1}-2m_{N+1}} 
$$
\begin{equation}
\times \frac{1}{\pi}
\frac{d^{N-2}}{d \mu
^{N-2}} (2 \mu ) ^{\sum _{n=1}^{N+1} J_{n} -1} \mid \log \mu \mid
. 
\end{equation}
If we then put $p_{N+1}=0$ and use that the sum of all $m$'s must be
zero the result follows. 
This can also be checked by an explicit phase space calculation.

\section{Some Consistency Checks}

Let us now study the results of the previous section more
closely. In particular we will compare the two point correlation
functions with the results derived in the last chapter. 
These were correlation functions of pure powers of the
matrix eigenvalue, without any momentum powers. In terms of the W's they are
given by
\begin{equation}
O_{n}=\sum _{k=0}^{n} \left(\begin{array}{c}
                               n \\ k
                            \end{array} \right)
W^{n,k-n} \frac{1}{2^{n}} .
\end{equation}
From this follows that the two point function is given by
\begin{equation}
<O_{n}O_{m}>_{q} = \frac{1}{2^{n+m}}\sum _{k=0}^{n} \sum _{l=0}^{m}
\left(\begin{array}{c}
n \\ k
\end{array} \right)
\left(\begin{array}{c}
m \\ l
\end{array}\right)
<W^{n-k,k}_{q}W^{m-l,l}_{-q}> .
\end{equation}
Using (\ref{2pp}) and some simple algebra we find
\begin{equation}
\frac{-1}{2^{\frac{1}{2}(n+m)}} \frac{1}{\pi}
\mu ^{\frac{1}{2}(n+m)}\mid \log \mu \mid \sum _{k=0}^{n}
\left(\begin{array}{c}
n \\ k
\end{array}\right)
\left(\begin{array}{c}
m \\ \frac{n+m}{2} -k
\end{array}\right)
\frac{4(\frac{n}{2} -k)^{2}}{4(\frac{n}{2}-k)^{2}-q^{2}} .
\end{equation}
In precise agreement with \cite{art2} and the result (\ref{g0}) of the previous
chapter, recalling our convention $\alpha
'=1$.
We can now understand why the $O$ operators gave correlation functions with sets
of poles and were, depending on momentum, capable of exciting several special
states \cite{art2}. They were, in fact, linear combinations of all 
special operators of 
a given gravitational dimension i.e. spin $J$. The above construction with the 
generators (\ref{w}) of the $W_{\infty}$ disentangles the correlation functions.
This means that the matrix model operators to be identified with the
Liouville model special states are those defined in (\ref{h5}).

Let us briefly comment on the meaning of the momentum poles.
As emphasized in \cite{klre} we should not treat the poles
and the $\mid \log \mu  \mid$ 
in the correlation functions asymmetrically since the 
source of the $\mid \log \mu \mid$ is also a momentum pole.
In fact, all the poles should be thought of as cut off by $\mid \log
\mu \mid$. 
A general $N$ point function (without zero momentum operators) would
then have $\mid \log \mu \mid ^{N}$.
This proliferation of logarithms was also noted in \cite{frku}.   

Let us make some further tests of the correlation functions in (\ref{2pp}) and
(\ref{strul}). This will illustrate some very simple consequences of the
$W_{\infty}$ algebra.
In fact, the seemingly innocent representation of the puncture and the
dilaton as $\mu$ and $t_{2}$ derivatives respectively is a reflection
of the $W_{\infty}$. Let us give a formal argument for this. First the
puncture. Write the
$SU(2)$ quantum numbers of the puncture as $J$ and $m$, which both will
be taken to zero. Choose one of the operators in (\ref{strul}) to be a
puncture. We get
\begin{eqnarray}
<W_{1}W_{2}P>=-(J_{1}+J_{2})\frac{2m_{1}}{2m_{1}-p_{1}} \frac{2}{\pi} (2\mu
)^{J_{1}+J_{2}-1} \mid \log \mu \mid ,
\label{puder}
\end{eqnarray}
which, by comparing with the two point function (\ref{2p}), shows how the
puncture is represented as a $\mu$ derivative.
The case of the dilaton is equally simple.
Proceeding as
above we find
\begin{equation}
<W_{1}W_{2}D>=-\left[ (J_{1}+J_{2})\frac{2m_{1}}{2m_{1}-p_{1}} +
\frac{4m_{1}^{2}}{(2m_{1}-p_{1})^{2}}\right] \frac{2}{\pi}
(2\mu )^{J_{1}+J_{2}} \mid \log \mu
\mid . 
\label{3pD}
\end{equation}
If we introduce explicit $t_{2}$'s in the two point function we can
write it as
\begin{equation}
<W_{1}W_{2}>=\frac{-2m_{1}}{2m_{1}-p_{1}/(-2t_{2})^{1/2}} \frac{1}{\pi}
\frac{1}{(-2t_{2})^{1/2}}
\left( \frac{2 \mu}{(-2t_{2})^{1/2}}\right) ^{J_{1}+J_{2}} \mid \log \mu \mid 
. \label{2t2}
\end{equation}
Taking a $(-2t_{2})^{1/2}$ derivative, we arrive at
\begin{equation}
<W_{1}W_{2}D>=-\left[ (J_{1}+J_{2}+1)\frac{2m_{1}}{2m_{1}-p_{1}} +
\frac{2m_{1}p_{1}}{(2m_{1}-p_{1})^{2}}\right] \frac{2}{\pi}
(2\mu )^{J_{1}+J_{2}} \mid \log \mu
\mid 
, \label{3pD2}
\end{equation}
which indeed reproduces (\ref{3pD}).
We must now return to the issue of how 
precisely the dilaton is defined. Recall
the original matrix model action
\begin{equation}
\beta \int dt [p\dot{\lambda} -\frac{1}{2}p^{2}-t_{2} \lambda ^{2}]
\end{equation}
with $\beta$ dimensionless, $t_{2}$ having the dimension of energy squared,
and $p^{2}$ and $\lambda^{2}$ the dimensions of energy and one over energy
respectively. To obtain (\ref{2t2}) as a
generating functional for dilaton insertions with the above definition
of the dilaton we should rescale $\lambda$ and $p$ to make them
dimensionless. We find
\begin{equation}
\beta \int dt [p\dot{\lambda}-\frac{1}{\sqrt{2}}(-t_{2})^{1/2} (p^{2}-\lambda
^{2})] .
\end{equation}
Hence the matrix model dilaton should be represented by $(-2t_{2})^{1/2}$ 
derivatives. This is the rescaling eluded to in chapter seven, in the context
of the Wheeler de Witt equation.
 
Let us give some further illustrations in the case 
of puncture 
and dilaton 
insertions from the field theoretic point of view.
We begin with the puncture. Starting with a general correlation
function and inserting a puncture does not change the Veneziano-like
integral which has to be calculated. When we insert a puncture, we also
must remove one of the screening insertions. The only thing which
changes is the zero mode part of the calculation. 
We recall the result
\begin{equation}
\int d\phi e^{m\phi -\Delta e^{-\phi}} \sim \Gamma (-m) \Delta ^{m} .
\end{equation}
If we start with
$\Gamma (-m) \mu ^{m}$ we end up with $\Gamma (-m+1) \mu
^{m-1}=-m\Gamma(-m) \mu ^{m-1}$ when we remove a screening insertion. 
Comparing with (\ref{puder}) this shows the origin of the $W_{\infty}$
related factor in the tachyon correlation function. It is a consequence
of the changing number of puncture screening operators needed.

For the tachyons the issue of connected or 1PI amplitudes
is trivial for our case with just one tachyon of differing chirality.
There can not be any internal punctures just from kinematics. This is no
longer the case when we turn to the dilaton. 
The crucial point is that the dilaton can be represented
as a $t_{2}$ derivative. Usually, this is precisely equivalent to using
the ordinary free field contractions, giving Veneziano-like correlation
functions. Inserting a dilaton in some tachyon correlation function
means taking derivatives with respect to $t_{2}$ (i.e. $1/\alpha '$). 
For dimensional reasons, all tachyon momenta are accompanied by a
$t_{2}$. Without explicit $t_{2}$'s one could write
$k\frac{\partial}{\partial k}$ for the dilaton. For a dilaton insertion in a
nonzero momentum correlation function, the dominating pole contribution
comes from letting the $t_{2}$ derivative act directly on the poles. All
other terms are clearly less singular. This gives the second term in
(\ref{3pD2}). At zero momentum we must also consider the
dependence from the $t_{2}$'s which go together with
the $\mu$'s. The latter is a consequence of dealing with 1PI rather than
connected correlation functions. One could in fact obtain the result by
considering the explicit combinations of 1PI amplitudes into connected
ones. To be more precise, if we want to obtain the 1PI amplitude from
the connected amplitude, we must amputate external puncture legs but
also subtract of the diagrams with internal puncture propagators. In
particular, we need to subtract a diagram where a puncture goes off and
converts into a dilaton. This diagram, therefore, involves a puncture
insertion and gives a contribution corresponding to a $\mu$ derivative.
The procedure is illustrated in figure 9.3.
This is then simulated by an explicit $t_{2}$ accompanying the $\mu$'s to
assure the proper subtraction, which corresponds to the first term in
(\ref{3pD2}) and only becomes relevant compared to the second term 
at zero momentum. Otherwise we will get
one
power less of $\mid \log \mu \mid$'s. We need not restrict ourselves to a
dilaton among special tachyons. The same reasoning works for a dilaton
inserted in a general correlation function. 

\begin{figure}
\ifx\figflag\figI
\epsfbox{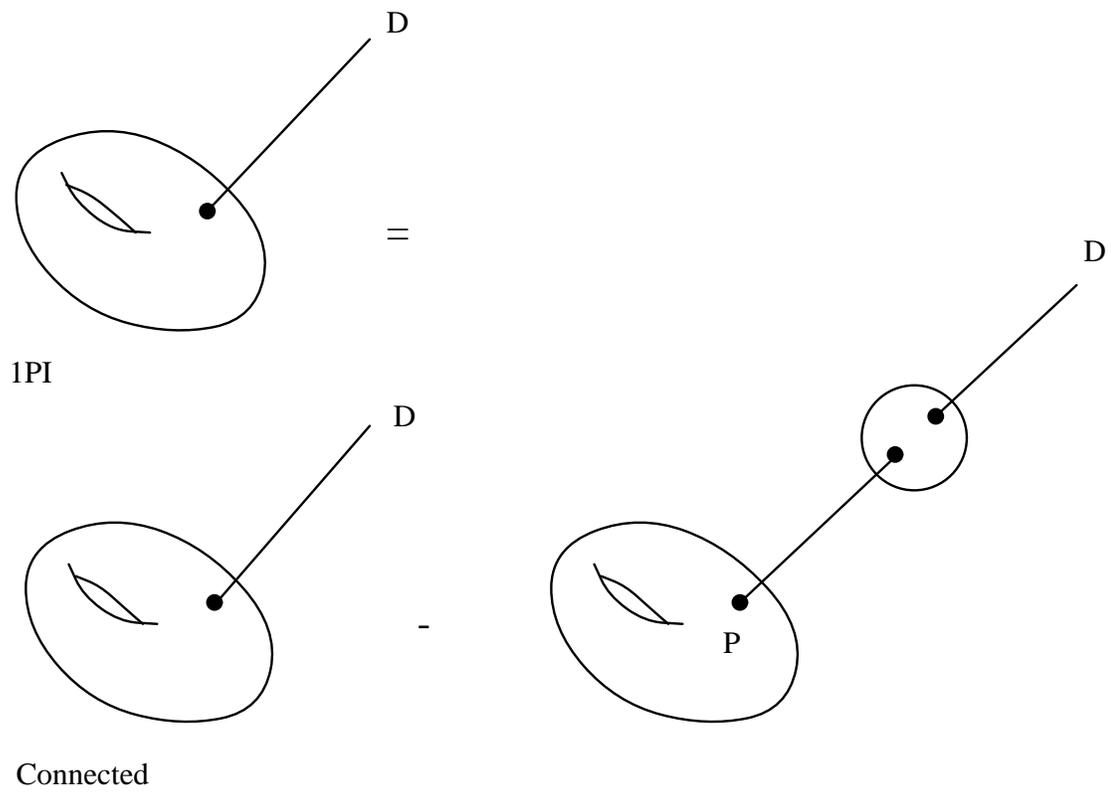}
\else
\vspace{6in}
\fi
\caption{Relation between dilaton 1PI and connected correlation
functions.}
\end{figure}

From these two examples we
can conclude that the 1PI nature of the correlation functions is very
important in the case of zero momentum.

Finally, it is useful to think of the correlation functions from a group
theoretic point of view.
The symmetry can be used to determine all correlation functions
given the special tachyon correlation functions which may be computed
by other means. The reason is that all $J$ and $m$ dependence of any
correlation function is given by some combination of Clebsch-Gordan 
coefficients. For given $J$'s we need the Clebsch-Gordan coefficients of
$SU(2)$, the $3j$ symbols, to get the $m$ dependence.
To be more explicit let us however look
at an example, the three point function, to see how the invariance
properties determine the correlation functions. 

The three point function
is 
obtained by considering
coupling $(J_{1},m_{1})$, $(J_{2},m_{2})$ and $(J_{3},m_{3})$ (with
$m_{1}+m_{2}+m_{3}=0$) to $(J_{1}+J_{2}+J_{3}-2,0)$. A complication is
that there are in general several different channels to sum over. This
is true already for the three point function. The reason, as we have
seen, is that we
actually should think of the three point function as a four point
function. The fourth leg carries the excess Liouville momentum, i.e. $J$
quantum number, into the vacuum. This is a consequence of the
nonconservation of Liouville momentum. Let us use the tachyon three
point function for normalization. It is given by
\begin{equation}
<T_{1}T_{2}T_{3}>=\frac{-8J_{1}J_{2}J_{3}}{(2m_{2}-p_{2})(2m_{3}-p_{3})} 
\frac{2}{\pi}
(2\mu ) ^{J_{1}+J_{2}+J_{3}-1}\mid \log \mu \mid , \label{t3}
\end{equation}
where we have kept the normalization choice of (\ref{tach}).
Tachyons 2 and 3 are of positive chirality, while tachyon 1 has negative
chirality.
There are two possible channels corresponding to either $p_{2}=2m_{2}$ or
$p_{3}=2m_{3}$, i.e. 1 and 2 coming together or 1 and 3 coming together.
The group theoretic factor in each case is simply proportional to a product of
$3j$ symbols. One for each vertex. For the 1-2 channel the factor is
\begin{equation}
\left(\begin{array}{c}
         J_{1} J_{2} J \\
         m_{1} m_{2} m_{3}
      \end{array} \right)
\left(\begin{array}{c}
         J J_{3} J^{\prime} \\
         -m_{3} m_{3} 0 
      \end{array} \right) ,
\end{equation}
where $J = J_{1}+J_{2}-1$ and $J^{\prime}=J_{1}+J_{2}+J_{3}-2$.
Just retaining the $m$ dependence and adding the two channels we find
\begin{eqnarray}
(2m_{2}-p_{2})(J_{1}m_{2}-J_{2}m_{1})m_{3}+(2m_{3}-p_{3})(J_{1}m_{3}-J_{3}m_{1})
m_{2} \nonumber \\
=-(2m_{1}-p_{1})J_{1}m_{2}m_{3}
-(2m_{2}-p_{2})J_{2}m_{1}m_{3}
-(2m_{3}-p_{3})J_{3}m_{1}m_{2}
\end{eqnarray}
which agrees with (25) after using (\ref{t3}) to fix the normalization and
$J$ dependence. This should come as no surprise since we have just
rephrased the previous calculation slightly.

Another convenient way to obtain more general correlation functions is
through factorization.
This is actually already implicit in
our previous calculations. In fact, if we look at (\ref{tok}), we see the
complete factorization of the tachyon correlation function into a
product of three point functions, each given by a $3j$ symbol times a
single zero momentum one point function. This last piece represents the
extra leg in any correlation function which absorbs excess Liouville
momenta. One may note that these three point functions in fact involve
states of the wrong dressing. This was also pointed out in \cite{tan}.
Strictly speaking, the expression in (\ref{tok}) is just for one
channel, the one where 1 fuses with 2 then with 3 etc. All channels,however,
give identical contributions and can not be distinguished. Clearly,
the tachyon correlation function is consistent with the single
$W_{\infty}$ factorization result.

\section{Summary}

In this chapter we have exploited the matrix model $W_{\infty}$ algebra
to derive correlation functions. The methods we have used provide a substantial simplification
compared to previous calculations. Also, the correct association of
special states and $W_{\infty}$ elements disentangles the correlation
functions in a neat way.

So far we have just looked at the sphere. In the next chapter we will
see how the picture generalizes to higher genus.

\chapter{Higher Genus Correlation Functions}

\section{Introduction}

In the previous chapter we showed how to use the $W_{\infty}$
algebra of the $c=1$ matrix model to derive correlation functions of
special states. In particular, the method is applicable to the special tachyons.
In this chapter we will  treat the higher genus case more explicitly. The
motivation is that by deriving the results using the $W_{\infty}$,
rather than just brute force calculations, the crucial physics will be
more apparent. Hopefully one can in this way obtain useful guidelines
for constructing a field theory approach to higher genus. In
\cite{das3}, similar results as in the previous chapter were obtained
independently. The authors combine the $W_{\infty}$ properties of
the system with the powerful path integral methods of \cite{mo1} to calculate
correlation functions. This is an useful approach for studying the
matrix model. We feel however, that the approach advocated in previous
chapters
and used
in this chapter, will prove very useful for field theory purposes.
Furthermore, there are certain subtle issues like ordering in the matrix
model definition of the special states which can not be addressed in the
path integral approach. 

An important feature of the matrix model higher genus calculations is a
deformation of the classical, genus zero $W_{\infty}$ algebra with
Planck's constant, i.e. the genus coupling, as parameter. This is simply
a result of replacing the Poisson brackets with commutators. 

As mentioned above, at higher genus we must also face the problem of
ordering when we define the special state operators. For the special
tachyons there is no ambiguities, but for the general special state
there are. However, in the very explicit calculations we will be doing,
the question of ordering will be important even in the case of tachyons.
Not for the final results, but for the intermediate steps. We will
choose Weyl ordering. Weyl ordering means that one takes an average over
all possible orderings with equal weights.
The reasons for this choice are several. It is from many points of
view the most ``natural'' ordering prescription. In a path integral
approach it would be the automatic choice. This is of course connected
with another reason, that there exist simple formulas for calculations
involving Weyl ordering. An important question is of course whether this
is the ordering prescription which will be relevant in the comparison
with field theory. It is conceivable that there will not exist an unique
choice but rather that we have to live with this ambiguity which, after
all, only amounts to a linear transformation.

\section{The Algebra}

We begin our discussion by considering the $W_{\infty}$ algebra of
the $c=1$ matrix model. Our conventions throughout this chapter are
such that $\alpha ' =1$. As explained in the previous
chapter, the theory is organized by a
$W_{\infty}$ algebra generated by the matrix eigenvalue and its
conjugate momentum. 
Classically, it is given through the
Poisson brackets by
\begin{equation}
\{ W_{J_{1},m_{1}},W_{J_{2},m_{2}}\} = 4i(m_{2}J_{1}-m_{1}J_{2})
W_{J_{1}+J_{2}-1, m_{1}+m_{2}}
\end{equation}
where
\begin{equation}
W_{J,m} = (\lambda -ip)^{J+m}(\lambda +ip)^{J-m} =2^{J} a^{\dagger J+m} 
a^{J-m} .
\end{equation}
For convenience, we have introduced the step up and down 
operators $a^{\dagger}$ 
and
$a$.
As shown in the previous chapter this algebra may be 
used to organize the special
state correlation functions in the theory. Parallel work has revealed
a similar structure in the field theory \cite{kl}, although the precise
connection between the two approaches is not yet clear. The great
advantage of the matrix model is of course that we may extend the
calculation to arbitrary genus. The main new ingredient at higher genus
is the deformation of the algebra mentioned in the introduction.
At higher genus, i.e. quantum mechanically, we must also be careful and
choose an ordering prescription. As explained above we choose Weyl
ordering where the new algebra is simply given by the Moyal brackets
\begin{equation}
\frac{1}{i \hbar} [ W_{1},W_{2}] =\{ W_{1},W_{2}\} _{M} =
\frac{2}{\hbar} \sin \frac{\hbar}{2} (\frac{\partial}{\partial p_{2}}
\frac{\partial}{\partial \lambda _{1}} - \frac{\partial}{\partial p
_{1}} \frac{\partial}{\partial \lambda _{2}}) W_{1}W_{2}
\end{equation}
which allow for simple calculations. The proof is elementary, but we
provide it for completeness. 

A Weyl ordered operator can be written as
\begin{equation}
\hat{A}(\hat{p},\hat{q}) _{W} = \frac{1}{(2\pi )^{2}}
\int A(p,q) e^{-i(\tau p +\theta q)} e^{i(\tau
\hat{p} + \theta \hat{q})} dpdqd\tau d\theta .
\end{equation}
If we take
the commutator of two such Weyl ordered operators, $\hat{A} _{W}$ and
$\hat{B} _{W}$, we encounter the expression
\begin{equation}
[e^{i(\tau _{1} \hat{p} + \theta _{1} \hat{q})}
,e^{i(\tau _{2} \hat{p} + \theta _{2} \hat{q})}]= 2i\sin \frac{\hbar}{2}
(\tau _{1} \theta
_{2} - \tau _{2} \theta _{1})
e^{i(\tau _{1}+\tau _{2}) \hat{p} + i(\theta _{1}+\theta _{2}) \hat{q})}
.
\label{neumann}
\end{equation}
The $\sin$ is then rewritten as
$2i\sin \frac{\hbar}{2} (\frac{\partial}{\partial p_{2}}
\frac{\partial}{\partial \lambda _{1}} - \frac{\partial}{\partial p
_{1}} \frac{\partial}{\partial \lambda _{2}})$ by partial integration. 
The $\tau _{1} -\tau
_{2}$ and $\theta _{1} - \theta _{2}$ integrations give $\delta$
functions which fixes $p_{1}=p_{2}$
and $q_{1}=q_{2}$ respectively. From this the result follows.

An equivalent way of exhibiting the quantum deformation is to focus on
(\ref{neumann}), 
which in fact is a very old
result, \cite{neu},
and inspired by this introduce a generalized
loop operator \cite{wad}. Instead of (\ref{loop}) we define
\begin{equation}
w(k,l)=e^{kp+l\lambda} ,
\end{equation}
where $p$ and $\lambda$ are the conjugate variables. 
We then have
\begin{equation}
[w(k_{1},l_{1}),w(k_{2},l_{2})]=2i \sin
\frac{\hbar}{2} (k_{1}l_{2}-l_{1}k_{1}) w(k_{1}+k_{2},l_{1}+l_{2}) \label{sun}
\end{equation}
with $\hbar \rightarrow 0$ giving back a $W_{\infty}$. Interestingly, it can
be shown \cite{bars} that (\ref{sun}) is a representation of $SU(N)$ with $\hbar
=1/N$. 
This is
reminiscent of the original unitary symmetry of the matrix model.

Let us give a simple example of a higher genus commutator. Using the
Moyal bracket it immediately follows that, for $\hbar =1$,
\begin{equation}
[a^{m},a^{\dagger n}]=\sum _{g=0}^{\infty} \frac{1}{2^{2g}(2g+1)!} \prod
_{k=0}^{2g} (m-k)(n-k) (a^{m-2g-1} a^{\dagger n-2g-1} )_{W} . \label{i6}
\end{equation}
The continuation to the upside down
harmonic oscillator gives an extra factor of $i$ and an extra
$(-1)^{g}$.
This result will be extensively used in a later section where we
calculate tachyon N-point functions.

For clarity, let us calculate explicitly a two point function to all
genus. We choose the correlator between spin $J=3/2$, $m=1/2$ and
$J=3/2$, $m=-1/2$. To do that we need to calculate 
$<W_{2,0}P>$. This is easy. We have
\begin{equation} 
<W_{2,0}P>=-\frac{1}{\pi} {\rm Im} \sum _{n=0}^{\infty}
\frac{(2E_{n})^{2}-1}{E_{n}-t_{0}} .
\end{equation}
The extra term $-1$ comes from Weyl ordering and will be thoroughly
discussed in the next section.
If we keep only terms nonanalytic in $t_{0}$ this reduces to
\begin{equation}
<W_{2,0}P>=(4t_{0}^{2} -1) <PP> . \label{1p}
\end{equation}
To evaluate our two point we use the Moyal bracket continued to the
upside down harmonic oscillator to calculate
\begin{equation}
[W_{3/2,1/2},W_{3/2,-1/2}]=i(6W_{2,0} +4W_{0,0})
\end{equation}
(\ref{kom2}) and (\ref{1p}) then finally give
\begin{equation}
<W_{3/2,1/2}W_{3/2,-1/2}P>=(24t_{0}^{2}-2)<PP>\frac{1}{1-p} .
\end{equation}
The same procedure may be used to calculate arbitrary correlation
functions.

In the previous chapter, 
it was shown how to calculate special state correlation functions
using this algebra. The number of operators in the correlation functions
are reduced one by one using the algebra. The last step, however, 
involves the calculation of
a remaining zero momentum one point function. This is where the
cosmological screening charges sit. Let us therefore turn to the issue
of calculating such objects.

\section{The One Point Functions}

To obtain the zero momentum special state one point functions at higher
genus, we need to be careful. From the algebra we understand
that they are given in terms 
of $(H^{n})_{W}$. These are just powers of the Hamiltonian. But it is
important to remember the Weyl ordering. This will make the one point
functions quite complicated. We need to rewrite the Weyl ordered powers
in terms of something simpler.
The objects with simple correlation functions
are clearly the $H^{n}$, which give
\begin{equation}
t^{n}_{0} <PP>
\end{equation}
since the resulting $E_{m}^{n}$ in the trace can be replaced by
$t_{0}^{n}$ up to terms analytic in $t_{0}$.
To obtain the $(H^{n})_{W}$ one point functions
we therefore need to perform the Weyl ordering and express the result
as a polynomial in $H$. One way to do this is to use the standard
formula
\begin{equation}
(a^{m}a^{\dagger n})_{W} = 2^{-m} \sum _{l=0}^{m} \left( \begin{array}{c}
                                                        n \\ l
                                                    \end{array} \right)
a^{\dagger m-l}a^{n}a^{\dagger l} .
\end{equation}
See e.g. \cite{lee}.
If we put $m=n$ and take the expectation value in the eigenstate $r$ we
get
\begin{equation}
2^{-n} \sum _{l=0}^{n} \left( \begin{array}{c}
                                n \\ l
                               \end{array} \right)
\prod _{k=0}^{n-1} (r+l-k) ,
\end{equation}
and finally
\begin{equation}
(H^{n})_{W} = 2^{-n} \sum _{l=0}^{n} \left( \begin{array}{c}
                                              n \\ l
                                            \end{array} \right)     
\prod _{k=0}^{n-1}
(H-\frac{1}{2} +l-k). \label{i16}
\end{equation}
This is unfortunately a very complicated polynomial. 
In the next section we will explicitly
calculate some tachyon correlation functions. To do so we
need explicit expression for the coefficients in (\ref{i16}). By some tedious
algebra, which is greatly simplified if {\it Mathematica} is used, the
following expression can be obtained 
\begin{equation}
\begin{array}{l}
(H^{n})_{W} = H^{n} + \frac{1}{24}(2n-1)n(n-1)H^{n-2} \\  \\
+\frac{1}{5760}(20n^{2}-48n+7)\prod_{k=0}^{3}(n-k) H^{n-4} \\ \\ 
+\frac{1}{2903040}(280n^{3}-1596n^{2}+1874n-93)\prod_{k=0}^{5}(n-k)
H^{n-6} +...  
\end{array} \label{i17}
\end{equation}
The continued case is obtained by sending $H^{n-2k} \rightarrow
(-1)^{k}H^{n-2k}$.
This is enough up to genus 3 for the general correlation function. In
appendix 10A, the expansion is shown up to genus 7.
These expressions are really remarkably complicated, and responsible for
the complicated structure of the higher genus correlation functions. The
piece coming from the $W_{\infty}$ algebra is much simpler.

For clarity, let us give an example. We choose $J=4$. We find
$$
\frac{1}{16}<W_{4,0}P>= <H^{4}P>-\frac{7}{2}<H^{2}P>+\frac{9}{16}<PP>
$$
\begin{equation}
=-\frac{1}{\pi} (t_{0}^{4} - \frac{7}{2}t_{0}^{2} +\frac{9}{16}) \log \mu +
O(1/t_{0}^{2}) .
\end{equation}
where we note resonance contributions up to genus 2, but no higher.
The same general structure
is clearly found for the tachyon correlation functions. Poles are
present only up to some genus. The precise value depends on the
scaling (i.e. momentum).

\section{Tachyon Correlation Functions}

We now need to combine these results with those given by the deformed
$W_{\infty}$ algebra to get the tachyon
correlation functions. This is, in principle, rather simple. 
In fact, one can write
down a set of diagrammatic rules which give the correlation functions.

By the variational procedure introduced in the last chapter
one can show that in the
case of $(N-1,1)$ amplitudes it is possible to arrange things in such a
way that only one basic graph is needed. There exist
other choices too, involving sums over several graphs, but this one is clearly 
the simplest. The graph, shown in figure 9.2, has a $T^{-}$ at one end and
the necessary one point function at the other. In between are the
$T^{+}$ distributed in some order. The order does not matter, but we
should pick just one ordering, no sum. To see this, one can start with
the two point function of a $T^{+}$ and a $T^{-}$. All subsequent
$T^{+}$ variations will only give a nonzero value when they hit the $-$
leg.
If we are considering a genus
$g$
correlation function, we must partition the handles in all possible 
ways among the
vertices and the one point function. Each such diagram is then given a
certain factor according to the following rules. 

For each vertex write
down a factor $f_{g_{k}}(p_{k})$ given by
\begin{equation}
f_{g_{k}}(p_{k})= \frac{(-1)^{g_{k}}}{2^{2g_{k}}(2g_{k}+1)!}
(p_{k}-1)(p_{k}-2)...(p_{k}-2g_{k}), \label{i19}
\end{equation}
where $p_{k}$ is the momentum of the $k$'th (positive) tachyon and
$g_{k}$ the genus of the corresponding vertex. 
This is obtained from (\ref{i6}).
Strictly speaking the
formulae are only valid when $p=2J$, i.e. for {\em special} tachyons.
The one point function is
given a factor $h_{\tilde{g}}$ according to
\begin{equation}
\begin{array}{l}
h_{0}(p) =1 \\ 
h_{1}(p) =\frac{-1}{24} (2p-1) \\ 
h_{2}(p) =\frac{1}{5760} (20p^{2}-48p+7) \\
h_{3}(p) =\frac{-1}{2903040} (280 p^{3} -1596 p^{2} +1874 p -93) ,
\label{i20}
\end{array}
\end{equation}
where the argument $p$ is
\begin{equation}
\frac{1}{2} \sum _{k=1}^{N} \mid p_{k} \mid -N +2(\tilde{g}-g)+1 
= \mid p_{1} \mid
-N + 2(\tilde{g} -g) +1 .
\end{equation}
$\tilde{g}$ is the genus associated with the one point function. Clearly,
$\sum _{k=2}^{N} g_{k} + \tilde{g} = g$. Some additional $h_{\tilde{g}}$
polynomials may be read off from appendix 10A while the 
general case is generated
by (\ref{i16}). In both (\ref{i19}) and (\ref{i20}) 
we have extracted a common factor depending only on
the momentum of the single negative chirality tachyon.

Using this technique, the two point function would be calculated as
$$
p^{2}f_{0}(p)h_{0}(p-1) \frac{t_{0}^{p-1}}{p-1}
+ p^{2}
(p-1)(p-2)
$$
$$
\times \left[ f_{1}(p)h_{0}(p-3)+f_{0}(p)h_{1}(p-1)\right]
\frac{t_{0}^{p-3}}{p-3}
+p^{2}(p-1)(p-2)(p-3)(p-4)
$$
\begin{equation}
\times \left[ f_{2}(p)h_{0}(p-5)+f_{1}(p)h_{1}(p-3)+
f_{0}(p)h_{2}(p-1)\right]
\frac{t_{0}^{p-5}}{p-5}
+ ...
\end{equation}
up to genus $2$.
In particular, if we concentrate on genus 2 the relevant calculation
involves
$$
p^{2}(p-1)...(p-4)
\left[ \frac{1}{1920} (p-1)...(p-4)+ \right.
$$
$$ \left.
\frac{1}{576} (p-1)(p-2)(2p-7) +\frac{1}{5760}(20(p-1)^{2}-48(p-1)+7)
\right]
$$
\begin{equation}
=p^{2}(p-1)...(p-4)
\frac{1}{5760} (3p^{4}-10p^{3}-5p^{2}+12p+7) .
\end{equation}

Although the algebra is somewhat tedious, 
the generalization to higher genus and N-point functions is
straightforward. As an example the three point function at 
genus 3 is computed in appendix 10B. 
Here, we will simply state the results for N-point functions up
to genus $3$.

$$
\frac{\partial ^{N-3}}{\partial t_{0} ^{N-3}}
t_{0}^{p_{1}-1} -\frac{1}{24} 
(\sum _{i=2}^{N} p_{i}^{2}-p_{1}-1)
\frac{\partial ^{N-1}}{\partial t_{0} ^{N-1}}
t_{0}^{p_{1}-1}
$$
$$
+\frac{1}{5760}
(3\sum _{i=2}^{N} p_{i}^{4}+
10\sum_{
        i,j=2,i>j
       }^{N} p_{i}^{2}p_{j}^{2}
-10p_{1}\sum _{i=2}^{N} p_{i}^{2}
-5(\sum _{
        i,j=2,i>j
       }^{N} (p_{i}-p_{j})^{2} -(N-3)
\sum _{i=2}^{N} p_{i}^{2}) 
$$
$$
+ 12p_{1}+7)
\frac{\partial ^{N+1}}{\partial t_{0} ^{N+1}}
t_{0}^{p_{1}-1} 
$$
$$
- \frac{1}{2903040}
(9\sum _{i=2}^{N} p_{i}^{6}+63
\sum_{
        i,j=2,i>j
       }^{N} p_{i}^{2}p_{j}^{4}
+210\sum ^{N}_{
         i,j,l=2,i>j>l
       } p_{i}^{2} p_{j}^{2} p_{l}^{2}
$$
$$
-p_{1}(63\sum _{i=2}^{N} p_{i}^{4}
+210\sum_{
        i,j=2,i>j
       }^{N} p_{i}^{2}p_{j}^{2})
+42\sum _{i=2}^{N} p_{i}^{4}
+210\sum_{
        i,j=2,i>j
       }^{N} p_{i}^{3}p_{j}
$$
$$
+210\sum ^{N}_{
         i,j,l,i>j
       } p_{i} p_{j} p_{l}^{2}
+p_{1}(217\sum _{i=2}^{N} p_{i}^{2}
-70\sum_{
        i,j=2,i>j
       }^{N} p_{i}p_{j})
-294\sum_{
        i,j=2,i>j
       }^{N} p_{i}p_{j}
$$
\begin{equation}
-205p_{1}-93)
\frac{\partial ^{N+3}}{\partial t_{0} ^{N+3}}
t_{0}^{p_{1}-1} . \label{ujejuj}
\end{equation}
This is to be multiplied with the appropriate factorized pole factor.
The result may be compared to other calculations of the correlation
functions. Clearly this is a very complicated expression which
practical application is probably very limited. 
The main point is, however, not this expression in itself, but
rather the way in which we have derived it.

\section{Summary}

We have seen how to calculate special tachyon correlation functions in
the matrix model using the deformed $W_{\infty}$ algebra. The algebra
nicely organizes the correlation functions even at higher genus.  

A natural challenge to the field theory approach is to reproduce this
deformation of the algebra, which we have described and found to be
important for the structure of the correlation functions. 
One would expect that the deformation is
related to contour integrals of the currents which when pulled back over
the surface, away from any insertions, fail to yield zero since they get
caught up around handles and cannot be contracted to points.

\newpage

\section*{Appendix 10A}

Weyl ordering of powers of the Hamiltonian for computations up to
genus 7:

$$
\begin{array}{l}
(H^{n})_{W} = H^{n} + \frac{1}{24}(2n-1)n(n-1)H^{n-2} \\  \\
+\frac{1}{5760}(20n^{2}-48n+7)\prod_{k=0}^{3}(n-k) H^{n-4} \\ \\ 
+\frac{1}{2903040}(280n^{3}-1596n^{2}+1874n-93)\prod_{k=0}^{5}(n-k)
H^{n-6} \\ \\  
+\frac{1}{1393459200}(2800n^{4}-29120n^{3}+85544n^{2}-68368n+1143) 
\prod_{k=0}^{7}(n-k) H^{n-8} \\ \\ 
+ \frac{1}{367873228800} (12320n^{5}-203280n^{4}+1100528n^{3}-
2255352n^{2} \\ \\ +1422434n-7665) \prod_{k=0}^{9}(n-k)
H^{n-10} \\ \\  
+ \frac{1}{24103053950976000} (11211200n^{6}-269068800n^{5}
+2312950640n^{4}-8743967232n^{3} \\ \\ 
+14194825268n^{2}-7658714592n-12730293) \prod_{k=0}^{11} (n-k)
H^{n-12} \\ \\  
+ \frac{1}{578473294823424000} (3203200n^{7}-105385280n^{6}+
1315394080n^{5}-7873930064n^{4} \\ \\  +23465896792n^{3}
-32333229596n^{2}+
15636417798n-7740495) \prod_{k=0}^{13} (n-k)
H^{n-14} +...
\end{array} 
$$

\newpage

\section*{Appendix 10B}

In this appendix we calculate the tachyon three point function at genus
3. As explained in the main text we should write down all possible
partitions of the three handles among the vertices and the one point.
This gives
$$           
\left[ f_{3}(p_{3})f_{0}(p_{2})
+f_{0}(p_{3})f_{3}(p_{2})
+f_{2}(p_{3})f_{1}(p_{2})
+f_{1}(p_{3})f_{2}(p_{2})\right] h_{0}(p_{1}-8)
$$
$$
+\left[ f_{2}(p_{3})f_{0}(p_{2})
+f_{0}(p_{3})f_{2}(p_{2})
+f_{1}(p_{3})f_{1}(p_{2})\right] h_{1}(p_{1}-6)
$$
\begin{equation}
+\left[ f_{1}(p_{3})f_{0}(p_{2})
+f_{0}(p_{3})f_{1}(p_{2})\right] h_{2}(p_{1}-4)
+f_{0}(p_{3})f_{0}(p_{2})h_{3}(p_{1}-2) .
\end{equation}
Using the expressions (\ref{i19}) and (\ref{i20}) we find
$$
\frac{1}{2^{6}7!} \left[ (p_{3}-1)...(p_{3}-6)+(p_{2}-1)...(p_{2}-6) \right]
$$
$$
+\frac{1}{2^{6}5!3!} \left[ (p_{3}-1)...(p_{3}-4)(p_{2}-1)(p_{2}-2) 
+(p_{2}-1)...(p_{2}-4)(p_{3}-1)(p_{3}-2) \right]
$$ 
$$
+( \frac{1}{2^{4}5!} \left[ (p_{3}-1)...(p_{3}-4) 
+(p_{2}-1)...(p_{2}-4) \right] 
$$
$$
+\frac{1}{2^{4}3!3!} \left[ (p_{3}-1)(p_{3}-2)(p_{2}-1)(p_{2}-2) \right]
\frac{1}{24}(2p_{1}-13)
$$
$$
+\frac{1}{2^{2}3!} \left[ (p_{3}-1)(p_{3}-2)+(p_{2}-1)(p_{2}-2) \right] 
\frac{1}{5760}
\left[ 20(p_{1}-4)^{2}-48(p_{1}-4)+7 \right]
$$
\begin{equation}
+\frac{1}{2903040}(280(p_{1}-2)^{3}-1596(p_{1}-2)^{2}+1874(p_{1}-2)-93),
\end{equation}
which is nothing else than, if we use momentum conservation,
$$
\frac{1}{2903040} (9(p_{2}^{6}+p_{3}^{6})
+63(p_{2}^{2}p_{3}^{4}+p_{2}^{4}p_{3}^{2})
-p_{1}(63(p_{2}^{4}+p_{3}^{4})+210p_{2}^{2}p_{3}^{2})
$$
$$
+42(p_{2}^{4}+p_{3}^{4})+210(p_{2}p_{3}^{3}+p_{2}^{3}p_{3})+
p_{1}(217(p_{2}^{2}+p_{3}^{2})-70p_{2}p_{3})
$$
\begin{equation}
-294p_{2}p_{3} -205p_{1} -93),
\end{equation}
which indeed is a special case of (\ref{ujejuj}).

\chapter{Comparison with Liouville Theory}

\section{Introduction}

In this chapter we will make some simple comparisons between the matrix
model and field theory calculations. The first and most obvious
observation is of course the presence of the special states in both
theories. This is in fact one of the main results of the previous
chapters. Furthermore, there is a $W_{\infty}$-algebra organizing the
correlation functions in both cases. All of this lend support to the
accepted view that the theories are, in fact, identical. A very nice and
explicit connection at the level of the algebra was given in \cite{wi1}.

Clearly, one would also like to have a precise comparison of the
correlation functions. As far as tachyons go, we have seen the agreement
to be perfect. At least on the sphere, the only case for which field
theory calculations in general exist. When comparing results for the
special states, additional care must be taken since we are automatically
sitting on the discrete momentum poles. This can be especially tricky in
the field theory approach.

An important issue to decide is whether the correlation functions one is
calculating are correlation functions of states
\begin{equation}
W(z)\bar{W}(\bar{z}),
\end{equation}
or currents
\begin{equation}
W(z)\bar{O}(\bar{z})+O(z)\bar{W}(\bar{z}).
\end{equation}
The $O$'s are ghost number zero spin zero fields needed for well defined
closed string currents. They build up the ``ground ring''. For a
discussion see \cite{wi1,zwi1}. 
As for $(\lambda \pm p )^{2J}$ there cannot be any disagreement. Clearly,
they give correlation functions of tachyon {\it states}. After all, as
explained in \cite{wi1}, there are no currents corresponding to the special
tachyons! The needed elements $O$ of the ground ring do not exist.
With this identification for the tachyons it would be very unnatural to
associate anything else than special {\it states} to the other
correlation functions of chapter 8.

Let us try to be more explicit. We will first look at a very simple
example involving the dilaton, and then proceed to a more general
case.

\section{A Dilaton Comparison}

There are some very simple examples of correlation functions easily
computable just using Liouville notions and no matrix model techniques.
These are correlation functions involving the dilaton. We will in fact
be able to obtain some results to all genus simply from dimensional
arguments. Consider the Liouville partition function (or space time free
energy)
\begin{equation}
E(\Delta ) = \lim _{R \rightarrow \infty} \frac{1}{R} \int {\cal D} X 
{\cal D} \phi
e^{-\frac{1}{4\pi}\int  (-2t_{2} \partial ^{a} X \partial _{a} X +
\frac{1}{2}\partial ^{a} \phi
\partial _{a} \phi -\frac{1}{2}QR\phi +\Delta e^{\alpha \phi})}, \label{path}
\end{equation}
where $t_{2}=-\frac{1}{2\alpha '}$, $Q=2\sqrt{2}$, $\alpha =-\sqrt{2}$ and $R$ is
the radius of the target space for the matter field $X$.
$\Delta$ is the world sheet cosmological constant, dimensionless from
the point of view of space time. The only dimensionful quantities are
$R$ and $\alpha '$. In the noncompact case we have in fact only $\alpha '$
at our disposal. From dimensional grounds and KPZ scaling we must have
\begin{equation}
E(\Delta )_{g} \sim (-t_{2})^{1/2} \Delta ^{2(1-g)} \label{E}
\end{equation}
at genus g. $E(\Delta )$ is the generator of connected amplitudes (in
space time). Let us perform a Legendre transform, as we did in chapter
seven, to obtain a generating
functional for 1PI amplitudes with respect to the puncture, i.e. the
zero momentum tachyon. This means taking away any pinches. We have
\begin{equation}
E(\Delta ) = \Delta \mu - \Gamma (\mu ),
\end{equation}
with 
\begin{equation}
\mu = \frac{\partial E}{\partial \Delta} = (-t_{2})^{1/2} (\Delta +...)
.
\end{equation}
$\mu$ has the dimensions of energy. Hence
\begin{equation}
\Gamma (\mu ) \sim (-t_{2})^{\frac{1}{2}(2g-1)} \mu ^{2-2g} \label{leg}
\end{equation}
In the 1PI generator $\Gamma$ we should of course regard $\mu$ as independent
of $t_{2}$.
Since $t_{2}$ derivatives should generate dilaton insertions we find the
following 1PI amplitude relation
\begin{equation}
<O_{2}...O_{2}>_{g} = \frac{1}{(2t_{2})^{n}} \prod _{p=1}^{n} (2g+1
-2p)<>_{g},
\end{equation}
which is identical to what was obtained in chapter seven, using the matrix model
recursion relations (generalizations of the zero momentum Wheeler de
Witt equation). As noted there the dilaton one point function
involves a factor $2g-1$ rather than the expected $2g-2$. From the above it
is clear that this discrepancy is simply due to including the overall
$(-t_{2})^{1/2}$ in (\ref{E}). 

In general we cannot get away with a simple trick like this, more work
is needed!

\section{More General Correlation Functions} 

To obtain the general special operator correlation function in the
Liouville theory one would like to use the group theoretic 
information provided by
the $W_{\infty}$ or, for given spin $J$, the $SU(2)$ symmetry.
The states in the Liouville theory are given by combinations
$W(z)\bar{W}(\bar{z})$ of the Liouville theory version of the special
states $W(z)$. Given this it is tempting to
believe that we have a representation of a 
$W_{\infty} \times W_{\infty}$ symmetry
(left times right). However, this is in general not correct. 
In the uncompactified
case the left and the right moving states must be the same.
The symmetry group is
broken down to just the diagonal subgroup.
This is achieved in two steps. First the gravitational dressing must be
the same for left and right, otherwise we would be unable to screen
using the cosmological constant which treats left and right in the same
way. This means that we always must have the same spin $J$ for left and
right. We get a reduction to the diagonal of the piece transverse to
$SU(2) \times SU(2)$. This is true even for the compactified case.
If we in addition are considering the
uncompactified case, the left and right moving momenta must be the same
and therefore also the $m$ quantum numbers. Consequently, we just have a
representation of the diagonal $W_{\infty}$. 
This is in precise agreement with the
matrix model, where we indeed only see one $W_{\infty}$. 

There is
however an apparent paradox here. If we would use the free field
contractions in computing the correlation functions the results would
seem to disagree since from this point of view left and right are still
independent. 
From such a
calculation you would expect to get a different result, all
Clebsch-Gordan coefficients squared, one from the left and one from the
right. We will however show that the results in the end turn out to be
consistent. 

Let us begin by considering the
tachyon correlation function as computed in \cite{frku}.
As we have seen the
result is in complete agreement with the matrix model results. On the
other hand we have seen how the matrix model organizes its correlation
functions using a {\em single} $W_{\infty}$. Let us consider the
Liouville calculation more carefully.
The result of \cite{frku} 
is obtained
through arguments of
analyticity and symmetry. In particular the by now well known factorized
product of gamma functions is found \cite{frku,gr5} 
with a certain unknown coefficient
independent of the particular momenta. This coefficient is then
determined by sending all the momenta, except three, to zero. This
reduces the expression to a three point function with $N-3$ extra
punctures. Since the three point function is possible to evaluate
directly, the
general result follows. The extra $N-3$ punctures is simply represented
as $\mu$ derivatives. This is the Liouville derivation of the expression
(\ref{tach}). 
The important point is that the use of a $\mu$ derivative for inserting
a puncture is a consequence of having just one $W_{\infty}$, see 
(\ref{puder})! This means
that the calculation in \cite{frku} automatically incorporates this
feature.

For the more general case with nontachyonic special states, we return to
the recursion relation (\ref{rek}). Let us redefine the fields according
to
\begin{equation}
W_{J,m}=\frac{4J}{2m-p} \tilde{W}_{J,m} .
\end{equation}
The recursion relation then takes the form
\begin{equation}
<\tilde{T}_{J,J}\tilde{W}_{J_{1},m_{1}}\prod _{i=2}^{N} 
\tilde{T}_{J_{i},J_{i}}>= 
\frac{(J_{1}-m_{1})(J+J_{1}-1)}{J_{1}} 
<\tilde{W}_{J+J_{1}-1,J+m_{1}}\prod_{i=2}^{N}
\tilde{T}_{J_{i},J_{i}}>  \label{tjohej}
\end{equation}
In \cite{kit} these very same recursion relations were obtained in the case
$J=m=1/2$ using Liouville methods. The coefficient in front of the right
hand side  was shown to be of the form $(2J_{1}-1)C^{2}$, where $C$
stands for the appropriate Clebsch-Gordan coefficient. The first factor
comes from comparing with the purely tachyonic case where the answer is
obtained from a simple Veneziano like integral. 
To see the agreement one uses the
Clebsch-Gordan coefficients of the special operator algebra as 
obtained in \cite{kle1}.
\begin{equation} \label{cg}
C_{J_{1},m_{1},J_{2},m_{2}}^{J_{3},m_{3}} = \frac{A(J_{3},m_{3})}
{A(J_{1},m_{1})A(J_{2},m_{2})}(J_{1}m_{2}-J_{2}m_{1}) ,
\end{equation} 
where
\begin{equation}
A(J,m)=-\frac{1}{2}\left[ (2J)!(J+m)!(J-m)!\right] ^{1/2} .
\end{equation}
At $J=m=1/2$ one finds $C^{2} = \frac{J_{1}-m_{1}}{2J_{1}}$ which then
leads to (\ref{tjohej}).
This is an important check on the equivalence between the Liouville and
matrix model approaches. An everywhere present difficulty in these
comparisons is, however, the fact that we are really sitting right on 
the momentum
poles. Clearly one needs to carefully regularize all expressions.

\section{Summary}

We have investigated the structure of special operator correlation
functions in $c=1$ quantum gravity. Due to the presence of a
$W_{\infty}$ symmetry the calculations become very simple. We have
also investigated the connection between the Liouville and the matrix
model, indicating the agreement for the correlation functions. 

An important point is the existence in the uncompactified Liouville as
well as matrix model formulation of c=1 of just {\em one} $W_{\infty}$.
From the Liouville point of view this is somewhat obscure since the
operator product expansion and its Clebsch-Gordan coefficients seem to
give a structure corresponding to {\em two} $W_{\infty}$'s. Fortunately,
the final outcome of the explicit calculations are identical. Further
work is however needed to establish the full equivalence. 

A part of the problem is that many of the calculations are so ill
defined.  The reason is that we are sitting right on the discrete
momentum poles. Especially in the Liouville theory this is a big
technical problem.
Often we must rely on guesswork concerning ill defined
analytical continuations. It is very doubtful if many of the results
would have been obtained correctly without knowing the answers in
advance, given by the much more powerful matrix model.

\chapter{Conclusions}

In this thesis we have studied the two dimensional noncritical string.
It is a remarkable example of a string theory where we are capable of
calculating everything. This is, of course, thanks to the matrix models.

The study of the two dimensional string theory has so far been of a
quite technical nature. The matrix model has been used at length to
extract all possible correlation functions and to study the spectrum.
One of the main results of this thesis is in fact a clear verification of
the presence of the special states in the matrix model.

Clearly, this should not be a goal in itself. After all, the reason why
we are studying this model in the first place, is that we want to learn
something about string theory and quantum gravity. There are several
different ways to do that. So far, the most important contribution of
the matrix model has been rather indirect. It has been to act as a
source of inspiration for new developments in field theory. As is clear
from many recent works, \cite{kac1,kl,ver1,wi1,zwi1}, 
there has been considerable progress in the
understanding of the field theory techniques. Perhaps, eventually, this
will help us in the study of higher dimensional theories. A somewhat
rebellious mind would perhaps argue that it is instead time to abandon
the old fashioned customs and stick with the matrix model. Unfortunately,
this is clearly not possible since the matrix models seem to have a
rather limited range of applicability. In particular there is the $c=1$
limit. It is however an important lesson to be learned, that there, in
some cases, can exist totally novel techniques which might lead to
exciting breakthroughs.

On the other hand, in the study of the specific two dimensional string
theory, remarkably little has been done with respect to the
understanding of the target space picture. Not only is the relation
between the Liouville mode and the matrix model eigenvalue rather
obscure (recall the relation through the loop length in the Wheeler de
Witt equation) we have in addition very little control over the target
space background. In the field theory approach we can choose to put in a
black hole etc. We would like to be able to exercise a similar control
in the matrix model. Only after we have learnt how to do this, we can
truly use the matrix model to learn about stringy quantum gravity.

There are also several issues regarding the back reaction of the
background geometry against the loop corrections to the
vacuum energy. The na\"{\i}ve loop corrected $\beta$ functions clearly
indicate that there should be a highly nontrivial influence.
From this point of view it is not clear how a flat space time can
remain a solution at loop level.
However, arguments in \cite{sei2} seem to indicate that the two
dimensional background geometry is quite rigid and not influenced in
this way, but rather frozen. If this is true, there is a risk that the
things we might learn which are relevant for higher dimensions, where
this certainly is {\it not} the case, are rather limited.

Other work point in a perhaps more promising direction,
\cite{horn,cghs,haw2,rus,suss}. There, a one loop corrected string
inspired field theory is studied for black hole solutions with
subsequent Hawking evaporation. An interesting elaboration would clearly
be to assume an underlying string theory and to study its consequences.
A better understanding of the matrix model could be of invaluable help
in this respect.

Clearly, there are still a lot of things which remain to be done in the
matrix model approach to the two dimensional string. We need to take
advantage of its simplicity and learn everything we can. But it is also
important to look forward, to more physical models. Although the matrix
model techniques as they are now seem to be insufficient for such a task,
there could clearly exist generalizations which might help us in our
quest.

The most important lesson we might learn from the matrix models is
the importance of keeping an open mind. The field theory
formulation might not be the only way of dealing with strings or
whatever it is that we feel deserves our attention. Finally, however, it
is important to keep the remarks made in the introduction in mind. We
should not become too carried away with the progress made. Even if our
mathematical capacity is continuously increasing, this does not mean
that our understanding of {\it physics} is any deeper.

String theory is an exciting possibility, but whether it is the right
answer only nature itself can tell.

\end{document}